\documentclass[11pt]{article}
\usepackage{axodraw}
\usepackage{epsf}
\newskip\humongous \humongous=0pt plus 1000pt minus 100pt
\def\caja{\mathsurround=0pt}
\def\eqalign#1{\,\vcenter{\openup1\jot \caja
       \ialign{\strut \hfil$\displaystyle{##}$&$
        \displaystyle{{}##}$\hfil\crcr#1\crcr}}\,}
\newif\ifdtup
\def\panorama{\global\dtuptrue \openup1\jot \caja
        \everycr{\noalign{\ifdtup \global\dtupfalse
        \vskip-\lineskiplimit \vskip\normallineskiplimit
        \else \penalty\interdisplaylinepenalty \fi}}}
\def\eqalignno#1{\panorama \tabskip=\humongous
        \halign to\displaywidth{\hfil$\displaystyle{##}$
        \tabskip=0pt&$\displaystyle{{}##}$\hfil
        \tabskip=\humongous&\llap{$##$}\tabskip=0pt
        \crcr#1\crcr}}
\newcounter{eqnumber}[section]
\renewcommand{\theeqnumber}{\thesection.\arabic{eqnumber}}
\def\equn{
\refstepcounter{eqnumber}
\eqno({\rm \theeqnumber})
}
\def\equnno{\refstepcounter{eqnumber}
({\rm \theeqnumber})}
\def\eqn#1{eq.~{(\ref{#1})}}

\def\eqns#1#2{eqs.~{(\ref{#1})} and {(\ref{#2})}}
\def\sect#1{sect.~{\ref{#1}}}

\def\fig#1{fig.~{\ref{#1}}}

\def\figs#1#2{figs.~{\ref{#1}} and {\ref{#2}}}

\def\hf{{\textstyle{1\over2}}}
\def\eps{\epsilon}

\def\Ord{{\cal O}}
\def\I{{\cal I}}
\def\A{{\cal A}}
\def\C{{\cal C}}
\def\R{{\cal R}}
\def\L{{\cal L}}

\def\bowtie{{\rm bow\mbox{-}tie}}
\def\D{{\rm D}}
\def\P{{\rm P}}
\def\Nf{n_{\! f}}
\def\NP{{\rm NP}}

\def\SUSY{{\rm SUSY}}
\def\QCD{{\rm QCD}}
\def\ggtogg{{gg \to gg}}
\def\tree{{\rm tree}}
\def\oneloop{{1 \mbox{-} \rm loop}}
\def\twoloop{{2 \mbox{-} \rm loop}}
\def\mud{\lambda}
\def\qb{{\bar q}}
\def\Tr{\, {\rm Tr}}

\def\la{\langle}
\def\ra{\rangle}
\def\e{\epsilon}

\def\tg{\lambda}

\def\SYM{{\scriptscriptstyle\rm SYM}}

\def\bom#1{{\mbox{\boldmath $#1$}}}

\def\f{\tilde f}
\def\spa#1.#2{\left\langle#1\,#2\right\rangle}
\def\spb#1.#2{\left[#1\,#2\right]}

\def\pol{\varepsilon}
\def\si{{\sigma}}

\def\ta{\tilde{a}}
\def\tb{\tilde{b}}
\def\MSbar{{\overline{\rm MS}}}
\def\DRbar{{\overline{\rm DR}}}
\def\FDHbar{{\overline{\rm FDH}}}
\def\yukawa{{\rm Yukawa}}
\def\tildeF{{\tilde F}}
\def\tildeS{{\tilde S}}

\begin{document}
\begin{titlepage}
\begin{flushright}
          hep-ph/0202271\\
          SLAC--PUB--9104\\
          UCLA/02/TEP/2\\
          February, 2002
\end{flushright}
\begin{center}
\begin{Large}
{\bf Supersymmetric Regularization, Two-Loop QCD Amplitudes \\ and
Coupling Shifts}
\end{Large}

\vskip 1.0cm

{Z. Bern$^\star$ \\
\it Department of Physics and Astronomy \\
UCLA, Los Angeles, CA 90095-1547}

\vskip 0.6cm

{A. De Freitas$^\star$\\
\it Department of Physics and Astronomy \\
UCLA, Los Angeles, CA 90095-1547}

\vskip 0.6cm

{L. Dixon$^\dagger$\\
\it Stanford Linear Accelerator Center\\
Stanford University\\
Stanford, CA 94309}

\vskip 0.4cm
and \vskip 0.4cm

{H.L. Wong$^\star$\\
\it Department of Physics and Astronomy \\
UCLA, Los Angeles, CA 90095-1547}
\end{center}

\vskip0.7cm
\begin{abstract}
We present a definition of the four-dimensional helicity (FDH)
regularization scheme valid for two or more loops.  This scheme
was previously defined and utilized at one loop. It amounts to a
variation on the standard 't~Hooft-Veltman scheme and is designed
to be compatible with the use of helicity states for ``observed''
particles.  It is similar to dimensional reduction in that it
maintains an equal number of bosonic and fermionic states, as
required for preserving supersymmetry.  Supersymmetry Ward
identities relate different helicity amplitudes in supersymmetric
theories. As a check that the FDH scheme preserves supersymmetry,
at least through two loops, we explicitly verify a number of these
identities for gluon-gluon scattering ($\ggtogg$) in
supersymmetric QCD. These results also cross-check recent
non-trivial two-loop calculations in ordinary QCD.  Finally, we
compute the two-loop shift between the FDH coupling and the
standard $\MSbar$ coupling, $\alpha_s$. The FDH shift is identical
to the one for dimensional reduction. The two-loop coupling shifts
are then used to obtain the three-loop QCD $\beta$ function in
the FDH and dimensional reduction schemes.
\end{abstract}


\vskip .3cm
\begin{center}
{\sl Submitted to Physical Review D}
\end{center}

\vfill \noindent\hrule width 3.6in\hfil\break
\begin{small}
${}^{\star}$Research supported by the US Department of Energy under grant
DE-FG03-91ER40662.\hfil\break
${}^{\dagger}$Research supported by the US Department of Energy under contract
DE-AC03-76SF00515.\hfil\break
\end{small}

\end{titlepage}

\baselineskip 16pt


\renewcommand{\thefootnote}{\arabic{footnote}}
\setcounter{footnote}{0}

\section{Introduction}

It is of great utility to regularize the divergences of quantum
field theory in a way that manifestly preserves the symmetries of
the theory.  The most widely utilized technique for preserving
gauge symmetry is dimensional regularization, which can
simultaneously handle the ultraviolet and infrared divergences of
massless gauge theory.  There are actually several common variants
of dimensional regularization, differing in their treatment of the
Lorentz vector indices associated with gauge particles. 
The different variants have advantages and disadvantages, depending on
the application.  The conventional dimensional regularization
scheme (CDR)~\cite{CDR} is conceptually the simplest and most
widely used variant.  In this scheme one uniformly continues all
momenta and vector polarizations to $D=4-2\e$ dimensions.  The
't~Hooft-Veltman (HV) scheme~\cite{HV} is closely related,
differing only in the treatment of ``observed'' states, which remain
in four dimensions. The HV scheme is especially convenient for
computing helicity amplitudes --- as has proven very useful in QCD
calculations~\cite{TreeHelicity,ManganoReview,BKgggg,Review} ---
because only the amplitudes with the four-dimensional helicity
values $\pm1$ (or 0, for massive vector bosons) have to be
evaluated.  For supersymmetric theories, the most commonly used
regularization scheme is Siegel's dimensional reduction (DR)
scheme~\cite{DR}, since it preserves supersymmetry. This scheme
has also found use in certain non-supersymmetric
calculations~\cite{NonSusyDimRed,JJR}.  Other useful schemes for
supersymmetric theories include the holomorphic and NSVZ
schemes~\cite{NSVZ}.

Another scheme that has been used at one loop is the
four-dimensional helicity (FDH) scheme\cite{BKgggg}. The FDH
scheme seeks to combine the natural use of helicity states (as in
the HV scheme) with the preservation of supersymmetry.  It is
similar to the DR scheme in maintaining the number of physical
states at their four-dimensional values.  In the FDH scheme,
however, the algebraic rules are defined with the notion that
$D>4$, but with an analytic continuation to bring the number of
physical states back to their $D=4$ values.  This may be
contrasted with the DR scheme, where one takes $D<4$, viewed as a
dimensional compactification.  The distinction between $D<4$ and $D>4$
is relevant only when an explicit basis of external states is 
required for a calculation; inside the loops in either case 
all indices on fields are treated as four-dimensional.
At one loop, the relationship between the FDH and DR schemes has been 
previously discussed in ref.~\cite{KSTfourparton}.

Preserving supersymmetry is useful even for higher order
calculations in QCD.  Although QCD itself is not a supersymmetric
theory, one can slightly modify QCD by altering the color
representations and multiplicities so that it becomes
supersymmetric.  The amplitudes in such a theory are closely
related to those in QCD, yet they must satisfy non-trivial
supersymmetry Ward identities~\cite{SWIOld,SWINew}, as long as the
regulator preserves supersymmetry.  These identities can therefore
provide an independent means for checking a non-trivial QCD
calculation.  The supersymmetry identities on scattering
amplitudes are phrased in terms of the helicity basis; thus the
FDH scheme is ideal for using supersymmetry in this way. Of course
one would also want to have available a simple way to convert
amplitudes computed in the standard variants of dimensional
regularization, such as the 't~Hooft-Veltman (HV) or conventional
dimensional regularization (CDR) schemes, into those computed in
the FDH scheme, or vice-versa.

Though we do not have a proof that the FDH scheme preserves
supersymmetry to all orders of perturbation theory, it inherently
maintains the number of physical states at their four-dimensional
values.  Moreover, from explicit calculations the scheme is known
to preserve supersymmetry at one loop~\cite{KSTfourparton}. In
this paper we shall explicitly verify its preservation at two
loops, by checking various supersymmetry identities involving
$\ggtogg$ helicity amplitudes.

The issue of higher-loop regularization is timely in light of the
recent substantial progress in the calculation of two-loop
scattering amplitudes.  Until recently, no such amplitudes
depending on more than a single kinematic variable were known. Now
several such computations have appeared.  The first of these were
gluon-gluon scattering amplitudes in the special cases of maximal
($N=4$) supersymmetry~\cite{BRY}, and for a particular gluon
helicity configuration in pure Yang-Mills
theory~\cite{AllPlusTwo}. Subsequently, complete calculations have
been performed for Bhabha scattering~\cite{Bhabha}, general
$2\rightarrow 2$ parton scattering in
QCD~\cite{GOTY2to2,GOTYgggg,ggggpaper}, the di-photon background
to Higgs production at the LHC~\cite{GGGamGam}, and light-by-light
scattering~\cite{Lbyl}.

Important technical breakthroughs, which allowed the more general
calculations to proceed, included the reduction of the two loop
momentum integrals appearing in the all-massless $2\to2$ processes
to a set of master integrals, and the evaluation of those master
integrals as a Laurent series in
$\e$~\cite{PBScalar,NPBScalar,PBReduction,NPBReduction,%
GRReduction,AGO}.  Even more recently, the corresponding integrals
where one of the four external legs is massive have been
evaluated~\cite{TwoloopOneMassIntegrals}, enabling the computation
of the two-loop amplitudes for $e^+e^-$ annihilation into three
partons~\cite{TwoLoopee3Jets}.

In our explicit verification of supersymmetry identities, we
investigate the $\ggtogg$ helicity amplitudes
$$
\A_4(1_g^\pm,2_g^+,3_g^+,4_g^+)\,,
\equn\label{MHVAmplitudes}
$$
where the subscript labels the particle species ($g$ for gluon),
and the superscript denotes the sign of the helicity. We use an
``all-outgoing'' helicity convention:  if a given leg is incoming,
then the actual helicity is the opposite of the superscript label.
We study the helicity amplitudes~(\ref{MHVAmplitudes}) because
they are relatively simple (for two-loop amplitudes), and because
supersymmetry is especially constraining. For both the ++++ and
$-$+++ helicity configurations, we have computed all the gluon and
fermion loop contributions that appear in QCD.  For the ++++
configuration, we have also included scalars with both gauge and
Yukawa interactions, to allow for a more extensive test of
supersymmetry identities.

It is important to be able to convert helicity amplitudes computed
in the FDH scheme to the more standard CDR and HV schemes, and
vice versa. A given dimensional regularization scheme has
implications for regularization of both ultraviolet and infrared
singularities. Both the ultraviolet and infrared aspects of scheme
conversion have been extensively discussed at one
loop~\cite{BKgggg,KSTfourparton,CST}, and to some degree at two
loops~\cite{ggggpaper}. The infrared aspects are not yet
understood for arbitrary processes, as only the $\ggtogg$ process
was studied in ref.~\cite{ggggpaper}.  Conversion from one scheme
to another in the ultraviolet, though, is a process-independent
procedure.  It just amounts to relating the two different
renormalized couplings implied by the two schemes, to a
sufficiently high accuracy in perturbation theory.  The FDH and
$\DRbar$ schemes behave the same in the ultraviolet; their
couplings are identical.  The relation between the $\DRbar$ (or
FDH) and $\MSbar$ versions of $\alpha_s$ in QCD has been known to
one-loop accuracy for some
time~\cite{NonSusyDimRed,AKT,MartinVaughn,KSTfourparton}. Here we
extend the relation to two-loop accuracy.

The first two coefficients of the QCD $\beta$ function are
scheme-independent (for analytic redefinitions of the coupling),
but the three-loop coefficient, $b_2$, depends on the scheme.
However, knowing the value of $b_2$ in the $\MSbar$
scheme~\cite{MSbarb2}, and the two-loop relation between
couplings, we can easily obtain the value of $b_2$ in the $\DRbar$
(or FDH) scheme.

This paper is organized as follows. In \sect{SWISection} we review
the supersymmetry Ward identities.  We then present the rules for
the FDH scheme in \sect{SusyRegularizationSection}.  In performing the
explicit two-loop calculations, we did not use Feynman diagrams directly.
Instead we computed (generalized) unitarity cuts of the amplitudes in all
channels and to all order in $\e$.  Then the amplitude was reconstructed
from the cuts.  This cutting method is described 
in~\sect{CuttingSection}.  In \sect{ColorSection} we perform a color
decomposition of the two-loop amplitude, into color structures multiplied
by ``primitive'' functions which have been stripped of color.  This
decomposition makes it convenient to obtain either QCD or supersymmetric
amplitudes.  The primitive amplitudes used in our explicit check
of the two-loop supersymmetry Ward identities are presented in
\sect{AmplitudesSection}.  We verify the identities in
\sect{SuperVerificationSection}.  In \sect{RenormalizationSection}
we discuss ultraviolet renormalization, the two-loop shifts in the
couplings and amplitudes, and their implications for the
three-loop $\beta$ function.

\section{Supersymmetry Ward identities}
\label{SWISection}

Helicity amplitudes in supersymmetric theories are subject to a
set of stringent conditions imposed by the super-algebra: the
$S$-matrix supersymmetry Ward identities
(SWI)~\cite{SWIOld,SWINew}. These identities allow us to distill
the information contained in the supersymmetry algebra and apply
it directly to on-shell $S$-matrix elements. They hold in any
supersymmetric theory.  They also lead to relations among
different components of amplitudes in any non-supersymmetric
theory, such as QCD, for which a re-adjustment of color
representations and/or multiplicities makes the theory
supersymmetric.  At tree level and at one loop, supersymmetry Ward
identities have been applied to QCD, either as checks or as
computational
aids~\cite{SWINew,TasiZvi,KSTfourparton,AllNSusy,Review}.  As we
discuss in this paper, the same ideas can be applied at two loops.

\subsection{Derivation}
\label{SWIDerivationSubsection}

The derivation of the supersymmetry Ward identities from the
super-algebra has been discussed in the literature in a number of
articles and reviews~\cite{SWIOld,ManganoReview}, so we describe
it only briefly. Since we are interested in applications to
non-supersymmetric theories we phrase the supersymmetry identities
in terms of the component fields.  The $N=1$ super-algebra
describing the action of the super-charge on the component gluon
field $g$ and gluino field $\tg$ comprising the vector
supermultiplet is
$$
[Q(p), g^\pm(k)] = \mp \Gamma^\pm(k,p)\, \tg^\pm(k) \,,
\hskip2cm
[Q(p), \tg^\pm(k)] = \mp \Gamma^\mp(k,p)\, g^\pm(k) \,,
\equn\label{Commutator}
$$
where $k$ is the light-like momentum carried by the field, $Q(p)$
is the super-charge contracted with a spinor for the (arbitrary)
light-like vector $p$, and
$$
\Gamma^+ (k,p) = \bar\theta \spb{p}.{k} \,,
\hskip2cm
\Gamma^-(k,p) = \theta \spa{p}.{k} \,.
$$
Because $\Gamma^-$ is proportional to a Grassmann variable
$\theta$ and $\Gamma^+$ is proportional to $\bar \theta$, the
coefficients $\Gamma^+$ and $\Gamma^-$ in a supersymmetry Ward
identity are independent. We use the notation $\la k_i^- |k_j^+\ra
= \spa{i}.j$ and $\la k_i^+| k_j^-\ra=\spb{i}.j$, where
$|k_i^{\pm}\ra$ are massless Weyl spinors with momentum $k_i$,
labeled with the sign of the helicity and normalized by $\spa{i}.j
\spb{j}.i = s_{ij} = 2k_i\cdot k_j$.  Note that $\spa{i}.{i} =
\spb{i}.{i} = 0$.

Since the super-charge $Q(p)$ annihilates the vacuum,
one can construct a typical $N=1$ supersymmetry
Ward identity in the following way:
$$
\eqalign{
0 = \langle 0| [Q,g_1^\pm g_2^+ \tg_3^+ g_4^+] |0\rangle
= & \mp \Gamma^\pm(k_1,p)
\A_4^\SUSY(1_\tg^\pm,2_g^+,3_\tg^+,4_g^+) 
- \Gamma^+(k_2,p)\A_4^\SUSY(1_g^\pm ,2_\tg^+,3_\tg^+,4_g^+) 
\cr &
- \Gamma^-(k_3,p) \A_4^\SUSY(1_g^\pm,2_g^+,3_g^+,4_g^+) 
+ \Gamma^+(k_4,p) \A_4^\SUSY(1_g^\pm,2_g^+,3_\tg^+,4_\tg^+) \,,
}\equn\label{LongSWI}
$$
where $\A_4^\SUSY$ is a four-point amplitude in a supersymmetric
theory, the integers refer to the leg labels, and the subscripts
$g$ and $\tg$ to the particle species of the specified leg. Since
the coefficients of $\Gamma^+$ and $\Gamma^-$ are independent, the
sum of the terms with $\Gamma^-$ prefactors must vanish independently. 
(The $\Gamma^+$ terms in~\eqn{LongSWI} also vanish using gluino 
helicity conservation.)  By choosing $p=k_1$ we obtain the identity,
$$
\A_4^\SUSY(1_g^\pm,2_g^+,3_g^+,4_g^+) = 0\,,
\equn\label{AllPlusSusyIdentity}
$$
which is the main SWI that we will investigate at two loops in this
paper.  

Using the super-algebra, one may systematically derive other
identities.  The nonvanishing amplitudes with external gluons only are 
related to amplitudes containing external gluinos, {\it e.g.},
$$
\eqalign{
\A_4^\SUSY(1_g^-,2_\tg^-,3_\tg^+,4_g^+)\ =&\ {\spa1.3 \over \spa1.2} \,
\A_4^\SUSY(1_g^-,2_g^-,3_g^+,4_g^+) \,, \cr
\A_4^\SUSY(1_\tg^-,2_\tg^+,3_\tg^-,4_\tg^+)\ =&\ {\spa2.4 \over \spa1.3} \,
\A_4^\SUSY(1_g^-,2_g^+,3_g^-,4_g^+) \,. \cr}
\equn\label{Susymmpp}
$$
The amplitudes on the right-hand side of~\eqn{Susymmpp} have recently 
been computed at two loops~\cite{ggggpaper}.  The amplitudes on the
left-hand side are related to the QCD amplitudes for $q\bar{q} \to gg$,
$qg\to qg$, $q\bar{q} \to q\bar{q}$, and $qq\to qq$, but these amplitudes
have not yet been computed at two loops in the helicity 
formalism,\footnote{They have been computed, and the SWI have been
verified, at one loop~\cite{KSTfourparton}.  At two loops, the
interference with the tree amplitude, summed over all external colors and
helicities, has been computed~\cite{GOTY2to2}, but the conversion to
a supersymmetric amplitude has not yet been performed.} and we shall 
not do so here.

\subsection{Lagrangian}
\label{LagrangianSubsection}

To allow us to separate out different supersymmetric combinations of 
amplitudes, we consider $N=1$ supersymmetric $SU(N_c)$ gauge theory
with two different types of matter content:
\begin{enumerate}
\item $\Nf$ identical chiral matter multiplets $Q_i$, $i=1,\ldots,\Nf$,
transforming in the {\bf fundamental} $N_c$ representation of the gauge group,
and their $\bar{N}_c$ partners $\tilde{Q}_i$, with vanishing superpotential,
$W = 0$.
\item A single chiral matter multiplet $\Phi \equiv \Phi^a T^a$
transforming in the {\bf adjoint} representation, with superpotential 
$W = {1\over3} g \xi \Tr \Phi^3$, where $\Tr$ is an $SU(N_c)$ trace. 
For convenience, we write the Yukawa coupling as $g \xi$, and take $\xi$ 
to be independent of the gauge coupling $g$.
\end{enumerate}
Our $SU(N_c)$ generators are normalized by $\Tr(T^aT^b)=\delta^{ab}$.  
Because of the gauge symmetry, gluons couple to all other particles via
the standard gauge theory interaction for fermions or scalars, with 
gauge coupling $g$. The couplings of gluinos or matter fermions may 
be conveniently extracted from the component expansion of the interaction 
Lagrangian~\cite{SuperLagrangian}.

The interaction terms in the case of fundamental matter are
$$
\eqalign{
\L_g^{\rm fund} =&\   i g \sum_{i=1}^{\Nf} \biggl(
    ( A_i^* T^a \psi_i -\tilde{\psi}_i  T^a \tilde{A}_i^* ) \lambda^a 
  - \bar\lambda^a ( A_i T^a \bar\psi_i 
                  -  \bar{\tilde{\psi}}_i T^a \tilde{A}_i ) \biggr) \cr
& -  {g^2 \over 4} \biggl( \sum_{i=1}^{\Nf} 
              (  A_i^* T^a A_i - \tilde{A}_i T^a \tilde{A}_i^* ) \biggr)^2
\,, \cr}
\equn\label{LgFund}
$$
following the notation of Wess and Bagger~\cite{WessBagger}, 
chapter 7.  Here $\lambda \equiv \lambda^a T^a$ is the gluino; 
$A_i$ and $\psi_i$ are, respectively, the scalar and fermionic 
components of $Q_i$,  while $\tilde{A}_i$ and $\tilde{\psi}_i$ are 
the corresponding components of $\tilde{Q}_i$.

In the case of adjoint matter, the interaction terms proportional to the
gauge coupling are,
$$
\eqalign{
\L_g^{\rm adjoint} & = 
      - ig \, \Tr\bigl( [A^* ,\psi] \lambda \bigr)
      + ig \, \Tr\bigl( [A,\bar\psi] \bar\lambda \bigr)
      - {g^2\over4} \sum_a \Tr \bigl( [A,  A^*] T^a \bigr)^2 \cr
  & = - ig \, \Tr\bigl( [A^* ,\psi] \lambda \bigr)
      + ig \, \Tr\bigl( [A,\bar\psi] \bar\lambda \bigr)
       - {g^2\over2} \Tr \bigl( A  A^* A  A^* -  A^2 {A^*}^2 \bigr)
 \,. \cr}
\equn\label{Lg}
$$
The terms arising from the superpotential $W$ are
$$
\eqalign{
\L_\xi^{\rm adjoint} &= - g \xi \, \Tr \psi\psi A
             - g \xi^* \, \Tr \bar\psi\bar\psi A^*
             - g^2 |\xi|^2 \sum_a \Tr (A^2 T^a) \Tr ({A^*}^2 T^a) \cr
 &= - g \xi \, \Tr \psi\psi A
             - g \xi^* \, \Tr \bar\psi\bar\psi A^*
             - g^2 |\xi|^2 \biggl[ \Tr A^2 {A^*}^2
          - {1\over N_c} \Tr A^2 \Tr {A^*}^2 \biggr] \,. \cr}
\equn\label{Lxi}
$$

\subsection{Amplitude decomposition by particle content}
\label{ParticleContentSubsection}

Supersymmetry can be exploited in non-supersymmetric theories such as
QCD~\cite{SWINew,ManganoReview,Review} by observing that appropriate
linear combinations of quantities occurring in QCD amplitudes are in
fact supersymmetric.  As an especially simple example, at tree level
an $n$-gluon amplitude is automatically
supersymmetric~\cite{SWINew}. Since the fermions do not appear in
intermediate states, the fermions in the theory might as well be in
the adjoint representation, {\it i.e.}, the theory might as well be 
pure $N=1$ super-Yang-Mills theory. Thus, by virtue
of~\eqn{AllPlusSusyIdentity}, the four-gluon tree amplitude in QCD 
satisfies,
$$
\A_4^{\QCD\ \rm tree}(1_g^\pm, 2_g^+, 3_g^+, 4_g^+) = 0 \,.
\equn\label{QCDTreeSusyIdentity}
$$

At loop level the application of supersymmetry in QCD is clearly more
intricate, because all particles in the theory can circulate in the
loops.  Nevertheless, one can still use supersymmetry to relate
different contributions.  

\begin{figure}[ht]
\begin{center}
\begin{picture}(295,85)(0,0)
\Gluon(05,05)(25,25){2.5}{3}
\Gluon(80,05)(60,25){2.5}{3}
\Gluon(80,80)(60,60){2.5}{3}
\Gluon(05,80)(25,60){2.5}{3}
\Gluon(60,25)(25,25){2.5}{3}
\Gluon(60,60)(60,25){2.5}{3}
\Gluon(25,60)(60,60){2.5}{3}
\Gluon(25,25)(25,60){2.5}{3}
\Text(43,0)[c]{\large (a)}
\ArrowLine(165,25)(130,25)
\ArrowLine(165,60)(165,25)
\ArrowLine(130,60)(165,60)
\ArrowLine(130,25)(130,60)
\Gluon(110,05)(130,25){2.5}{3}
\Gluon(185,05)(165,25){2.5}{3}
\Gluon(185,80)(165,60){2.5}{3}
\Gluon(110,80)(130,60){2.5}{3}
\Text(147,0)[c]{\large (b)}
\DashLine(270,25)(235,25){3.1}
\DashLine(270,60)(270,25){3.1}
\DashLine(235,60)(270,60){3.1}
\DashLine(235,25)(235,60){3.1}
\Gluon(215,05)(235,25){2.5}{3}
\Gluon(290,05)(270,25){2.5}{3}
\Gluon(290,80)(270,60){2.5}{3}
\Gluon(215,80)(235,60){2.5}{3}
\Text(251,0)[c]{\large (c)}
\end{picture}
\caption[]{
\label{OneLoopSusyFigure}
\small Sample diagrams contributing to: (a) a gluon circulating
in the loop, (b) a fermion circulating in the loop, and (c) a scalar
circulating in the loop.}
\end{center}
\end{figure}
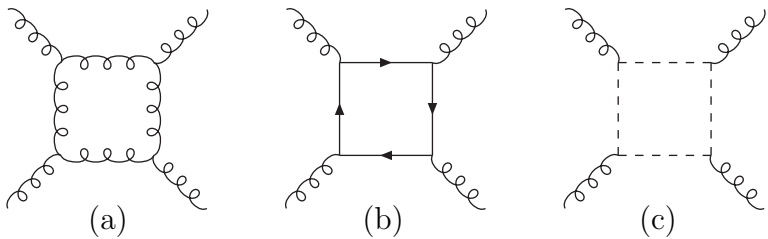

At one loop, the four-gluon amplitude in supersymmetric QCD with $\Nf$
matter multiplets --- case 1 in \sect{LagrangianSubsection} ---
has the structure,
$$
\A^{N=1}_4 = \A_4^{\rm vector} +  \Nf \A_4^{\rm matter} \,, 
\equn\label{N=1SchematicAmplitudeOneLoop}
$$
where the contribution $\A_4^{\rm vector}$ is that of a vector multiplet
consisting of a gluon and gluino, and $\A_4^{\rm matter}$ is that of
a matter multiplet consisting of a quark and squark.
Identities may be obtained from each term in this 
decomposition.  For example, by setting $\Nf=0$ 
in~\eqn{N=1SchematicAmplitudeOneLoop}, corresponding to 
pure $N=1$ super-Yang-Mills theory, from the SWI~(\ref{AllPlusSusyIdentity}) 
we have,
$$
\A^{\rm gluon\ loop}_{4}(1_g^\pm, 2_g^+, 3_g^+, 4_g^+)  =
 - \A^{\rm gluino\ loop}_{4}(1_g^\pm, 2_g^+, 3_g^+, 4_g^+) \,,
\equn
$$
which relates the gluon loop contribution depicted in
\fig{OneLoopSusyFigure}(a) to the fermion loop contribution depicted
in \fig{OneLoopSusyFigure}(b).  To apply this identity to QCD,
one uses group theory to relate the contribution of an
adjoint representation fermion (gluino) in the loop to that of a
fundamental representation fermion (quark), as we shall describe
further in section~\ref{ColorTreeOneLoopSubsection}.

Similarly, by considering the contribution from an $N=1$ matter
multiplet, consisting of a fermion and a scalar transforming in the same
representation of the gauge group, {\it i.e.}, the $\Nf$-dependent term
in~\eqn{N=1SchematicAmplitudeOneLoop}, we obtain
$$
\A^{\rm scalar\ loop}_{4}(1_g^\pm, 2_g^+, 3_g^+, 4_g^+)  =
- \A^{\rm fermion\ loop}_{4}(1_g^\pm, 2_g^+, 3_g^+, 4_g^+) \,,
\equn\label{ColorfulScalarFermion}
$$
which relates the fermion loop contribution in \fig{OneLoopSusyFigure}(b)
to the scalar loop contribution in \fig{OneLoopSusyFigure}(c).
By supersymmetry, the number of physical states in the scalar loop
matches that in the fermion loop.  For a scalar in the adjoint representation 
of $SU(N_c)$, paired with a Majorana gluino (in $N=2$ supersymmetry, say),
this amounts to $2 \times (N_c^2-1)$ states.  For a fundamental 
representation scalar, paired with a quark (a Dirac fermion),
it amounts to $4 \times N_c$ states.  (At the one-loop level, 
considering adjoint matter with a superpotential --- case 2 in
\sect{LagrangianSubsection} --- leads to no new identities from 
the four-gluon amplitude.)

%
\vskip .2 cm

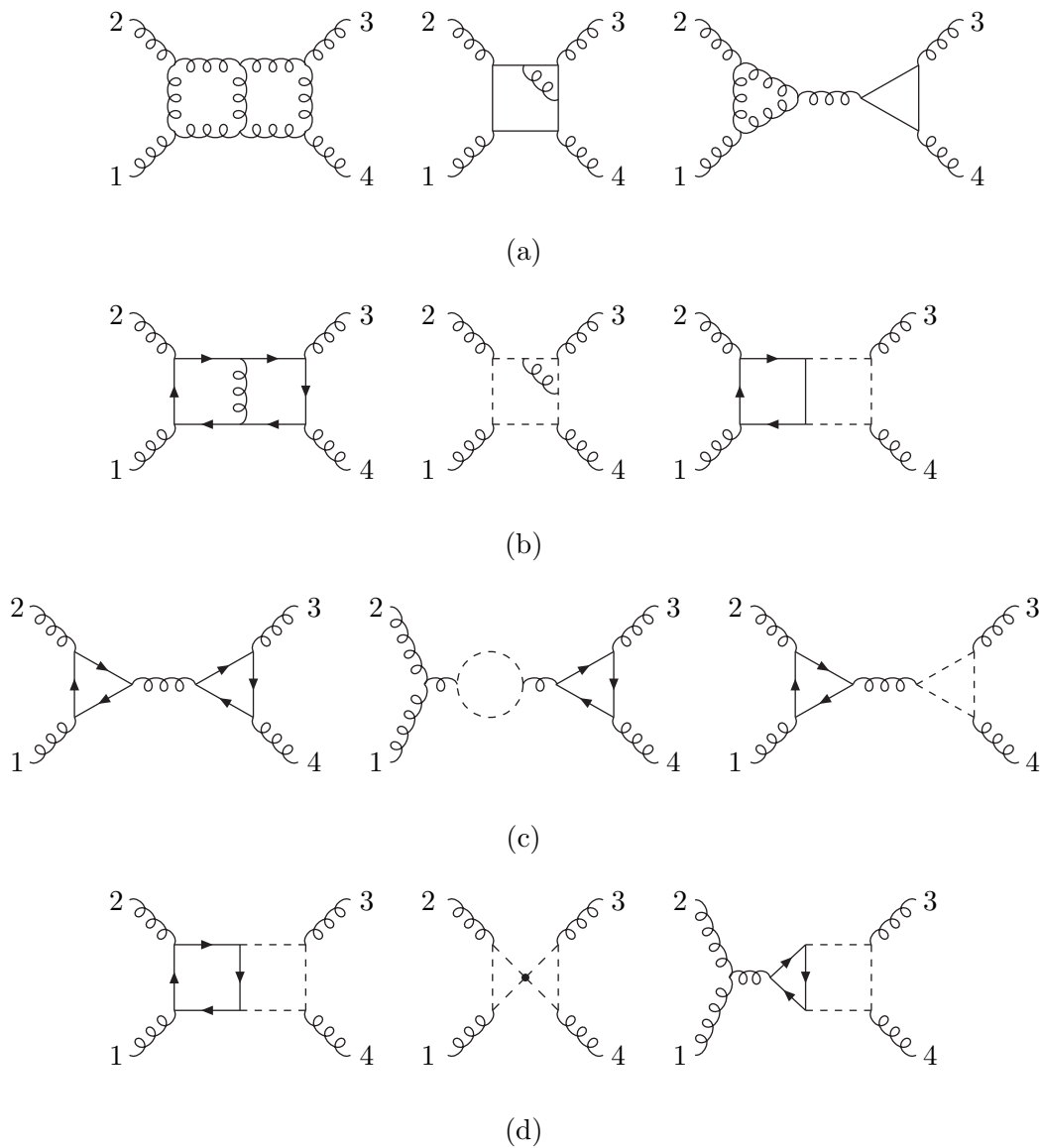
\begin{figure}[ht]
\begin{center}
\begin{picture}(394,69)(0,0)
\Gluon(102,22)(77,22){2.5}{3} \Gluon(77,47)(102,47){2.5}{3}
\Gluon(77,22)(52,22){2.5}{3}\Gluon(52,47)(77,47){2.5}{3}
\Gluon(52,22)(52,47){2.5}{3} \Gluon(77,47)(77,22){2.5}{3}
  \Gluon(102,47)(102,22){2.5}{3}
\Gluon(52,22)(35,5){2.5}{3} \Gluon(35,64)(52,47){2.5}{3}
\Gluon(119,5)(102,22){2.5}{3} \Gluon(102,47)(119,64){2.5}{3}
\Text(33,5)[r]{1} \Text(33,64)[r]{2}
\Text(123,64)[l]{3} \Text(123,5)[l]{4}
\Line(198,22)(173,22)
 \Line(173,47)(198,47)
 \Line(173,22)(173,47)
  \Line(198,47)(198,22)
\Gluon(198,34)(185,47){2.5}{2}
\Gluon(173,22)(156,5){2.5}{3} \Gluon(156,64)(173,47){2.5}{3}
\Gluon(215,5)(198,22){2.5}{3} \Gluon(198,47)(215,64){2.5}{3}
\Text(152,5)[r]{1} \Text(152,64)[r]{2}
\Text(219,64)[l]{3} \Text(219,5)[l]{4}
\Text(248,5)[r]{1} \Text(248,64)[r]{2}
\Text(356,64)[l]{3} \Text(356,5)[l]{4}
\Gluon(267,22)(267,47){2.5}{3} \Gluon(267,47)(289,34.5){2.5}{3}
\Gluon(289,34.5)(267,22){2.5}{3}
\Line(335,47)(335,22) \Line(313,34.5)(335,22)
 \Line(335,47)(313,34.5)
\Gluon(289,34.5)(313,34.5){2.5}{3}
\Gluon(267,22)(250,5){2.5}{3} \Gluon(250,64)(267,47){2.5}{3}
\Gluon(352,5)(335,22){2.5}{3} \Gluon(335,47)(352,64){2.5}{3}
\end{picture}
\\
 \null \hskip -.8 cm (a)
\end{center}
\begin{center}
\begin{picture}(394,69)(0,0)
\ArrowLine(102,22)(77,22) \ArrowLine(77,47)(102,47)
\ArrowLine(77,22)(52,22) \ArrowLine(52,47)(77,47)
\ArrowLine(52,22)(52,47) \Gluon(77,47)(77,22){2.5}{3}
  \ArrowLine(102,47)(102,22)
\Gluon(52,22)(35,5){2.5}{3} \Gluon(35,64)(52,47){2.5}{3}
\Gluon(119,5)(102,22){2.5}{3} \Gluon(102,47)(119,64){2.5}{3}
\Text(33,5)[r]{1} \Text(33,64)[r]{2}
\Text(123,64)[l]{3} \Text(123,5)[l]{4}
\DashLine(198,22)(173,22){3.1}
 \DashLine(173,47)(198,47){3.1}
 \DashLine(173,22)(173,47){3.1}
  \DashLine(198,47)(198,22){3.1}
\Gluon(198,34)(185,47){2.5}{2}
\Gluon(173,22)(156,5){2.5}{3} \Gluon(156,64)(173,47){2.5}{3}
\Gluon(215,5)(198,22){2.5}{3} \Gluon(198,47)(215,64){2.5}{3}
\Text(152,5)[r]{1} \Text(152,64)[r]{2}
\Text(219,64)[l]{3} \Text(219,5)[l]{4}
\ArrowLine(292,22)(267,22) \ArrowLine(267,47)(292,47)
\DashLine(292,22)(317,22){3.1} \DashLine(292,47)(317,47){3.1}
\ArrowLine(267,22)(267,47) \Line(292,22)(292,47)
  \DashLine(317,47)(317,22){3.1}
\Gluon(267,22)(250,5){2.5}{3} \Gluon(250,64)(267,47){2.5}{3}
\Gluon(334,5)(317,22){2.5}{3} \Gluon(317,47)(334,64){2.5}{3}
\Text(248,5)[r]{1} \Text(248,64)[r]{2}
\Text(338,64)[l]{3} \Text(338,5)[l]{4}
\end{picture}
\\ \null \hskip -.8 cm (b)
\end{center}
\begin{center}
\begin{picture}(470,69)(0,0)
\Text(33,5)[r]{1} \Text(33,64)[r]{2}
\Text(141,64)[l]{3} \Text(141,5)[l]{4}
\ArrowLine(52,22)(52,47) \ArrowLine(52,47)(74,34.5)
\ArrowLine(74,34.5)(52,22)
\ArrowLine(120,47)(120,22)\ArrowLine(120,22)(98,34.5)
 \ArrowLine(98,34.5)(120,47)
\Gluon(74,34.5)(98,34.5){2.5}{3}
\Gluon(52,22)(35,5){2.5}{3} \Gluon(35,64)(52,47){2.5}{3}
\Gluon(137,5)(120,22){2.5}{3} \Gluon(120,47)(137,64){2.5}{3}
\Text(170,5)[r]{1} \Text(170,64)[r]{2}
\Text(278,64)[l]{3} \Text(278,5)[l]{4}
\ArrowLine(257,47)(257,22) \ArrowLine(257,22)(235,34.5)
\ArrowLine(235,34.5)(257,47)
\Gluon(222.5,34.5)(235.5,34.5){2.5}{1}
\Gluon(185,34.5)(197.5,34.5){2.5}{1}
\DashCArc(210,34.5)(12.5,0,360){3}
\Gluon(185,34.5)(172,5){2.5}{4} \Gluon(172,64)(185,34.5){2.5}{4}
\Gluon(274,5)(257,22){2.5}{3} \Gluon(257,47)(274,64){2.5}{3}
\Text(307,5)[r]{1} \Text(307,64)[r]{2}
\Text(415,64)[l]{3} \Text(415,5)[l]{4}
\ArrowLine(326,22)(326,47)  \ArrowLine(326,47)(348,34.5)
\ArrowLine(348,34.5)(326,22)
\DashLine(394,47)(394,22){3} \DashLine(372,34.5)(394,22){3}
 \DashLine(394,47)(372,34.5){3}
\Gluon(348,34.5)(372,34.5){2.5}{3}
\Gluon(326,22)(309,5){2.5}{3} \Gluon(309,64)(326,47){2.5}{3}
\Gluon(411,5)(394,22){2.5}{3} \Gluon(394,47)(411,64){2.5}{3}
\end{picture}
\\  \null \hskip -.8 cm (c)
\end{center}
\begin{center}
\begin{picture}(394,69)(0,0)
\DashLine(102,22)(77,22){3} \DashLine(77,47)(102,47){3}
\ArrowLine(77,22)(52,22)\ArrowLine(52,47)(77,47)
\ArrowLine(52,22)(52,47) \ArrowLine(77,47)(77,22)
  \DashLine(102,47)(102,22){3}
\Gluon(52,22)(35,5){2.5}{3} \Gluon(35,64)(52,47){2.5}{3}
\Gluon(119,5)(102,22){2.5}{3} \Gluon(102,47)(119,64){2.5}{3}
\Text(33,5)[r]{1} \Text(33,64)[r]{2}
\Text(123,64)[l]{3} \Text(123,5)[l]{4}
\DashLine(173,22)(173,47){3}
\DashLine(198,47)(198,22){3}
\DashLine(198,47)(185.5,34.5){3.5}
\DashLine(173,22)(185.5,34.5){3.5}
\DashLine(173,47)(185.5,34.5){3.5}
\DashLine(198,22)(185.5,34.5){3.5}
\Vertex(185.5,34.5){1.5}
\Gluon(173,22)(156,5){2.5}{3} \Gluon(156,64)(173,47){2.5}{3}
\Gluon(215,5)(198,22){2.5}{3} \Gluon(198,47)(215,64){2.5}{3}
\Text(152,5)[r]{1} \Text(152,64)[r]{2}
\Text(219,64)[l]{3} \Text(219,5)[l]{4}
\Text(248,5)[r]{1} \Text(248,64)[r]{2}
\Text(338,64)[l]{3} \Text(338,5)[l]{4}
\DashLine(292,22)(317,22){3.1} \DashLine(292,47)(317,47){3.1}
\DashLine(317,22)(317,47){3.1}
\ArrowLine(292,47)(292,22)
\ArrowLine(278.5,34.5)(292,47)
\ArrowLine(292,22)(278.5,34.5)
\Gluon(263,34.5)(278.5,34.5){2.5}{2}
\Gluon(263,34.5)(250,5){2.5}{4} \Gluon(250,64)(263,34.5){2.5}{4}
\Gluon(334,5)(317,22){2.5}{3} \Gluon(317,47)(334,64){2.5}{3}
\end{picture}
\\ \null \hskip -.8 cm (d)
\caption[]{ \label{TwoLoopSusyFigure} \small Representative
diagrams contributing to the supersymmetric amplitudes (a)
$\A_4^{\rm vector}$, (b) $\A_4^{\rm matter (1)}$, 
(c) $\A_4^{\rm matter (2)}$, and (d) $\A_4^\yukawa$ 
in~\eqns{N=1fundSchematicAmplitude}{N=1adjSchematicAmplitude}. 
The curly, solid without arrows, solid with arrows, and dotted lines 
represent gluons, gluinos, quarks, and scalars, respectively.}
\end{center}
\end{figure}

These considerations extend to two loops.
In the case of $\Nf$ matter multiplets in the fundamental representation
--- case 1 in \sect{LagrangianSubsection} ---
the four-gluon amplitude takes the form,
$$
\A^{N=1,\,{\rm fund}}_4 = \A_4^{\rm vector} +  \Nf \A_4^{\rm matter (1)} 
+ \Nf^2 \A_4^{\rm matter (2)} \,.
\equn\label{N=1fundSchematicAmplitude}
$$
In the case of one adjoint matter multiplet with a superpotential
--- case 2 in \sect{LagrangianSubsection} ---
the four-gluon amplitude takes the form,
$$
\A^{N=1\,{\rm adj}}_4 = \A_4^{\rm vector} +  \A_4^{\rm matter (1)} 
+ \A_4^{\rm matter (2)} + |\xi|^2 \A_4^\yukawa \,. 
\equn\label{N=1adjSchematicAmplitude}
$$
The quantities $\A_4^{{\rm matter}(i)}$ are ``dressed'' differently with
color in the two cases, as will be explained in~\sect{ColorSection}.
We have subdivided the contributions according to their dependence on 
the number of matter multiplets and on the couplings.  
Identities may be obtained from each term in the 
decomposition~(\ref{N=1fundSchematicAmplitude}), and from the
$\xi$-dependent term in \eqn{N=1adjSchematicAmplitude},
because $\Nf$ and $\xi$ are independent parameters.  
Representative diagrams contributing 
to each of the four independent supersymmetric components 
$\A_4^{\rm vector}$, $\A_4^{\rm matter (1)}$, $\A_4^{\rm matter (2)}$ and 
$\A_4^\yukawa$ are depicted in figs.~\ref{TwoLoopSusyFigure}(a)--(d).

For the helicity configurations ++++ and $-$+++, 
four independent identities are obtained by
setting~\eqns{N=1fundSchematicAmplitude}{N=1adjSchematicAmplitude}
to vanish using the 
SWI~(\ref{AllPlusSusyIdentity}).  Each such identity actually generates
several equations, once the amplitudes are further decomposed according to
the colors of the external gluons.  The equations stemming from 
$\A^{\rm vector}(\pm,+,+,+) = 0$, which only involve gluons and 
gluinos, can be applied to QCD once one understands the group theory
relations at two loops between the contribution of fundamental and adjoint
representation fermions.  The remaining identities have scalar 
particles propagating in the loops as well.
A systematic discussion of all these relations,
based on ``primitive'' or ``color-stripped'' amplitudes, will be presented in
sections~\ref{ColorSection}-\ref{SuperVerificationSection}.

\section{Supersymmetric regularization}
\label{SusyRegularizationSection}

For supersymmetry identities to hold directly, a necessary condition is
that the regularization should not alter the number of bosonic states
relative to the number of fermionic states.  The conventional and
't~Hooft-Veltman variants of dimensional regularization~\cite{CDR,HV} are
incompatible with supersymmetry precisely because they alter the balance
of bosonic and fermionic degrees of freedom.  In the four-dimensional
limit, manifest supersymmetry will not generally be recovered in these
schemes due to divergences in the amplitudes.   An $\Ord(\eps)$ discrepancy
between the number of bosonic and fermion states can be multiplied by an
$1/\eps$ singularity, leaving a supersymmetry-violating remainder even as
$\eps \rightarrow 0$.  If the only divergences are ultraviolet in nature,
it is possible to repair such violations by adding suitable finite
counterterms, order by order in perturbation theory 
(see {\it e.g.} ref.~\cite{MartinVaughn}).  However, it is
clearly desirable to avoid this situation if possible, particularly in
theories with a large number of coupling constants.  Also, we are 
interested in on-shell scattering amplitudes for theories with severe 
infrared divergences, and here local counterterms will not suffice.

\subsection{Dimensional reduction scheme}
\label{DimredSubsection}

The most widely used scheme for preserving supersymmetry is the
dimensional reduction (DR)~\cite{DR} variant of dimensional 
regularization.\footnote{As is customary in essentially all variants of 
dimensional regularization, we treat the fermions as four-dimensional
by letting the Dirac trace of the identity be ${\rm tr}(1) = 4$.} 
The rules for dimensional reduction follow from viewing it as a
compactification of a four-dimensional theory to $D<4$.  The rules 
for dimensional reduction are~\cite{Superspace}:
\begin{itemize}

\item As in ordinary dimensional regularization, all momentum integrals
are integrated over $D$-component momenta. Any Kronecker
$\delta_\mu^{~\nu}$'s resulting from the integration are $D$-dimensional.
(This is necessary for maintaining gauge invariance.)

\item All indices on the fields, and on corresponding matrices
coming from the action, are treated as four-dimensional indices.

\item Since $D<4$ always, any four-dimensional Kronecker
$\delta_\mu^{(4) \nu}$ contracted with a $D$-dimension momentum
$p_\nu^{(D)}$ yields a $D$-dimensional momentum, $\delta_\mu^{(4)\nu}
p_\nu^{(D)} = p_\mu^{(D)}$.  Similarly, for any four-dimensional
vector, $\pol^\mu_{(4)}$, the dot product with a $D$-dimensional
vector yields a $D$-dimensional dot product, $\pol^\mu_{(4)}
p^{(D)}_\mu = \pol^\mu_{(D)} p^{(D)}_\mu$.  In general, dot products
of four-dimensional vectors with $2\eps$-dimensional ones will not
vanish.

\end{itemize}

The first rule is necessary for preserving gauge invariance and the
second for preserving supersymmetry.  It is the third rule which
defines the regularization as dimensional reduction.
In the DR scheme a four-dimensional vector may be
viewed as a combination of the $D=4-2\eps$ vector plus a set of
$2\eps$ scalars.  In non-supersymmetric theories in the DR scheme
it is especially important to keep track of the distinction between
the vectors and the $2\eps$ scalars because of their differing
renormalization properties.  Moreover, in the non-supersymmetric case
it is essential to keep evanescent couplings and operators~\cite{JJR}.

The rule $\pol^\mu_{(4)} p^{(D)}_\mu = \pol^\mu_{(D)} p^{(D)}_\mu$
is also awkward to handle in the presence of explicit
four-dimensional polarization vectors $\pol^\mu_{(4)}$. Such vectors are
encountered when evaluating helicity amplitudes in (for example) the 
spinor helicity formalism~\cite{TreeHelicity}.  Intuitively, for $D<4$,
there are less than 2 spatial directions transverse to the gluon
direction, so one cannot really perform a rotation in this transverse 
plane, as required to define a helicity eigenstate.

\subsection{The four-dimensional helicity scheme}
\label{FDHSubsection}

Motivated by the desire of having a supersymmetric scheme whose
rules are more closely related to the more conventional
't~Hooft-Veltman scheme, and more compatible with the helicity formalism,
we define the {\it four-dimensional helicity} scheme, which has already 
been used in a number of one-loop
calculations~\cite{BKgggg,KSTfourparton,FiveParton,Z4Partons}. The
essential difference between the DR scheme and the FDH scheme is
that in the former case the rules for dot products follow from taking 
$D<4$ while in the latter one they follow from taking $D>4$.

Our rules for  extending the four-dimensional helicity scheme to 
    two loops are as follows:
\begin{itemize}

\item As in ordinary dimensional regularization, all momentum integrals
are integrated over $D$-component momenta. Any Kronecker
$\delta_\mu^{~\nu}$'s resulting from the integration are $D$-dimensional.
(This is necessary for maintaining gauge invariance.)

\item All ``observed'' external states are left in four dimensions; their
momenta are also four-dimensional.  (In QCD the ``observed'' states refer to
the external states appearing in the hard part of the process described by
Feynman diagrams, ignoring any subsequent hadronization.) Because $D>4$,
we may view this rule as choosing momenta and polarizations to lie solely 
in a four-dimensional subspace.  In this way it is natural to use 
helicity states for ``observed'' particles.

\item All ``unobserved'' internal states are treated as $D_s$ dimensional,
where $D_s \ge D$ in all intermediate steps.  The ``unobserved'' states
include virtual states in loops, virtual intermediate states in trees
(which may be attached to loops), as well as any external states which are
in collinear or soft parts of phase space. Any explicit factors of
dimension arising from the Lorentz and $\gamma$-matrix algebra should be
labeled as $D_s$, and should be kept distinct from the dimension $D$ 
describing the number of components of the loop momenta.

\item Since $D>4$, for any four-dimensional vector (such as an
``observed'' polarization vector or momentum), $\pol^\mu_{(4)}$, the dot
product with a $D$-dimensional vector yields a four-dimensional dot
product, $\pol^\mu_{(4)} p^{(D)}_\mu = \pol^\mu_{(4)} p^{(4)}_\mu$.
In general, dot products of four-dimensional vectors with
$(-2\eps)$-dimensional ones always vanish.

\end{itemize}
Though the rules for various dot products are constructed with $D>4$
in mind, at the end the expressions are analytic functions of $D$ and
$D_s$ which can be continued to any desired region.  By setting $D_s =
D$ these rules are {\it precisely} the ones for the 't~Hooft-Veltman
scheme.  The FDH scheme is specified by taking the parameter $D_s
\rightarrow 4$, after all Lorentz and $\gamma$-matrix algebra has been
performed. In performing the $\gamma$-matrix algebra, the `t
Hooft-Veltman prescription~[2] for $\gamma_5$ should be used
($\gamma_5$ {\it commutes} with $\gamma_\mu$ when the index $\mu$ lies
outside of four dimensions).

A feature of the FDH scheme that makes it useful in QCD amplitude
computations is its simple relation to the widely used 't~Hooft-Veltman
scheme.  By keeping track of the $D_s$ parameter when performing
calculations one can easily switch between the 't Hooft-Veltman and the
FDH scheme. In section~\ref{AmplitudesSection}, we will quote results
in both the FDH and the 't Hooft-Veltman scheme, leaving $D_s$ as a
free parameter.  In fact, one can define a continuous class of schemes 
by setting $D_s = 4-2\e \, \delta_R$, as in~\eqn{deltaR} below, though 
we see no particular utility to schemes other than FDH ($\delta_R = 0$) 
and HV ($\delta_R = 1$).

For this approach to be sensible, the coefficients of each power of $D_s$
must be separately gauge invariant.  In any theory with either gauge or
Yukawa interactions, the terms with $D_s$ in a given Feynman diagram can
be mapped to another diagram where (fictitious) scalar lines replace some
of the gluon lines in the diagram.  After summing over diagrams, the terms
with $D_s$ to a certain power are proportional to a sum of diagrams 
containing a certain number of fictitious scalar loops; 
the sum is gauge invariant because it corresponds to a physical amplitude.
  
At one loop, the supersymmetry preservation properties of the FDH
scheme have been verified in a number of
papers~\cite{KSTfourparton,AllNSusy}.
One of the aims of this paper is to provide explicit examples 
demonstrating that the FDH variant of dimensional regularization 
preserves the SWI~(\ref{AllPlusSusyIdentity}) through at least 
two loops.

\section{Cutting method}
\label{CuttingSection}

As mentioned in the introduction, we did not use Feynman diagrams directly
to compute the explicit two-loop helicity amplitudes.  Instead we used
a cutting method~\cite{AllNSusy,BernMorgan,OtherSewing}, which has been
applied previously to a number of one-loop calculations, including the
corrections to $e^+ e^- \rightarrow 4$ partons~\cite{Z4Partons}, and more
theoretical studies, such as the construction of infinite sequences of
maximally helicity violating amplitudes~\cite{AllNSusy,OtherSewing} and
the investigation of the divergence structure of
supergravity~\cite{N8Susy}.  More recently, it has also been used to
produce two-loop $2\to2$ scattering amplitudes in
super-Yang-Mills theory, QCD and
QED~\cite{BRY,AllPlusTwo,GGGamGam,Lbyl,ggggpaper}.
In~\sect{AmplitudesSection} we will present the four-gluon amplitude for
the ++++ helicity configuration, in both QCD and supersymmetric theories,
obtained via the cutting method.  We use this amplitude to investigate
the supersymmetry identities and associated regularization issues.  We
have also computed, by the same techniques, the $-$+++ amplitude components
which do not involve scalars; they too satisfy the supersymmetry
identities. The cutting method can help clarify the unitarity and gauge
invariance of the regularization procedure, because the basic building
blocks for loop amplitudes are gauge invariant tree-level $S$-matrix
elements.

The cutting method amounts to an extension of traditional
unitarity methods~\cite{OldCutting}.  Traditional applications
of unitarity in four dimensions, via dispersion relations, often
suffer from subtraction ambiguities.  These ambiguities are
related to the appearance of rational functions with vanishing
imaginary parts, $R(S_{i})$, where $S_{i}= \{s, t, u, \ldots\}$ are
the kinematic variables for the amplitude.  However,
dimensionally-regulated amplitudes for massless particles, as we
consider here, necessarily acquire a factor of
$(-S_{i})^{-\e}$ for each loop, from the loop integration measure 
$\int d^{4-2\e} L$ and dimensional analysis.  
For small $\e$, we expand $(-S_{i})^{-\e} \, R(S_{i}) =
R(S_{i}) - \e \, \ln (-S_{i}) \, R(S_{i}) + \Ord(\e^2)$, so every term
has an imaginary part (for some $S_{i}>0$), though not necessarily in
those terms which survive as $\e\rightarrow 0$.  Thus, the unitarity
cuts evaluated to $\Ord(\e)$ provide sufficient information for the
complete reconstruction of an amplitude through $\Ord(\e^0)$, subject
only to the usual prescription dependence associated with
renormalization.  The subtraction ambiguities
that arise in traditional dispersion relations are related to the
non-convergence of dispersion integrals.  A dimensional regulator
makes such integrals well-defined and correspondingly eliminates the
subtraction ambiguities. In a sense, we use dimensional regularization
as a calculational tool, beyond its usual role as an infrared and
ultraviolet regulator.

It is useful to view the unitarity-based technique as an alternate way
of evaluating sets of ordinary Feynman diagrams.  It does this by
collecting together gauge-invariant sets of terms which correspond to
the residues of poles in the integrands. The poles are those of the
propagators of the cut lines. This corresponds to a region of
loop-momentum integration where the cut loop momenta go on shell and
the corresponding internal lines become the intermediate states in a
unitarity relation.  From this point of view, we may consider even
more restricted regions of loop momentum integration, where additional
internal lines go on shell (and, if they are gluons, become transverse
as well).  This amounts to imposing cut conditions on additional
internal lines.

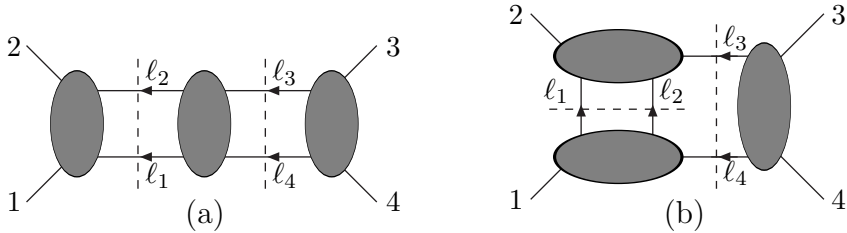
\begin{figure}[ht]
\begin{center}
\begin{picture}(350,85)(0,0)
\Text(10,5)[r]{1}  \Text(10,64)[r]{2}
\Text(148,64)[l]{3} \Text(148,5)[l]{4}
\Text(72,0)[l]{\large (a)}
\Text(57,15)[l]{$\ell_1$}   \Text(57,54)[l]{$\ell_2$}
\ArrowLine(82,22)(32,22) \ArrowLine(82,47)(32,47)
\DashLine(54,10)(54,59){3.1}
\Text(105,15)[l]{$\ell_4$}   \Text(105,54)[l]{$\ell_3$}
\ArrowLine(130,22)(80,22) \ArrowLine(130,47)(80,47)
\DashLine(102,10)(102,59){3.1}
\Line(29,22)(12,5) \Line(12,64)(29,47)
\Line(144,5)(127,22) \Line(127,47)(144,64)
\GOval(31,34.5)(20,10)(0){0.5}
\GOval(79,34.5)(20,10)(0){0.5}
\GOval(127,34.5)(20,10)(0){0.5}

\Text(200,6)[r]{1} \Text(200,76)[r]{2}
\Text(316.5,76)[l]{3} \Text(316.5,6)[l]{4}
\Text(253,0)[l]{\large (b)}
\Text(217,47)[r]{$\ell_1$} \Text(251.5,47)[l]{$\ell_2$}
\Text(275.5,67)[l]{$\ell_3$} \Text(275.5,16)[l]{$\ell_4$}
\ArrowLine(221,19)(221,60) \ArrowLine(248,19)(248,60)
\Line(290,22)(247,22) \Line(290,60)(247,60)
\ArrowLine(280,22)(270,22) \ArrowLine(280,60)(270,60)
\Line(290,28.5)(312.5,6) \Line(290,53.5)(312.5,76)
\Line(218,22)(202,6) \Line(218,60)(202,76)
\GOval(235,22)(10,24)(0){0.5}
\GOval(235,60)(10,24)(0){0.5}
\GOval(290,41)(24,10)(0){0.5}
\DashLine(209,40)(260,40){3.1} \DashLine(272,10)(272,72){3.1}

\end{picture}
\caption[]{
\label{DoubleDoubleAmplFigure}
\small Two examples of $s$-channel double two-particle cuts of a two-loop
amplitude, which separate it into a product of three tree amplitudes.  The
dashed lines represent the generalized cuts. }
\end{center}
\end{figure}

Besides the more traditional two- and three-particle cuts one can define
``double'' two-particle generalized cuts~\cite{AllPlusTwo}
for a two-loop four-point amplitude.  An example of this
quantity is  illustrated in \fig{DoubleDoubleAmplFigure}(a), and written
in terms of on-shell tree amplitudes as,
$$
\A_4^{\twoloop}  \Bigr|_{\rm 2\times2\hbox{\small -} cut} =
\sum_{\rm physical \atop states} \A_4^\tree(1,2, -\ell_2, -\ell_1)
\times \A_4^\tree(\ell_1, \ell_2, -\ell_3, -\ell_4) \times
\A_4^\tree(\ell_4, \ell_3, 3, 4) \,, \equn\label{DoubleDoubleCut}
$$
where the on-shell conditions $\ell_i^2 = 0$ are imposed on the $\ell_i$,
$i=1,2,3,4$ appearing on the right-hand side.  This equation should
{\it not\/} be interpreted as trying to take ``the imaginary part of an
imaginary part''. Rather it should be understood in the sense of the
previous paragraph as supplying information about the integrand of the
two-loop amplitude.  It supplies only part of the information
contained in the usual two-particle cut, which effectively imposes
only two kinematic constraints on the intermediate lines.  However, it
is simpler to evaluate because it is composed only of tree amplitudes.
There are, of course, other ways to cut the two-loop amplitude to
obtain trees.  For example, in \fig{DoubleDoubleAmplFigure}(b) a
different arrangement of the cut trees is shown,
$$
\A_4^{\twoloop}  \Bigr|_{\rm hv\hbox{\small -} cut} =
\sum_{\rm physical \atop states} \A_4^\tree(1,\ell_1, \ell_2, -\ell_4)
\times \A_4^\tree(2, -\ell_3, -\ell_2, -\ell_1) \times
\A_4^\tree(\ell_4, \ell_3, 3, 4) \,. \equn\label{HVCut}
$$
The combined set of double two-particle cuts provides all information
present in the ordinary two-particle cuts (where a single pair of lines is
cut, and a loop amplitude is still present), thus obviating the need to
evaluate such cuts.

%
\vskip .2 cm

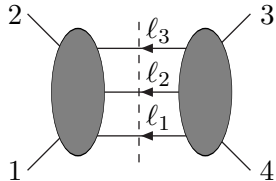
\begin{figure}[ht]
\begin{center}
\begin{picture}(108,69)(0,0)
\Text(10,5)[r]{1}  \Text(10,64)[r]{2}
\Text(100,64)[l]{3} \Text(100,5)[l]{4}
\Text(57,25)[l]{$\ell_1$}   \Text(57,41.5)[l]{$\ell_2$}
\Text(57,58)[l]{$\ell_3$}
\ArrowLine(82,18)(32,18) \ArrowLine(82,34.5)(32,34.5)
\ArrowLine(82,51)(32,51)
\DashLine(54,8)(54,61){3.1}
\Line(29,22)(12,5) \Line(12,64)(29,47)
\Line(96,5)(79,22) \Line(79,47)(96,64)
\GOval(31,34.5)(24,10)(0){0.5}
\GOval(79,34.5)(24,10)(0){0.5}
\end{picture}
\caption[]{
\label{TripleAmplFigure}
\small The standard three-particle cut of a two-loop amplitude.}
\end{center}
\end{figure}

The full amplitude, including all color factors, may be obtained by
combining a suitable set of generalized cuts into a single expression 
whose cuts match the explicitly calculated cuts.  At two loops,
it is sufficient to evaluate all the double two-particle cuts with the
topologies shown in~\fig{DoubleDoubleAmplFigure}, plus the ``standard''
three-particle cut, shown in~\fig{TripleAmplFigure} and given by
$$
\A_4^{\twoloop}  \Bigr|_{\rm 3{ \hbox{-}} cut} =
\sum_{\rm physical \atop states} \A_5^\tree(1,2,-\ell_3,-\ell_2,-\ell_1)
\times \A_5^\tree(\ell_1,\ell_2,\ell_3,3,4) \,. \equn\label{ThreeCut}
$$
For the identical-helicity amplitudes discussed in this paper,
the integrands are sufficiently simple that we can combine all double
two-particle cuts and three-particle cuts into compact integrands
containing no cut restrictions. We present these compact integrands in
\sect{AmplitudesSection}.

%
\vskip .2 cm

To obtain and verify the compact representations of the amplitudes
presented in this paper, it proved useful to compare numerically, at a
number of random kinematic points, two different representations of the
cut integrands (``raw'' and simplified), before performing any loop
integrations.  This comparison is only simple to implement when the number
of dimensions $D$ is an integer.  Integer values for $D$ may seem at odds
with dimensional regularization, which requires expressions to be
evaluated in non-integer numbers of dimensions in order to analytically
continue to $D=4$.  However, a cut integrand contains no explicit
dependence on the dimension $D$, only that implicit in the dot products of
loop momentum vectors.  It is therefore sufficient to verify the cut
integrands numerically for integer values of $D$.  In so doing, we keep
$D_s$ an analytic parameter, independent of $D$; {\it i.e.}, we verify the
cut integrands for all values of $D_s$.  One should take care that the
number of dimensions is no smaller than the number of independent vectors
in the problem.  At two loops, there are two independent loop momenta with
non-vanishing components in the extra dimensions. Hence the
extra-dimensional subspace must be at least two-dimensional, for a minimum
total dimension of $D=6$.  If two cut integrands agree numerically in six
or more dimensions (with $D_s$ left arbitrary), then the lack of any
explicit $D$ dependence ensures that they are identical for any value of
$D$.

\section{Color decomposition of QCD amplitudes}
\label{ColorSection}

Gauge theory scattering amplitudes have a rich color structure.
A number of different color decompositions have been used to organize
this structure, particularly at the tree and one-loop 
level~\cite{TreeColor,ManganoReview,OneloopColor,Review,DDFColor}.
In general, an amplitude is color decomposed by writing it as a sum of
terms.  Each term is the product of a ``color structure'' and a 
``primitive'' (or ``color-stripped'') amplitude.  A color structure
is a tensor in color space, but is independent of the momenta
and polarizations of the external states.  A primitive amplitude,
on the other hand, contains no color indices or group theory information;
it is a function only of the kinematic variables.  

For gluon scattering amplitudes in $SU(N_c)$ gauge theory, ``trace-based''
decompositions have been used
frequently~\cite{TreeColor,ManganoReview,OneloopColor,Review}.  Here the
color structures have the form $\Tr(T^{a_{i_1}} \cdots T^{a_{i_n}})$,\ \
$\Tr(T^{a_{i_1}}\ldots T^{a_{i_m}}) \, \Tr(T^{a_{i_{m+1}}}\ldots
T^{a_{i_n}})$, {\it etc.}, where $T^a$ is a generator in the fundamental
representation.  However, a color decomposition based on the Lie algebra
structure constants $f^{abc}$~\cite{DDFColor} is more suitable for our
purposes.  Note that the set of color structures may be complete, that is,
it may form a linearly independent basis; or it may be overcomplete, with
its elements obeying linear relations.  (Usually the completeness is
defined with respect to an arbitrarily large value of $N_c$; for small
$N_c$ it may degenerate.)  If the set is complete, then the primitive
amplitudes are uniquely defined; if it is overcomplete, then there is some
freedom in defining them, although there may still be a natural, symmetric 
way to do it.

The utility of primitive amplitudes in the context of supersymmetry is
that they can serve as building blocks for amplitudes in disparate
theories, where the matter transforms in different representations of the
gauge group.  If one of the theories is supersymmetric, one can express
supersymmetry relations in terms of primitive amplitudes.  Then the
supersymmetric properties of the primitive amplitudes can be applied to
non-supersymmetric theories such as QCD.

In this section, we first review tree and one-loop color decompositions,
and remind the reader how they allow supersymmetry Ward identities to be
applied to non-supersymmetric theories.  Next we proceed to two loops.  We
organize the color and kinematics of the two-loop $\ggtogg$ amplitudes in
a representation convenient for discussing the supersymmetry
identity~(\ref{AllPlusSusyIdentity}).  For the case where all particles in
the loops are in the adjoint representation, the two-loop color
decomposition was given previously~\cite{N8Susy,AllPlusTwo}.  Here we
review this result and extend it to the case of matter in the 
fundamental representation.  The sets of color structures used in both
decompositions are overcomplete, and therefore the primitive amplitudes
they define are not unique.  However, the overcompleteness gives us the
freedom to find very symmetric and compact forms for the primitive
amplitudes, which do obey simple supersymmetry relations, at least for the
special case of the ++++ helicity configuration.

\subsection{Color organization of tree and one-loop amplitudes}
\label{ColorTreeOneLoopSubsection}

In order to explain the utility of primitive amplitudes in applying
supersymmetry identities to QCD, first consider tree amplitudes.  We employ
an $f^{abc}$-based (not a trace-based) color decomposition~\cite{DDFColor},
because it most closely matches the one we shall use at two loops.  Using
the Jacobi identity for the structure constants, the four-gluon tree amplitude
can be decomposed as a sum of two terms,
$$
\A_4^\tree(1_g, 2_g, 3_g, 4_g) =
g^2 \sum_{\sigma \in S_2} (F^{a_{\sigma(2)}} F^{a_{\sigma(3)}})_{a_1 a_4} \,
   A_4^\tree(1_g, \sigma(2_g), \sigma(3_g), 4_g) \,,
\equn\label{GluonTreeDecomposition}
$$
where the primitive amplitudes $A_4^\tree$ do not contain any color
information.  Here $S_2$ is the set of two permutations of the set
$\{2,3\}$, while $(F^a)_{bc} \equiv i \f^{bac}$ is an $SU(N_c)$ generator
for the adjoint representation,
$$
\f^{abc} \equiv \sqrt{2} f^{abc} = - i \Tr\bigl( [T^a,T^b] T^c \bigr) \,,
\equn\label{ftildedef}
$$
where $f^{abc}$ are the usual $SU(N_c)$ structure constants, and
$T^a$ are generators for the fundamental representation, normalized so
that $\Tr(T^a T^b) = \delta^{ab}$.  Because the two color structures
in~\eqn{GluonTreeDecomposition} are linearly independent, the
SWI~(\ref{QCDTreeSusyIdentity}) for $\A_4^{\QCD\ \rm \tree}$ applies
separately to each primitive amplitude in~\eqn{GluonTreeDecomposition}.

Next consider the two-gluino two-gluon tree amplitude in a supersymmetric
theory.  Because gluinos are in the adjoint representation, this amplitude
satisfies a color decomposition equivalent to~\eqn{GluonTreeDecomposition},
$$
\A_4^\tree(1_\tg, 2_g, 3_g, 4_\tg) =
g^2 \sum_{\sigma \in S_2} (F^{a_{\sigma(2)}}\, F^{a_{\sigma(3)}})_{a_1 a_4}
                 A_4^\tree(1_\tg, \sigma(2_g), \sigma(3_g), 4_\tg) \,.
\equn\label{GluinoTreeDecomposition}
$$
Compare this decomposition to the one for the two-quark two-gluon 
amplitude in QCD,
$$
\A_4^\tree(1_q, 2_g, 3_g, 4_\qb) =
g^2 \sum_{\sigma \in S_2} (T^{a_{\sigma(2)}}
                              T^{a_{\sigma(3)}})_{i_1}{}^{\bar \imath_4}
   A_4^\tree(1_q,\sigma(2_g),\sigma(3_g),4_\qb)\,,
\equn\label{QuarkTreeDecomposition}
$$
where the quarks, and $T^{a_i}$, are in the fundamental
representation.  The crucial point is that the $A_4^\tree$ appearing
in \eqn{GluinoTreeDecomposition} are in fact identical to the
$A_4^\tree$ appearing in \eqn{QuarkTreeDecomposition}, as can easily
be verified using Feynman diagrams.  (In the double-line formalism
for $SU(N_c)$, the color line running between the two gluinos is merely
``stripped off'' to obtain the quark amplitudes.) Thus, there is no real
distinction between gluinos and quarks as far as the primitive
amplitudes are concerned.  Any SWI that holds for the primitive
gluino amplitudes will hold for the quark ones.

Similar color decompositions hold at one loop.  Consider the
four-gluon amplitude.  When the particle circulating in the loop is in
the adjoint representation, {\it e.g.} a gluon or gluino, the
decomposition is~\cite{DDFColor}
$$
\A_{4}^{\rm adjoint\ loop}(1,2,3,4) = g^4 \Biggl[
\sum_{\sigma \in S_4/Z_4/\R}
 \Tr(F^{a_{\si(1)}}F^{a_{\si(2)}} F^{a_{\si(3)}} F^{a_{\si(4)}})
    A^\oneloop_{4} (\si(1), \si(2), \si(3),\si(4)) \Biggl] \,.
\equn\label{LoopColorAdjoint}
$$
Here we have suppressed the gluon $g$ labels on the external legs;
$\sigma$ runs over the set of permutations of $\{1,2,3,4\}$, after
removing those equivalent under a cyclic $Z_4$ permutation or the
reflection $\R$: $\{1,2,3,4\} \to \{4,3,2,1\}$.
In comparison, the contribution of a fundamental representation particle 
in the loop, {\it e.g.} a quark, is decomposed as
$$
\A_{4}^{\rm fund.\ loop}(1,2,3,4) = g^4 \Biggl[
\sum_{\sigma \in S_4/Z_4} \Tr(T^{a_{\si(1)}}\ldots
T^{a_{\si(4)}})
  \,  A_{4}^\oneloop (\si(1), \si(2), \si(3),\si(4)) \Biggr] \,.
\equn\label{LoopColorFund}
$$
Note that in both \eqns{LoopColorAdjoint}{LoopColorFund}, the color
factors may be read off of the appropriate one-loop ``parent'' 
diagram shown in \fig{OneLoopSusyFigure}.  One simply assigns a group 
theory factor of $\tilde{f}^{abc}$ or $(T^a)^{i}_{\bar{\jmath}}$ 
for each vertex, and a factor of $\delta^{ab}$ or 
$\delta^{\bar{\jmath}}_{i}$ 
for each internal line, then performs the index contractions.
We shall use this diagrammatic representation of the 
color factors at two loops in sections~\ref{AdjTwoLoopColorSubsection}
and \ref{FundTwoLoopColorSubsection}.

Once again, using Feynman diagrams, it is straightforward to
demonstrate that the primitive amplitude $A_{4}^\oneloop$
appearing in the fundamental representation case~(\ref{LoopColorFund})
is exactly the same object appearing in the adjoint representation
case (\ref{LoopColorAdjoint}), provided that the type of particle in
the loop (fermion or scalar) is the same in both cases.

Because of this identification, QCD quark loop primitive
amplitudes obey SWI.  For example, let us apply \eqn{AllPlusSusyIdentity}
to pure $N=1$ supersymmetric Yang-Mills theory (SYM).
Linear independence of the color factors appearing in
\eqn{LoopColorAdjoint} implies that the primitive amplitudes satisfy
$$
A^{\SYM\ \oneloop}_{4}(1^\pm, 2^+, 3^+, 4^+)
= A^{\rm gluon\ loop}_{4}(1^\pm, 2^+, 3^+, 4^+)
  + A^{\rm fermion\ loop}_{4}(1^\pm, 2^+, 3^+, 4^+) = 0 \,,
\equn\label{GluonPlusGluino}
$$
where $A^{\rm gluon\ loop}_{4}$ represents the gluon
loop contribution to $A_4^\oneloop$ in \eqn{LoopColorAdjoint}, shown in
\fig{OneLoopSusyFigure}a, and $A^{\rm fermion\ loop}_{4}$ represents the
gluino loop contribution to $A_4^\oneloop$, shown in
\fig{OneLoopSusyFigure}b.  Because $A_{4}^{\rm fermion\ loop}$ is the same
for an adjoint or a fundamental representation fermion (although it
gets inserted into a different color decomposition formula, 
(\ref{LoopColorAdjoint}) or (\ref{LoopColorFund}), in the two cases), 
the identity
$$
A^{\rm  gluon\ loop}_{4}(1^\pm, 2^+, 3^+, 4^+)
= - A^{\rm fermion\ loop}_{4}(1^\pm, 2^+, 3^+, 4^+)
\equn\label{OneLoopGluonFermionSWI}
$$
holds even in QCD, where the fermion would be a quark.  Continuing
along these lines, consider the contribution of a chiral multiplet in
super-QCD, consisting of a fermion and a scalar,
to~\eqn{AllPlusSusyIdentity}.  One obtains
$$
 A^{\rm scalar\ loop}_{4}(1^\pm, 2^+, 3^+, 4^+)
= - A^{\rm fermion\ loop}_{4}(1^\pm, 2^+, 3^+, 4^+)\,,
\equn\label{OneLoopFermionScalarSWI}
$$
even for a non-supersymmetric gauge theory. As discussed
in section~\ref{ParticleContentSubsection}, one must take the number 
of scalar states to match the number of fermion states on the two sides
of~\eqn{OneLoopFermionScalarSWI}.

It is not difficult to verify that the SWI~(\ref{OneLoopGluonFermionSWI})
and (\ref{OneLoopFermionScalarSWI}) are indeed satisfied.  
For the identical-helicity case (++++), the one-loop amplitudes
have a compact representation~\cite{BernMorgan,OneLoopAllPlus}, similar to the 
one we shall use at two loops,
$$
\eqalign{
& A_{4}^{\rm gluon\ loop}(1^+,2^+,3^+,4^+) =
 (D_s - 2) \, { \rho \over i } \,
\I_4^{\rm \oneloop}[\mud_p^4] \,, \cr
& A_{4}^{\rm fermion\ loop}(1^+,2^+,3^+,4^+) =
-2 \, { \rho \over i } \,
\I_4^{\rm \oneloop}[\mud_p^4] \,, \cr
& A_{4}^{\rm scalar \ loop}(1^+,2^+,3^+,4^+) =
2 \, { \rho \over i } \,
\I_4^{\oneloop}[\mud_p^4] \,, \cr}
\equn\label{OneLoopAllPlusAmplitudes}
$$
corresponding to the three representative diagrams in 
\fig{OneLoopSusyFigure}.  Here
$$
\rho\ \equiv\ i \, {\spb1.2 \spb3.4 \over \spa1.2 \spa3.4}
\equn\label{RhoDef}
$$
is a ubiquitous spinor product prefactor, which is a pure phase,
and is totally symmetric under permutations of the external legs 
$\{1,2,3,4\}$. The one-loop integral appearing in the amplitudes is
$$
\eqalign{
\I_4^{\rm \oneloop}[\mud_p^4] & =
\int {d^D p \over (2\pi)^D} \,
{\mud_p^4 \over p^2 (p-k_1)^2 (p-k_1-k_2)^2 (p+k_4)^2} \cr
& =
- {i\over 6} {1\over (4 \pi)^{2}} + \Ord(\eps) \,,
}
\equn
$$
where we have split the loop momentum $p = p_{[4]} + \vec{\mud}_p$
into its four-dimensional components $p_{[4]}$ and its
$(-2\eps)$-dimensional components $\vec{\mud}_p$, with 
$\mud_p^4 \equiv (\vec{\mud}_p \cdot \vec{\mud}_p)^2$.
(The value of $\I_4^{\rm \oneloop}[\mud_p^4]$ is known through 
$\Ord(\eps^2)$~\cite{AllPlusTwo}.)
Full amplitudes, including all color factors, are obtained
by substituting the primitive amplitudes in
\eqn{OneLoopAllPlusAmplitudes} into eqs.~(\ref{LoopColorAdjoint}) and
(\ref{LoopColorFund}).

{}From \eqn{OneLoopAllPlusAmplitudes} we see that the 
SWI~(\ref{OneLoopGluonFermionSWI}) and (\ref{OneLoopFermionScalarSWI})
hold to all orders in $\e$ in the FDH scheme, where $D_s = 4$.
Note that in the HV scheme, where $D_s = 4-2\e$, the SWI do hold at one loop 
{\it in the limit} $\e \to 0$ for the ++++ (and $-$+++) helicity
configurations~\cite{BKgggg}; however, the identities for the $-$$-$++ 
and $-$+$-$+ amplitudes are known to be explicitly violated, even in this 
limit~\cite{KSTfourparton}.

\subsection{Purely adjoint representation two-loop color decomposition}
\label{AdjTwoLoopColorSubsection}

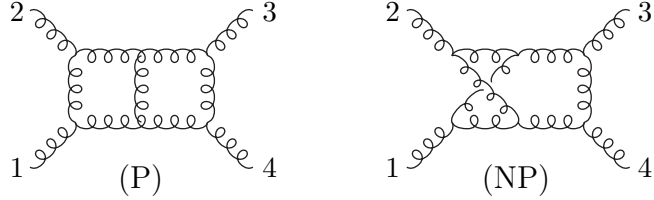
\begin{figure}[ht]
\begin{center}
\begin{picture}(250,80)(0,0)
\Text(10,5)[r]{1} \Text(10,64)[r]{2}
\Text(100,64)[l]{3} \Text(100,5)[l]{4}
\Gluon(79,22)(29,22){2.5}{7} \Gluon(29,47)(79,47){2.5}{7}
\Gluon(29,22)(29,47){2.5}{3} \Gluon(54,22)(54,47){2.5}{3}
\Gluon(79,47)(79,22){2.5}{3}
\Gluon(29,22)(12,5){2.5}{3} \Gluon(12,64)(29,47){2.5}{3}
\Gluon(96,5)(79,22){2.5}{3} \Gluon(79,47)(96,64){2.5}{3}
\Text(54,0)[c]{\large (\P)}
\Text(152,5)[r]{1} \Text(152,64)[r]{2}
\Text(242,64)[l]{3} \Text(242,5)[l]{4}
\Gluon(221,22)(196,22){2.5}{3}            \Gluon(196,22)(171,22){2.5}{2}
    \Gluon(171,47)(196,47){2.5}{2}     \Gluon(196,47)(221,47){2.5}{3}
\Gluon(171,47)(196,22){2.0}{4} \Gluon(171,22)(183,34){2.0}{1}
\Gluon(186,37)(196,47){2.0}{1}
\Gluon(221,47)(221,22){2.5}{3}
\Gluon(171,22)(154,5){2.5}{3} \Gluon(154,64)(171,47){2.5}{3}
\Gluon(238,5)(221,22){2.5}{3} \Gluon(221,47)(238,64){2.5}{3}
\Text(196,0)[c]{\large (\NP)}
\end{picture}
\caption[]{
\label{ParentGlueFigure}
\small Parent graphs for the pure gluon $G$ contributions.}
\end{center}
\end{figure}

\begin{figure}[ht]
\begin{center}
\begin{picture}(394,69)(0,0)
\Gluon(102,22)(77,22){2.5}{3} \Gluon(77,47)(102,47){2.5}{3}
\ArrowLine(77,22)(52,22)\ArrowLine(52,47)(77,47)
\ArrowLine(52,22)(52,47) \ArrowLine(77,47)(77,22)
  \Gluon(102,47)(102,22){2.5}{3}
\Gluon(52,22)(35,5){2.5}{3} \Gluon(35,64)(52,47){2.5}{3}
\Gluon(119,5)(102,22){2.5}{3} \Gluon(102,47)(119,64){2.5}{3}
\Text(33,5)[r]{1} \Text(33,64)[r]{2}
\Text(123,64)[l]{3} \Text(123,5)[l]{4}
\Text(77,0)[c]{$(\P_1)$}
\ArrowLine(221,22)(196,22) \Gluon(196,22)(171,22){2.5}{3}
\Gluon(171,47)(196,47){2.5}{3} \ArrowLine(196,47)(221,47)
\Gluon(171,22)(171,47){2.5}{3} \ArrowLine(196,22)(196,47)
  \ArrowLine(221,47)(221,22)
\Gluon(171,22)(154,5){2.5}{3} \Gluon(154,64)(171,47){2.5}{3}
\Gluon(238,5)(221,22){2.5}{3} \Gluon(221,47)(238,64){2.5}{3}
\Text(152,5)[r]{1} \Text(152,64)[r]{2}
\Text(242,64)[l]{3} \Text(242,5)[l]{4}
\Text(196,0)[c]{$(\P_2)$}
\ArrowLine(315,22)(290,22) \ArrowLine(290,47)(315,47)
\ArrowLine(340,22)(315,22)\ArrowLine(315,47)(340,47)
\ArrowLine(290,22)(290,47) \Gluon(315,22)(315,47){2.5}{3}
  \ArrowLine(340,47)(340,22)
\Gluon(290,22)(273,5){2.5}{3} \Gluon(273,64)(290,47){2.5}{3}
\Gluon(357,5)(340,22){2.5}{3} \Gluon(340,47)(357,64){2.5}{3}
\Text(271,5)[r]{1} \Text(271,64)[r]{2}
\Text(361,64)[l]{3} \Text(361,5)[l]{4}
\Text(315,0)[c]{$(\P_3)$}
\end{picture}
\end{center}
\begin{center}
\begin{picture}(250,80)(0,0)
\Text(46,0)[c]{$(\P_4)$}
\Text(-7,5)[r]{1} \Text(-7,64)[r]{2}
\Text(103,64)[l]{3} \Text(103,5)[l]{4}
\ArrowLine(12,22)(12,47) \ArrowLine(12,47)(34,34.5)
\ArrowLine(34,34.5)(12,22)
\ArrowLine(80,47)(80,22) \ArrowLine(80,22)(58,34.5)
\ArrowLine(58,34.5)(80,47)
\Gluon(34,34.5)(58,34.5){2.5}{3}
\Gluon(12,22)(-5,5){2.5}{3} \Gluon(-5,64)(12,47){2.5}{3}
\Gluon(97,5)(80,22){2.5}{3} \Gluon(80,47)(97,64){2.5}{3}
\end{picture}
\begin{picture}(394,100)(0,0)
\Text(33,5)[r]{1} \Text(33,64)[r]{2}
\Text(123,64)[l]{3} \Text(123,5)[l]{4}
\ArrowLine(102,22)(77,22) \ArrowLine(77,22)(52,22)
\Gluon(52,47)(77,47){2.5}{2} \ArrowLine(77,47)(102,47)
\ArrowLine(52,22)(64.5,34.5)    
\ArrowLine(67,37)(77,47) 

\Gluon(52,47)(77,22){2.5}{4}
\ArrowLine(102,47)(102,22)
\Gluon(52,22)(33,5){2.5}{3} \Gluon(33,64)(52,47){2.5}{3}
\Gluon(119,5)(102,22){2.5}{3} \Gluon(102,47)(119,64){2.5}{3}
\Text(77,0)[c]{($\NP_1$)}
\Text(152,5)[r]{1} \Text(152,64)[r]{2}
\Text(242,64)[l]{3} \Text(242,5)[l]{4}
\Gluon(196,22)(171,22){2.5}{2} \ArrowLine(221,22)(196,22)
\ArrowLine(171,47)(196,47) \ArrowLine(196,47)(221,47)
\Gluon(196,47)(171,22){2.5}{4}
\ArrowLine(196,22)(186.5,31.5)  
\ArrowLine(183.5,34.5)(171,47) 
\ArrowLine(221,47)(221,22)
\Gluon(171,22)(154,5){2.5}{3} \Gluon(154,64)(171,47){2.5}{3}
\Gluon(238,5)(221,22){2.5}{3} \Gluon(221,47)(238,64){2.5}{3}
\Text(196,0)[c]{($\NP_2$)}
\Text(271,5)[r]{1}  \Text(271,64)[r]{2}
\Text(361,64)[l]{3} \Text(361,5)[l]{4}
\Gluon(340,22)(315,22){2.5}{3} \Gluon(315,47)(340,47){2.5}{3}
\ArrowLine(290,22)(315,22) \ArrowLine(290,47)(315,47)
\ArrowLine(315,22)(304,33) \ArrowLine(301,36)(290,47) 
                 \ArrowLine(302.5,34.5)(290,22)\Line(315,47)(302.5,34.5)
\Gluon(340,47)(340,22){2.5}{3}
\Gluon(290,22)(273,5){2.5}{3} \Gluon(273,64)(290,47){2.5}{3}
\Gluon(357,5)(340,22){2.5}{3} \Gluon(340,47)(357,64){2.5}{3}
\Text(315,0)[c]{($\NP_3$)}
\end{picture}
\caption[]{
\label{ParentFermionFigure}
\small Parent diagrams for the fermion loop $F$ contributions.}
\end{center}
\end{figure}
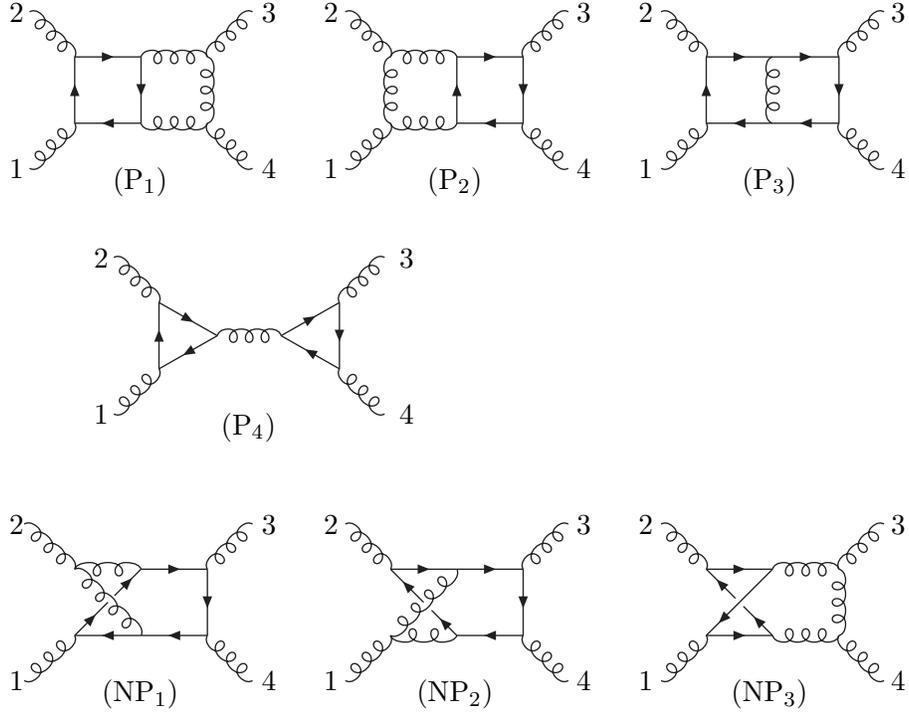

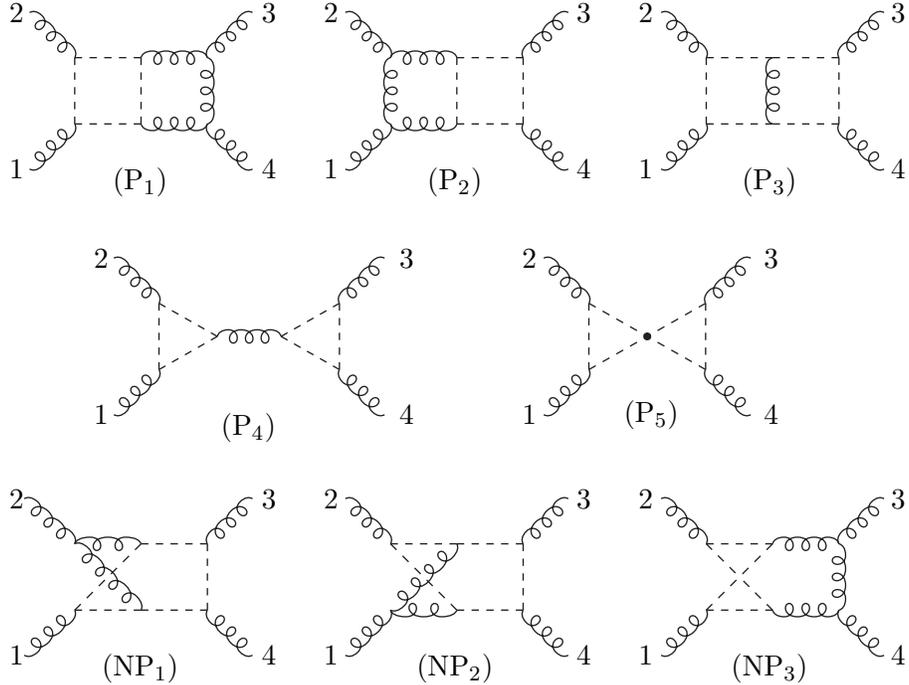
\begin{figure}[ht]
\begin{center}
\begin{picture}(394,69)(0,0)
\Gluon(102,22)(77,22){2.5}{3} \Gluon(77,47)(102,47){2.5}{3}
\DashLine(52,22)(77,22){3.1} \DashLine(52,47)(77,47){3.1}
\DashLine(52,22)(52,47){3.1} \DashLine(77,22)(77,47){3.1}
  \Gluon(102,47)(102,22){2.5}{3}
\Gluon(52,22)(35,5){2.5}{3} \Gluon(35,64)(52,47){2.5}{3}
\Gluon(119,5)(102,22){2.5}{3} \Gluon(102,47)(119,64){2.5}{3}
\Text(33,5)[r]{1} \Text(33,64)[r]{2}
\Text(123,64)[l]{3} \Text(123,5)[l]{4}
\Text(77,0)[c]{$(\P_1)$}
\DashLine(221,22)(196,22){3.1} \Gluon(196,22)(171,22){2.5}{3}
\Gluon(171,47)(196,47){2.5}{3} \DashLine(196,47)(221,47){3.1}
\Gluon(171,22)(171,47){2.5}{3} \DashLine(196,22)(196,47){3.1}
  \DashLine(221,47)(221,22){3.1}
\Gluon(171,22)(154,5){2.5}{3} \Gluon(154,64)(171,47){2.5}{3}
\Gluon(238,5)(221,22){2.5}{3} \Gluon(221,47)(238,64){2.5}{3}
\Text(152,5)[r]{1} \Text(152,64)[r]{2}
\Text(242,64)[l]{3} \Text(242,5)[l]{4}
\Text(196,0)[c]{$(\P_2)$}
\DashLine(315,22)(290,22){3.1} \DashLine(290,47)(315,47){3.1}
\DashLine(315,22)(340,22){3.1} \DashLine(315,47)(340,47){3.1}
\DashLine(290,22)(290,47){3.1} \Gluon(315,22)(315,47){2.5}{3}
  \DashLine(340,47)(340,22){3.1}
\Gluon(290,22)(273,5){2.5}{3} \Gluon(273,64)(290,47){2.5}{3}
\Gluon(357,5)(340,22){2.5}{3} \Gluon(340,47)(357,64){2.5}{3}
\Text(271,5)[r]{1} \Text(271,64)[r]{2}
\Text(361,64)[l]{3} \Text(361,5)[l]{4}
\Text(315,0)[c]{$(\P_3)$}
\end{picture}
\end{center}
\begin{center}
\begin{picture}(250,80)(0,0)
\Text(46,0)[c]{$(\P_4)$}
\Text(-7,5)[r]{1} \Text(-7,64)[r]{2}
\Text(103,64)[l]{3} \Text(103,5)[l]{4}
\DashLine(12,22)(12,47){3.1} \DashLine(12,47)(34,34.5){3.1}
\DashLine(34,34.5)(12,22){3.1}
\DashLine(80,47)(80,22){3.1} \DashLine(80,22)(58,34.5){3.1}
\DashLine(58,34.5)(80,47){3.1}
\Gluon(34,34.5)(58,34.5){2.5}{3}
\Gluon(12,22)(-5,5){2.5}{3} \Gluon(-5,64)(12,47){2.5}{3}
\Gluon(97,5)(80,22){2.5}{3} \Gluon(80,47)(97,64){2.5}{3}
\Text(188,5)[l]{{$(\P_5)$}}
\Text(155,5)[r]{1} \Text(155,64)[r]{2}
\Text(241,64)[l]{3} \Text(241,5)[l]{4}
\DashLine(174,22)(174,47){3.1} \DashLine(218,47)(218,22){3.1}
\DashLine(218,22)(174,47){3.1} \DashLine(174,22)(218,47){3.1}
\Gluon(174,22)(157,5){2.5}{3} \Gluon(157,64)(174,47){2.5}{3}
\Gluon(235,5)(218,22){2.5}{3} \Gluon(218,47)(235,64){2.5}{3}
\Vertex(196,34.5){1.5}
\end{picture}
\begin{picture}(394,90)(0,0)
\Text(33,5)[r]{1} \Text(33,64)[r]{2}
\Text(123,64)[l]{3} \Text(123,5)[l]{4}
\DashLine(102,22)(77,22){3.1} \DashLine(77,22)(52,22){3.1}
\Gluon(52,47)(77,47){2.5}{2} \DashLine(77,47)(102,47){3.1}
\DashLine(77,47)(52,22){3.1} \Gluon(52,47)(77,22){2.5}{4}
\DashLine(102,47)(102,22){3.1}
\Gluon(52,22)(33,5){2.5}{3} \Gluon(33,64)(52,47){2.5}{3}
\Gluon(119,5)(102,22){2.5}{3} \Gluon(102,47)(119,64){2.5}{3}
\Text(77,0)[c]{($\NP_1$)}
\Text(152,5)[r]{1} \Text(152,64)[r]{2}
\Text(242,64)[l]{3} \Text(242,5)[l]{4}
\Gluon(196,22)(171,22){2.5}{2} \DashLine(221,22)(196,22){3.1}
\DashLine(171,47)(196,47){3.1} \DashLine(196,47)(221,47){3.1}
\Gluon(196,47)(171,22){2.5}{4}
\DashLine(171,47)(196,22){3.1} \DashLine(221,47)(221,22){3.1}
\Gluon(171,22)(154,5){2.5}{3} \Gluon(154,64)(171,47){2.5}{3}
\Gluon(238,5)(221,22){2.5}{3} \Gluon(221,47)(238,64){2.5}{3}
\Text(196,0)[c]{($\NP_2$)}
\Text(271,5)[r]{1}  \Text(271,64)[r]{2}
\Text(361,64)[l]{3} \Text(361,5)[l]{4}
\Gluon(340,22)(315,22){2.5}{3} \Gluon(315,47)(340,47){2.5}{3}
\DashLine(290,22)(315,22){3.1} \DashLine(290,47)(315,47){3.1}
\DashLine(315,22)(304,33){3.1} \DashLine(301,36)(290,47){3.1}
\DashLine(290,22)(315,47){3.1}
\Gluon(340,47)(340,22){2.5}{3}
\Gluon(290,22)(273,5){2.5}{3} \Gluon(273,64)(290,47){2.5}{3}
\Gluon(357,5)(340,22){2.5}{3} \Gluon(340,47)(357,64){2.5}{3}
\Text(315,0)[c]{($\NP_3$)}
\end{picture}
\caption[]{
\label{ParentScalarFigure}
\small Parent diagrams for the scalar loop $S$ contributions.}
\end{center}
\end{figure}

\begin{figure}[ht]
\begin{center}
\begin{picture}(394,69)(0,0)
\Line(102,22)(77,22) \Line(77,47)(102,47)
\DashLine(52,22)(77,22){3.1} \DashLine(77,47)(52,47){3.1}
\DashLine(52,22)(52,47){3.1} \ArrowLine(77,47)(77,22)
  \Line(102,47)(102,22)
\Gluon(52,22)(35,5){2.5}{3} \Gluon(35,64)(52,47){2.5}{3}
\Gluon(119,5)(102,22){2.5}{3} \Gluon(102,47)(119,64){2.5}{3}
\Text(33,5)[r]{1} \Text(33,64)[r]{2}
\Text(123,64)[l]{3} \Text(123,5)[l]{4}
\Text(77,0)[c]{$(\P_1)$}
\DashLine(221,22)(196,22){3.1} \Line(196,22)(171,22)
\Line(171,47)(196,47) \DashLine(196,47)(221,47){3.1}
\Line(171,22)(171,47) \ArrowLine(196,22)(196,47)
  \DashLine(221,47)(221,22){3.1}
\Gluon(171,22)(154,5){2.5}{3} \Gluon(154,64)(171,47){2.5}{3}
\Gluon(238,5)(221,22){2.5}{3} \Gluon(221,47)(238,64){2.5}{3}
\Text(152,5)[r]{1} \Text(152,64)[r]{2}
\Text(242,64)[l]{3} \Text(242,5)[l]{4}
\Text(196,0)[c]{$(\P_2)$}
\ArrowLine(315,22)(290,22) \ArrowLine(290,47)(315,47)
\DashLine(315,22)(340,22){3.1} \DashLine(315,47)(340,47){3.1}
\ArrowLine(290,22)(290,47) \Line(315,22)(315,47)
  \DashLine(340,47)(340,22){3.1}
\Gluon(290,22)(273,5){2.5}{3} \Gluon(273,64)(290,47){2.5}{3}
\Gluon(357,5)(340,22){2.5}{3} \Gluon(340,47)(357,64){2.5}{3}
\Text(271,5)[r]{1} \Text(271,64)[r]{2}
\Text(361,64)[l]{3} \Text(361,5)[l]{4}
\Text(315,0)[c]{$(\P_3)$}
\end{picture}

\end{center}

\begin{center}
\begin{picture}(250,80)(0,0)
\Text(46,0)[c]{$(\P_4)$}
\Text(-7,5)[r]{1} \Text(-7,64)[r]{2}
\Text(103,64)[l]{3} \Text(103,5)[l]{4}
\ArrowLine(12,22)(12,47) \ArrowLine(12,47)(34,34.5) \ArrowLine(34,34.5)(12,22)
\DashLine(80,47)(80,22){3.1} \DashLine(80,22)(58,34.5){3.1}
\DashLine(58,34.5)(80,47){3.1}
\Gluon(34,34.5)(58,34.5){2.5}{3}
\Gluon(12,22)(-5,5){2.5}{3} \Gluon(-5,64)(12,47){2.5}{3}
\Gluon(97,5)(80,22){2.5}{3} \Gluon(80,47)(97,64){2.5}{3}
\end{picture}
\begin{picture}(394,90)(0,0)
\Text(33,5)[r]{1} \Text(33,64)[r]{2}
\Text(123,64)[l]{3} \Text(123,5)[l]{4}
\ArrowLine(102,22)(77,22) \DashLine(77,22)(52,22){3.1}
\Line(52,47)(77,47) \ArrowLine(77,47)(102,47)
\DashLine(77,47)(52,22){3.1} \Line(52,47)(77,22)
\ArrowLine(102,47)(102,22)
\Gluon(52,22)(33,5){2.5}{3} \Gluon(33,64)(52,47){2.5}{3}
\Gluon(119,5)(102,22){2.5}{3} \Gluon(102,47)(119,64){2.5}{3}
\Text(77,0)[c]{($\NP_1$)}
\Text(152,5)[r]{1} \Text(152,64)[r]{2}
\Text(242,64)[l]{3} \Text(242,5)[l]{4}
\Line(196,22)(171,22) \ArrowLine(221,22)(196,22)
\DashLine(171,47)(196,47){3.1} \ArrowLine(196,47)(221,47)
\Line(196,47)(171,22)
\DashLine(171,47)(196,22){3.1} \ArrowLine(221,47)(221,22)
\Gluon(171,22)(154,5){2.5}{3} \Gluon(154,64)(171,47){2.5}{3}
\Gluon(238,5)(221,22){2.5}{3} \Gluon(221,47)(238,64){2.5}{3}
\Text(196,0)[c]{($\NP_2$)}
\Text(271,5)[r]{1}  \Text(271,64)[r]{2}
\Text(361,64)[l]{3} \Text(361,5)[l]{4}
\Line(340,22)(315,22) \Line(315,47)(340,47)
\DashLine(290,22)(315,22){3.1} \ArrowLine(290,47)(315,47)
\ArrowLine(315,22)(304,33) \ArrowLine(301,36)(290,47) 
\DashLine(290,22)(315,47){3.1}
\Line(340,47)(340,22)
\Gluon(290,22)(273,5){2.5}{3} \Gluon(273,64)(290,47){2.5}{3}
\Gluon(357,5)(340,22){2.5}{3} \Gluon(340,47)(357,64){2.5}{3}
\Text(315,0)[c]{($\NP_3$)}
\end{picture}
\end{center}
\caption[a]{\small Parent graphs for the mixed $M$ fermion scalar loop
contributions.  Diagrams $\P_1$, $\P_2$, $\P_3$, $\NP_1$, $\NP_2$ and
$\NP_3$ contain Yukawa interactions and can contribute to either the
pure gauge coupling case, or the superpotential case proportional to
$|\xi|^2$; whereas diagram $\P_4$ contributes only to the pure gauge
coupling case.  In the pure gauge coupling case, the solid lines with
arrows represent matter fermions and the solid lines with no arrows
gluinos.  In each case there are additional contributions obtained
from swapping the scalar lines with the matter fermion lines, but 
this swap does not alter the color factor. }
\label{ParentMixedFigure}
\end{figure}
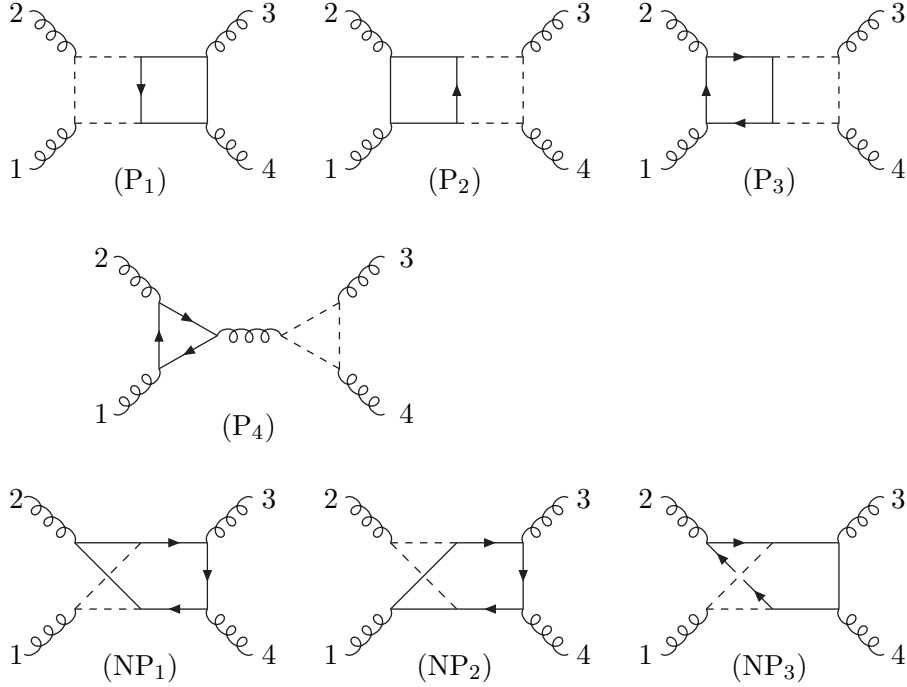

The two-loop color decomposition we use is similar to the tree and
one-loop decompositions described above.  For the case where all particles
are in the adjoint representation, the complete amplitude is given by
$$
\eqalign{ \A_{X}^{\rm adjoint} & = g^6 \sum_{\{\D_i\}}
\Bigl[C^{\D_i}_{1234} \, A^{\D_i}_{X1234} + C^{\D_i}_{3421} \,
A^{\D_i}_{X3421}
    +\ \C(234) \Bigr] \,. \cr}
\equn\label{FColor}
$$
Here $C^{\D_i}$ are color factors, while $A^{\D_i}$ are primitive
amplitudes.  The notation ``$+\ \C(234)$'' instructs one to add the two
non-trivial cyclic permutations of $\{2,3,4\}$.  To simplify the formulae, we
adopt a diagrammatic representation of the color factors, in terms of a
set of ``parent'' diagrams, depicted in
figs.~\ref{ParentGlueFigure}--\ref{ParentMixedFigure}.  The label
$X=G,S,F,M$ refers to whether the contribution is respectively from
gluons, adjoint scalars, adjoint fermions, or mixed fermion-scalar loop
contributions, corresponding to figs.~\ref{ParentGlueFigure},
\ref{ParentFermionFigure}, \ref{ParentScalarFigure}
and~\ref{ParentMixedFigure}, respectively.  Which of these contributions
appear in the full amplitude depends, of course, on the matter content of
the theory under consideration, through equations
like~(\ref{N=1fundSchematicAmplitude}) and (\ref{N=1adjSchematicAmplitude}). 

Each color factor $C^{\D_i}$ in \eqn{FColor} is specified by a parent
diagram, where the corresponding label $\D_i$ is shown in parentheses in
the figure.  For example, the fermion loop contribution color factor
$C^{\P_1}_{1234}$ is found from \fig{ParentFermionFigure}($\P_1$).  The
color factor is calculated from the parent diagram by associating each
vertex and internal line with a color tensor, and then performing the
internal index contractions.  For the case where all particles are in 
the adjoint representation, the internal lines are ``dressed'' with
factors of $\delta^{ab}$.  The vertex dressing rules are as follows:
\begin{itemize}
\item A gauge 3-point vertex is dressed with 
$i\tilde{f}^{abc} = \Tr([T^a,T^b]T^c)$.
\item A superpotential 3-point vertex is dressed with 
$d^{abc} \equiv \Tr(\{T^a,T^b\}T^c)$.
\item The 4-point vertex in the double-scalar loop diagram $P_5$
in \fig{ParentScalarFigure} is dressed with
\subitem{$\star$} ${1\over 2} i^2(\tilde{f}^{abe} \tilde{f}^{ecd} 
                     + \tilde{f}^{bce}\tilde{f}^{eda})$
in the case of a gauge coupling ($D$ term), and
\subitem{$\star$} $ d^{bce} d^{eda}$ in the case of a superpotential 
coupling ($F$ term).
\end{itemize}
\par\noindent
The structure of the 4-point vertices follows from the
Lagrangian terms~(\ref{Lg}) and~(\ref{Lxi}), respectively.  Color
factors with the legs ordered differently from the figures, such as
$C^{\P_1}_{3421}$, are obtained by appropriate relabeling of the
external legs.  In the fundamental representation case to be discussed
in section~\ref{FundTwoLoopColorSubsection}, the same color factor
rules will hold, after correcting for the different representations;
for example, $\delta^{\bar{\jmath}}_{i}$ should clearly be used for a
fundamental line, and $(T^a)^{i}_{\bar{\jmath}}$ for a three-point
vertex with two fundamental lines emanating from it.

The primitive amplitudes $A^{\D_i}$ are defined as the coefficients of the
color factors in \eqn{FColor}.  Actually, in contrast to the tree and
one-loop cases, this prescription does not completely specify the
primitive amplitudes, because the color factors in \eqn{FColor} are not
linearly independent (see below).  Also, the quantities $A_4^\tree$ 
and $A_4^\oneloop$ can be given gauge-invariant definitions, 
as sums of complete sets of color-ordered Feynman diagrams.
Such a definition fails here.  For example, the pure glue contributions
$A^\P_{G1234}$ and $A^\P_{G1432}$ both contribute at leading order in
$N_c$ to the cyclic color ordering 1234, and thus only the sum of them has
a gauge-invariant definition.  In any case, the primitive amplitudes
$A^{\D_i}$ that we present in \sect{AmplitudesSection} are arranged to
fulfill \eqn{FColor}.  In addition, each such $A^{\D_i}$ does include the
corresponding parent diagram from
figs.~\ref{ParentGlueFigure}--\ref{ParentMixedFigure} in its definition;
it also includes pieces of other ``daughter'' Feynman diagrams (not shown
in the figures), which are typically shared with other primitive
amplitudes.

For the purely adjoint color representation, in the gauge coupling case
for which all vertices are built out of factors of $\tilde{f}^{abc}$,
an equivalent but much simpler color decomposition has been 
given~\cite{N8Susy,AllPlusTwo}.  This decomposition is
in terms of just the color factors $\P$ and $\NP$ shown
in~\fig{ParentGlueFigure}.  Subdividing the parent diagrams $\D_i$ into 
planar ones $\P_i$ and non-planar ones $\NP_i$, we have
$$
\eqalign{ \A_{X}^{\rm adjoint} & = g^6 \biggl\{ 
 \sum_{\{\P_i\}}
\Bigl[C^{\P}_{1234} \, A^{\P_i}_{X1234} + C^{\P}_{3421} \,
A^{\P_i}_{X3421} \Bigr]
+ \sum_{\{\NP_i\}}
\Bigl[C^{\NP}_{1234} \, A^{\NP_i}_{X1234} + C^{\NP}_{3421} \,
  A^{\NP_i}_{X3421}
    \Bigr] \ +\ \C(234) \biggr\} \,. \cr}
\equn\label{NewFColor}
$$
In the pure gluon case, as indicated in~\fig{ParentGlueFigure}, 
we lump all of the $\P_i$ primitive amplitudes together, and similarly for 
the $\NP_i$ ones, so that the actual decomposition is
$$
\eqalign{ \A_{G}^{\rm adjoint} & = g^6 \biggl\{ 
  C^{\P}_{1234} \, A^{\P}_{G1234} + C^{\P}_{3421} \, A^{\P}_{G3421}
+ C^{\NP}_{1234} \, A^{\NP}_{G1234} + C^{\NP}_{3421} \, A^{\NP}_{G3421}
 \ +\ \C(234) \biggr\} \,. \cr}
\equn\label{GlueFColor}
$$

The equivalence of \eqns{FColor}{NewFColor} is not completely manifest.
While it is clear that
$$
C^{\P_i}_{1234} = C^\P_{1234} \,, \hskip 1 cm 
C^{\NP_i}_{1234} = C^\NP_{1234} \,, \hskip 1 cm i = 1,2,3,
\equn
$$
it is also true that 
$$
C^{\P_4}_{1234} \neq C^{\P}_{1234} \neq C^{\P_5}_{1234}.
$$
Nevertheless, the symmetry properties of the primitive amplitudes
presented in \sect{AmplitudesSection} are such that, after summing
over the permutations in \eqns{FColor}{NewFColor}, 
the two forms are equivalent.
The $\P_4$ primitive amplitudes, after loop integration, all turn out 
to be proportional to the same function, which is antisymmetric under
exchange of legs 1 and 2 (or 3 and 4).  In the permutation 
sum~(\ref{NewFColor}), this property induces an antisymmetric projection 
on the double planar box ($\P$) color factor, which removes certain unwanted
(subleading-color) terms, and renders it equivalent to the double 
triangle ($\P_4$) color factor~\cite{AllPlusTwo}.  Similarly, the 
double-scalar $\P_5$ primitive amplitude is symmetric under exchange 
of legs 1 and 2 (or 3 and 4), and the symmetric projection on its
$C^{\P_5}_{1234}$ color factor in~\eqn{FColor} renders it equivalent 
to $C^{\P}_{1234}$.
We emphasize that the simplified color decomposition~(\ref{NewFColor})
applies only to the pure gauge coupling, adjoint representation case, and 
not when either Yukawa couplings or fundamental representations are present.

The symmetries of the color factors $C^{\P}$ and $C^{\NP}$ can be read 
off the diagrams. The ones required in the sum~(\ref{NewFColor}) are
$$
C^\P_{4321} = C^\P_{1234} \,, \hskip 2 cm
C^\NP_{1243} = C^\NP_{1234} \,.
\equn\label{Csymmetries}
$$
The corresponding planar and non-planar primitive amplitudes share
the same symmetries with their associated color factors.  Due to
these symmetries, $\A_X^{\rm adjoint}(1^+,2^+,3^+,4^+)$ has the
required total ($S_4$) permutation symmetry, even though only six
permutations appear explicitly in \eqn{NewFColor}.  Although diagrams
with differing color factors contribute to each primitive
amplitude, roughly speaking, gauge invariance dictates that in the
final gauge invariant expression, all diagrams follow the lead of
the parent diagrams.  The final permutation sum then ensures that
diagrammatic contributions with seemingly incorrect color factors
receive the correct ones after assembly.

The color representation in \eqn{NewFColor} is not unique,
although it is a particularly symmetric one.  Indeed, the twelve color
factors appearing in \eqn{NewFColor} or~(\ref{GlueFColor})
satisfy a set of seven linear relations,
$$
\eqalign{
 C^\P_{1234} - C^\P_{2341}  = C^\NP_{1234} - C^\NP_{1423}\,, \hskip .4 cm
 C^\P_{1342} - C^\P_{3421} &= C^\NP_{1342} - C^\NP_{1234}\,, \hskip .4 cm
 C^\P_{1423} - C^\P_{4231}  = C^\NP_{1423} - C^\NP_{1342}\,, \cr
 C^\NP_{1234}  = C^\NP_{3421} \,, \hskip 3.1 cm
 C^\NP_{1342} &= C^\NP_{4231} \,, \hskip 3.1 cm
  C^\NP_{1423} = C^\NP_{2341} \,, \cr
 C^\NP_{1234} + C^\NP_{1342} + C^\NP_{1423} &= 0 \,, \cr }
\equn\label{Crelations}
$$
hence only five color factors are linearly independent.  The 
amplitude~(\ref{GlueFColor}) may be rewritten, for example, into 
the non-symmetric form,
$$
\eqalign{ \A_G^{\rm adjoint} &= g^6 \Bigl[ C^\P_{2341}
(A^\P_{G1234} + A^\P_{G2341}) + C^\P_{1342} (A^\P_{G1342} +
A^\P_{G3421}) + C^\P_{1423} (A^\P_{G1423} + A^\P_{G4231}) \cr &
+\; C^\NP_{1423}( A^\NP_{G1423} + A^\NP_{G2341} - A^\NP_{G1234}
                - A^\NP_{G3421}
                - A^\P_{G3421} - 2A^\P_{G1234} - A^\P_{G4231} ) \cr
& +\; C^\NP_{1342}( A^\NP_{G1342} + A^\NP_{G4231} - A^\NP_{G1234}
                - A^\NP_{G3421}
                - 2A^\P_{G3421} - A^\P_{G1234} + A^\P_{G4231} ) \Bigr] \,.\cr}
\equn\label{PrimitiveIndepDecom}
$$
where each color factor is linearly independent from the others for
$SU(N_c)$ with a generic value of $N_c$.
An advantage of a linearly independent form like~\eqn{PrimitiveIndepDecom}
is that the coefficient of each color factor must satisfy supersymmetry 
identities by itself.  The particular form~(\ref{PrimitiveIndepDecom}) is
convenient for analyzing the $s$-channel three-particle cuts shown 
in~\fig{TripleAmplFigure}.

We mention that although the original construction~\cite{N8Susy} of the
color organization~(\ref{GlueFColor}) made use of special properties of
$N=4$ supersymmetric amplitudes, the decomposition actually
holds for any purely adjoint two-loop amplitude, as long as all vertices
are built out of $\tilde{f}^{abc}$ factors.  In such cases, one can
make use of the Jacobi identity to rearrange the color factor of any
Feynman diagram into a combination of those appearing
in~\eqn{GlueFColor}~\cite{DDFColor}.

\subsection{Fundamental representation two-loop color decomposition}
\label{FundTwoLoopColorSubsection}

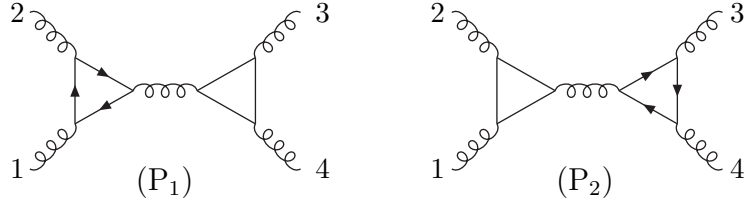
\begin{figure}[ht]
\begin{center}
\begin{picture}(250,70)(0,0)
\Text(46,0)[c]{\large ($\P_1$)}
\Text(-7,5)[r]{1} \Text(-7,64)[r]{2}
\Text(103,64)[l]{3} \Text(103,5)[l]{4}
\ArrowLine(12,22)(12,47) \ArrowLine(12,47)(34,34.5) \ArrowLine(34,34.5)(12,22)
\Line(80,47)(80,22) \Line(80,22)(58,34.5) \Line(58,34.5)(80,47)
\Gluon(34,34.5)(58,34.5){2.5}{3}
\Gluon(12,22)(-5,5){2.5}{3} \Gluon(-5,64)(12,47){2.5}{3}
\Gluon(97,5)(80,22){2.5}{3} \Gluon(80,47)(97,64){2.5}{3}
\Text(205,0)[c]{\large ($\P_2$)}
\Text(152,5)[r]{1} \Text(152,64)[r]{2}
\Text(260,64)[l]{3} \Text(260,5)[l]{4}
\Line(171,47)(171,22) \Line(193,34.5)(171,47) \Line(171,22)(193,34.5)
\ArrowLine(239,47)(239,22) \ArrowLine(239,22)(217,34.5)
\ArrowLine(217,34.5)(239,47)
\Gluon(193,34.5)(217,34.5){2.5}{3}
\Gluon(171,22)(154,5){2.5}{3} \Gluon(154,64)(171,47){2.5}{3}
\Gluon(256,5)(239,22){2.5}{3} \Gluon(239,47)(256,64){2.5}{3}
\end{picture}
\caption[a]{\label{ClosedFermionGluinoFigure}
\small The $\tildeF$ contributions with a closed fundamental 
representation matter fermion loop and a closed gluino loop. The lines with 
the arrows represent matter fermions.}
\end{center}
\end{figure}

\begin{figure}[ht]
\begin{center}
\begin{picture}(250,70)(0,0)
\Text(46,0)[c]{\large ($\P_1$)}
\Text(-7,5)[r]{1} \Text(-7,64)[r]{2}
\Text(103,64)[l]{3} \Text(103,5)[l]{4}
\DashLine(12,22)(12,47){3} \DashLine(12,47)(34,34.5){3} 
          \DashLine(34,34.5)(12,22){3}
\Line(80,47)(80,22) \Line(80,22)(58,34.5) \Line(58,34.5)(80,47)
\Gluon(34,34.5)(58,34.5){2.5}{3}
\Gluon(12,22)(-5,5){2.5}{3} \Gluon(-5,64)(12,47){2.5}{3}
\Gluon(97,5)(80,22){2.5}{3} \Gluon(80,47)(97,64){2.5}{3}
\Text(205,0)[c]{\large ($\P_2$)}
\Text(152,5)[r]{1} \Text(152,64)[r]{2}
\Text(260,64)[l]{3} \Text(260,5)[l]{4}
\Line(171,47)(171,22) \Line(193,34.5)(171,47) \Line(171,22)(193,34.5)
\DashLine(239,22)(239,47){3} \DashLine(217,34.5)(239,22){3}
\DashLine(239,47)(217,34.5){3}
\Gluon(193,34.5)(217,34.5){2.5}{3}
\Gluon(171,22)(154,5){2.5}{3} \Gluon(154,64)(171,47){2.5}{3}
\Gluon(256,5)(239,22){2.5}{3} \Gluon(239,47)(256,64){2.5}{3}
\end{picture}
\caption[a]{\label{ClosedScalarGluinoFigure}
\small The $\tildeS$ contributions with a closed fundamental 
representation matter scalar loop and a closed gluino loop. }
\end{center}
\end{figure}
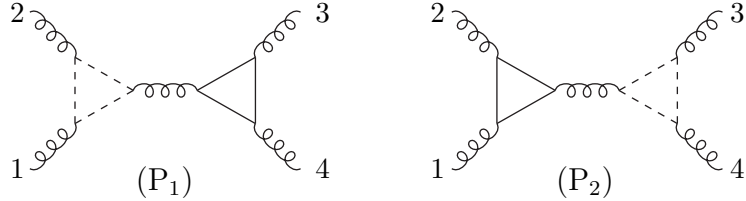

The color decomposition for four-gluon two-loop amplitudes
containing fundamental representation particles in the loops is similar
to the purely adjoint case~(\ref{FColor}),
$$
\eqalign{ \A^{\rm fund}_X = g^6 \sum_{\{\D_i\}} \Bigl[
\Bigl(F^{\D_i}_{1234} \, A^{\D_i}_{X1234} + F^{\D_i}_{3421} \,
A^{\D_i}_{X3421} \Bigr) + \C(234) \Bigr] \,,\cr} \equn
\label{FundamentalColor}
$$
where each term in the sum corresponds to a parent diagram in
figs.~\ref{ParentFermionFigure}--\ref{ClosedScalarGluinoFigure}.  In
this case, it is useful to include separate $\tildeS$ and $\tildeF$
contributions with a closed gluino loop, and either a closed matter fermion
or a closed scalar loop, as depicted in
\figs{ClosedFermionGluinoFigure}{ClosedScalarGluinoFigure}.  Again the
label $X \in \nobreak\{S,F,M,\tildeS,\tildeF\}$ specifies the figure
containing the parent diagram.  The values of the $F^{\D_i}$
fundamental representation color coefficients are read off from these
figures by dressing the parent diagrams with standard Feynman rule
color factors, as in~\sect{AdjTwoLoopColorSubsection} but taking into
account the different representations here.  For example, for the
fermion loop $F$ amplitudes, the color factor associated with each
primitive amplitude is obtained by assigning an $\tilde{f}^{abc}$ to
each three-gluon vertex and a $(T^a)^{i}_{\bar{\jmath}}$ to each
quark-anti-quark-gluon vertex, and summing over the two directions of
the fermion arrows.

The color factors obtained by dressing the $\P_1$ and $\P_2$ diagrams in
\figs{ClosedFermionGluinoFigure}{ClosedScalarGluinoFigure} with 
$\tilde f^{abc}$ and $(T^a)^{i}_{\bar{\jmath}}$ factors are not equal
to the color factors $F^{\P_1}_{1234}$ or $F^{\P_2}_{1234}$ obtained from
figs.~\ref{ParentFermionFigure}--\ref{ParentMixedFigure}; they differ
in the $\Tr (T^{a_1}T^{a_2})\Tr (T^{a_3}T^{a_4})$ term.
Nevertheless, after the permutation sum in \eqn{FundamentalColor}, this
discrepancy cancels, in the same way that the discrepancy between 
$C^{\P_4}_{1234}$ and $C^\P_{1234}$ cancelled in the
purely adjoint representation case.  This feature allows us to maintain a
uniform set of color factors for all contributions with fundamental
representation matter; {\it i.e.}, in \eqn{FundamentalColor} there is
no need to include an $X$ label on the color factors.

For the fundamental representation contributions to the 
four-gluon amplitudes, the explicit values of the color factors are
$$
\eqalign{
F^{\P_1}_{1234} &= F^{\P_2}_{1234}  =
N_c\left[ \Tr (T^{a_1}T^{a_2}T^{a_3}T^{a_4}) +
      \Tr (T^{a_4}T^{a_3}T^{a_2}T^{a_1}) \right] +
             2\Tr (T^{a_1}T^{a_2})\Tr (T^{a_3}T^{a_4})\,, \cr
F^{\P_3}_{1234} &=  F^{\P_5}_{1234} 
  =  -\frac{1}{N_c}\left[ \Tr (T^{a_1}T^{a_2}T^{a_3}T^{a_4}) +
                       \Tr (T^{a_4}T^{a_3}T^{a_2}T^{a_1}) \right] +
             2\Tr (T^{a_1}T^{a_2})\Tr (T^{a_3}T^{a_4}) \,,\cr
F^{\P_4}_{1234} &=  \Tr (T^{a_1}T^{a_2}T^{a_3}T^{a_4}) +
       \Tr (T^{a_4}T^{a_3}T^{a_2}T^{a_1})
        -\frac{2}{N_c}\Tr (T^{a_1}T^{a_2})\Tr (T^{a_3}T^{a_4})\,, \cr
F^{\NP_1}_{1234} &= F^{\NP_2}_{1234}  =
2\Tr (T^{a_1}T^{a_2})\Tr (T^{a_3}T^{a_4})\,, \cr
F^{\NP_3}_{1234} &= -2\left[ \Tr (T^{a_1}T^{a_3})\Tr (T^{a_2}T^{a_4})+
\Tr (T^{a_1}T^{a_4})\Tr(T^{a_2}T^{a_3})\right] \,. \cr}
\equn\label{ExplicitColorFundGluons}
$$
We have dropped terms containing $\Tr(T^{a_i})$ 
from~\eqn{ExplicitColorFundGluons}, since they vanish for $SU(N_c)$ gluons. 
Such terms would have to be restored in order to describe two-loop 
amplitudes with a mixture of external gluons and photons, as in 
ref.~\cite{GGGamGam}.

As in the case of the adjoint representation color 
decomposition~(\ref{NewFColor}), the 
decomposition~(\ref{FundamentalColor}) is
not unique;  there are linear relations between the various color
factors.  In particular, in the case of four external gluons, after
eliminating redundant color factors, using the explicit representation
of the color factors in terms of color traces
(\ref{ExplicitColorFundGluons}), one finds a total of six independent
color factors.  Also note that the separation into primitive
amplitudes is in many cases artificial.  For example, since
$F^{\P_1}_{1234} = F^{\P_2}_{1234}$, the distinction between
$A^{\P_1}_{1234}$ and $A^{\P_2}_{1234}$ is moot, when all four 
external legs are gluons.

\subsection{Assembly of complete amplitudes}
\label{AssemblySubsection}

We can specify the amplitudes in various theories in terms of the
primitive amplitudes introduced above, and presented
in~\sect{AmplitudesSection} for the ++++ helicity configuration.
For example, the four-gluon 
amplitude in QCD with $\Nf$ massless quark flavors is,
$$
\A_{\rm QCD} = \A_G^{\rm adjoint} + \Nf \A_F^{\rm fund (1)} 
 + \Nf^2 \A_F^{\rm fund (2)} \,.
\equn\label{QCDFullAmplitude}
$$
The pure glue contribution $\A_G^{\rm adjoint}$ is given in terms
of primitive amplitudes via \eqn{GlueFColor}, where the sum in that
equation runs over the two diagrams in \fig{ParentGlueFigure}.
Similarly, the single quark loop contribution $\A_F^{\rm fund (1)}$ is
given via \eqn{FundamentalColor}, where the sum runs over all
diagrams in \fig{ParentFermionFigure} except ($\P_4$), which gives the
double quark loop contribution $\A_F^{\rm fund (2)}$.

Using the supersymmetric Lagrangians in \sect{SWISection} as a guide,
we may combine the primitive amplitudes given in
\sect{AmplitudesSection} into supersymmetric combinations, which must
satisfy the SWI (\ref{AllPlusSusyIdentity}).
The amplitude for $N=1$ super-QCD with $\Nf$ matter
multiplets in the fundamental representation is,
$$
\eqalign{ \A_{N=1} & = \A_G^{\rm adjoint} + \A_F^{\rm adjoint} \cr
& +  \Nf \Bigl( \A_F^{\rm fund (1)} +
                \A_S^{\rm fund (1)} +
                \A_M^{\rm fund(1)} +
                \A_\tildeF^{\rm fund(1)} +
                \A_\tildeS^{\rm fund(1)} \Bigr) \cr
& + \Nf^2 \Bigl( \A_F^{\rm fund (2)} +
                 \A_S^{\rm fund (2)} +
                 \A_M^{\rm fund(2)} \Bigr) \,, \cr}
\equn\label{N=1FullAmplitude}
$$
which is expressed in terms of primitive amplitudes via
eqs.~(\ref{NewFColor}), (\ref{GlueFColor}), and (\ref{FundamentalColor}).  
The sum over parent diagrams of the form $A_{1234}$ for each contribution is
$$
\eqalign{ & \A_G^{\rm adjoint} : \,
\hbox{Fig.~\ref{ParentGlueFigure} }, \{D_i\} = \{\P, \NP\} \,, \cr
& \A_F^{\rm adjoint} : \, \hbox{Fig.~\ref{ParentFermionFigure} },
\{D_i\} = \{ \P_1, \P_2, \P_3, \P_4,
                                    \NP_1, \NP_2, \NP_3 \} \,, \cr
& \A_F^{\rm fund (1)} : \, \hbox{Fig.~\ref{ParentFermionFigure} },
\{D_i\} = \{ \P_1, \P_2,\P_3,
                                     \NP_1, \NP_2, \NP_3 \} \,, \cr
& \A_S^{\rm fund (1)} : \, \hbox{Fig.~\ref{ParentScalarFigure} },
\{D_i\} = \{ \P_1, \P_2, \P_3, \P_5,
                                     \NP_1, \NP_2, \NP_3 \} \,, \cr
& \A_M^{\rm fund (1)} : \, \hbox{Fig.~\ref{ParentMixedFigure} },
\{D_i\} = \{ \P_1, \P_2, \P_3,
                                      \NP_1, \NP_2, \NP_3 \} \,, \cr
& \A_\tildeF^{\rm fund (1)} : \, \hbox{Fig.~\ref{ClosedFermionGluinoFigure} },
\{D_i\} = \{ \P_1, \P_2 \} \,, \cr
& \A_\tildeS^{\rm fund (1)} : \, \hbox{Fig.~\ref{ClosedScalarGluinoFigure} },
\{D_i\} = \{ \P_1, \P_2 \} \,, \cr
& \A_F^{\rm fund (2)} : \, \hbox{Fig.~\ref{ParentFermionFigure} },
\{D_i\} = \{ \P_4 \} \,, \cr
& \A_S^{\rm fund (2)} : \, \hbox{Fig.~\ref{ParentScalarFigure} },
\{D_i\} = \{ \P_4 \} \,, \cr
& \A_M^{\rm fund (2)} : \, \hbox{Fig.~\ref{ParentMixedFigure} },
\{D_i\} = \{ \P_4 \} \,. \cr
}
\equn\label{PrimitiveContrib}
$$
Note that $\A_S^{\rm fund (2)}$ does not receive a contribution from 
$\P_5$ because of a cancellation between fields $A_i$ and 
$\tilde{A}_i$ circulating in one of the loops.


For the case of a matter multiplet in the adjoint representation and the
superpotential $W = {1\over3} g \xi \Tr \Phi^3$, the
contributions depending on $\xi$ are
$$
\A_{N=1}^\yukawa = |\xi|^2 \bigl( \A_S^\yukawa +  \A_M^\yukawa \bigr) \,,
\equn\label{YukawaAmplitude}
$$
where the contributing parent diagrams are,
$$
\eqalign{ & \A_S^\yukawa : \,
\hbox{Fig.~\ref{ParentScalarFigure} }, \{D_i\} = \{\P_5 \} \,, \cr
& \A_M^\yukawa : \, \hbox{Fig.~\ref{ParentMixedFigure} },
\{D_i\} = \{ \P_1, \P_2, \P_3, \NP_1, \NP_2, \NP_3 \} \,. \cr } 
\equn\label{YukawaPrimitiveContrib}
$$
In \eqn{YukawaAmplitude}, the implicit color factors appearing are
generated by the ``dressing rules'' given in 
\sect{AdjTwoLoopColorSubsection}.

One can also consider theories with a higher degree of supersymmetry.
For example, a pure $N=2$ super-Yang-Mills theory contains an 
$N=2$ vector supermultiplet composed of a vector, two gluinos and 
two scalar states.  This content can be viewed as an $N=1$ theory 
with one adjoint matter multiplet. Modifying \eqn{N=1FullAmplitude} for
$\Nf=1$, the pure $N=2$ amplitude is
$$
\eqalign{ \A_{N=2} & = \A_G^{\rm adjoint} + \A_F^{\rm adjoint} 
    + \A_F^{\rm adjoint (1)} + \A_S^{\rm adjoint (1)}
    + \A_M^{\rm adjoint (1)} \cr
& \hskip 1 cm
    + \A_\tildeF^{\rm adjoint (1)} + \A_\tildeS^{\rm adjoint (1)}
    + \A_F^{\rm adjoint (2)} + \A_S^{\rm adjoint (2)}
    + \A_M^{\rm adjoint (2)} \cr
& = \A_G^{\rm adjoint} +
       2 \, \A_F^{\rm adjoint (1)} +
                \A_S^{\rm adjoint (1)} +
                  \, \A_M^{\rm adjoint (1)}
    +    4 \, \A_F^{\rm adjoint (2)} +
                \A_S^{\rm adjoint (2)} +
                 2 \,\A_M^{\rm adjoint (2)} \,. \cr}
\equn\label{N=2FullAmplitude}
$$
In the second step we have used trivial identities for the 
adjoint representation,
$$
\eqalign{ \A_F^{\rm adjoint} &= \A_F^{\rm adjoint (1)} 
+ \A_F^{\rm adjoint (2)} \,, \cr
\A_\tildeF^{\rm adjoint (1)} &= 2 \, \A_F^{\rm adjoint (2)} \,, \cr
\A_\tildeS^{\rm adjoint (1)} &= \A_M^{\rm adjoint (2)} \,. \cr}
\equn\label{AdjId}
$$
The contributing parent diagrams for $\A_X^{\rm adjoint ({\it i})}$
are the same as for $\A_X^{\rm fund ({\it i})}$ in~\eqn{PrimitiveContrib}, 
except that the associated 
color factors are the ones for the adjoint representation, {\it i.e.},
$C^{\D_i}_{1234}$. 
(For the $N=2$ case, the arrows on the fermion lines in the figures
are not important, although in \fig{ParentMixedFigure} they can be
used to distinguish between the two species of gluinos.)

\section{Identical-helicity two-loop amplitudes}
\label{AmplitudesSection}

In this section we review the results previously obtained for the pure gluon
and scalar loop contributions to the ++++ helicity 
amplitude~\cite{AllPlusTwo}.  Then we present our results for the QCD 
fermion loop contributions, followed by the contributions involving Yukawa
couplings, which contribute to some of the supersymmetry Ward identities. 
The amplitudes presented in this section have not been renormalized;
in~\sect{RenormalizationSection} we discuss their renormalization.
The overall normalization of the primitive amplitudes below are
such that for the gauge case they can be inserted directly into 
\eqns{FColor}{FundamentalColor} without any additional combinatoric
factors.c

\subsection{Pure glue primitive amplitudes}
\label{PureGluePrimitiveSubsection}

Using \eqn{GlueFColor}, the pure gluon two-loop amplitudes are conveniently
expressed in terms of planar and non-planar primitive
amplitudes whose explicit values are~\cite{AllPlusTwo}:
$$
\eqalign{
A^\P_{G1234}
& =
\rho
\biggl\{
s_{12} \, \I_4^\P \Bigl[ (D_s-2) ( \mud_p^2 \, \mud_q^2
         + \mud_p^2 \, \mud_{p+q}^2  + \mud_q^2 \, \mud_{p+q}^2 )
       + 16 \Bigl( (\mud_p \cdot \mud_q)^2 - \mud_p^2 \, \mud_q^2 \Bigr)
                    \Bigr](s_{12},s_{23}) \cr
& \hskip 2 cm
+ 4 \, (D_s-2) \, \I_4^\bowtie[(\mud_p^2 + \mud_q^2)
                \, (\mud_p \cdot \mud_q) ] (s_{12}) \cr
& \hskip 2 cm
+ {(D_s-2)^2 \over s_{12}} \, \I_4^\bowtie\Bigl[
      \mud_p^2 \, \mud_q^2 \, ( (p+q)^2 + s_{12} ) \Bigr] (s_{12},s_{23})
              \biggr\}
                           \,, \cr}
\equn\label{PlanarPrimitiveGlue}
$$
$$
A^\NP_{G1234}
=
\rho \, s_{12}
\I_4^\NP \Bigl[ (D_s-2) ( \mud_p^2 \, \mud_q^2
                   + \mud_p^2 \, \mud_{p+q}^2
                   + \mud_q^2 \, \mud_{p+q}^2 )
+ 16 \Bigl( (\mud_p \cdot \mud_q)^2
           - \mud_p^2 \, \mud_q^2 \Bigr) \Bigr](s_{12},s_{23})
\,,
\equn \label{NonPlanarPrimitiveGlue}
$$
where the parent diagrams corresponding to these two primitive amplitudes
are displayed in \fig{ParentGlueFigure}, and $\rho$ is defined
in~\eqn{RhoDef}.  The other primitive amplitudes appearing in
\eqn{GlueFColor} are just relabelings of these two basic ones.

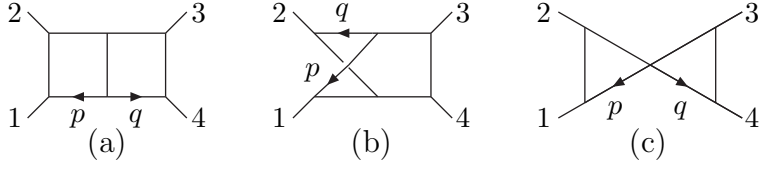
\begin{figure}[ht]
\begin{center}
\begin{picture}(280,69)(0,0)
\Text(3,10)[r]{1} \Text(3,50)[r]{2}
\Line(5,10)(13,18)  \Line(5,50)(13,42)
\Line(57,18)(65,10)  \Line(57,42)(65,50)
\ArrowLine(35,18)(13,18) \ArrowLine(35,18)(57,18)  \Line(13,42)(57,42)
\Line(13,18)(13,42) \Line(35,18)(35,42) \Line(57,18)(57,42)
\Text(67,50)[l]{3} \Text(67,10)[l]{4}
\Text(24,10)[c]{$p$}
\Text(46,10)[c]{$q$}
\Text(35,0)[c]{\large (a)}
\Text(103,10)[r]{1} \Text(103,50)[r]{2}
\Line(105,10)(113,18) \ArrowLine(126,30)(113,18) \Line(126,30)(137,42)  
\Line(105,50)(124,31) \Line(126,29)(137,18)
\Line(157,18)(165,10)  \Line(157,42)(165,50)
\Line(113,18)(157,18)  \Line(157,42)(135,42)
\ArrowLine(135,42)(113,42) \Line(157,18)(157,42)
\Text(167,50)[l]{3} \Text(167,10)[l]{4}
\Text(124,50)[c]{$q$}
\Text(110,27)[l]{$p$}
\Text(135,0)[c]{\large (b)}
\Text(203,10)[r]{1} \Text(203,50)[r]{2}
\Line(205,10)(274.3,50) \ArrowLine(239.65,30)(215,15.77) 
\Line(239.65,30)(274.3,50)   
\Line(205,50)(274.3,10) \ArrowLine(239.65,30)(264.3,15.77) 
\Line(264.3,15.77)(274.3,10)
\Line(215,15.77)(215,44.23) \Line(264.3,15.77)(264.3,44.23)
\Text(276.3,50)[l]{3} \Text(276.3,10)[l]{4}
\Text(227.3,13)[c]{$p$}
\Text(252,13)[c]{$q$}
\Text(239.65,0)[c]{\large (c)}
\end{picture}
\caption[]{
\label{ParentsFigure}
\small Integral topologies appearing in the two-loop identical-helicity 
amplitudes, and the loop momentum routings: (a) the planar double box
integral; (b) the non-planar double box integral; (c) the bow-tie
integral. The arrows here denote the direction of momentum flow.}
\end{center}
\end{figure}

The two-loop momentum integrals appearing 
in~\eqns{PlanarPrimitiveGlue}{NonPlanarPrimitiveGlue} are defined 
as follows.  The planar double box integral
$$
\eqalign{
\I_4^\P [{\cal P} & (\mud_i, p,q,k_i)] (s_{12},s_{23}) \cr
& \equiv \int
{d^{D}p\over (2\pi)^{D}} \,
{d^{D}q\over (2\pi)^{D}}\,
{ {\cal P} (\mud_i, p,q,k_i) \over
     p^2\, q^2\, (p+q)^2 (p - k_1)^2 \,(p - k_1 - k_2)^2 \,
        (q - k_4)^2 \, (q - k_3 - k_4)^2 }  \cr}
\equn\label{PlanarInt}
$$
is displayed in \fig{ParentsFigure}(a).  The numerator factor 
${\cal P} (\mud_i, p,q,k_i)$ is a polynomial in the momenta.
The vectors $\vec\mud_p$, $\vec\mud_q$ represent
the $(-2\eps)$-dimensional components of the loop momenta $p$ and $q$.
We also define
$\mud_p^2 \equiv \vec{\mud}_p \cdot \vec{\mud}_p \geq 0$, 
$\; \; \mud_q^2 \equiv \vec{\mud}_q \cdot \vec{\mud}_q$,
and $\mud_{p+q}^2 \equiv (\vec{\mud}_p + \vec{\mud}_q)^2 
= \mud_p^2 + \mud_q^2 + 2 \vec{\mud}_p \cdot \vec{\mud}_q$.  
The bow-tie integral $\I_4^\bowtie$ shown in \fig{ParentsFigure}(c) is
defined by
$$
\eqalign{
\I_4^{\bowtie} [{\cal P} & (\mud_i, p,q,k_i)] (s_{12}) \cr
& \equiv \int
{d^{D}p\over (2\pi)^{D}} \,
{d^{D}q\over (2\pi)^{D}}\,
{ {\cal P} (\mud_i, p,q,k_i) \over
     p^2\, q^2\, (p - k_1)^2 \,(p - k_1 - k_2)^2 \,
        (q - k_4)^2 \, (q - k_3 - k_4)^2 }\,. \cr}
\equn\label{BowTieInt}
$$
The non-planar double box integral, depicted in
\fig{ParentsFigure}(b), is given by
$$
\eqalign{
\I_4^\NP [{\cal P} (&\mud_i, p,q,k_i)] (s_{12},s_{23}) \cr
& \equiv \int \! {d^{D} p \over (2\pi)^{D}} \,
        {d^{D} q \over (2\pi)^{D}}\,
{{\cal P} (\mud_i, p, q, k_i) \over p^2\, q^2\, (p+q)^2 \,
         (p-k_1)^2 \,(q-k_2)^2\,
   (p+q+k_3)^2 \, (p+q+k_3+k_4)^2}\,. \cr}
\equn\label{NonPlanarInt}
$$
The explicit values of the integrals, as a Laurent series in $\e$ through
$\Ord(\e^0)$, and expressed in terms of polylogarithms~\cite{Lewin},
may be found in appendix A of ref.~\cite{AllPlusTwo}.  
We have checked that these values agree with results obtained using
general integration
methods~\cite{PBScalar,NPBScalar,PBReduction,NPBReduction,GRReduction,AGO}.
The integral
$$
\I_4^\bowtie[(\mud_p^2 + \mud_q^2)
                \, (\mud_p \cdot \mud_q) ] (s_{12})\,,
\equn
$$
in \eqn{PlanarPrimitiveGlue} vanishes identically, due to antisymmetry
of its integrand in $\vec{\mud}_p \to - \vec{\mud}_p$.  However, in 
constructing a consistent set of cut integrands it is useful to keep it
around, as well as related integrals found in the next subsections.

%

\subsection{Scalar loop primitive amplitudes}
\label{ScalarLoopPrimitiveSubsection}

The planar scalar loop amplitudes have already been presented in
ref.~\cite{AllPlusTwo}, where they were used as a guide for the
construction of the pure glue amplitudes discussed above.  Here we
reorganize the results a bit to be compatible with
\eqn{FundamentalColor} for the full amplitudes for the case of
fundamental representation scalars.  We also present the non-planar
contributions.

The scalar loop primitive amplitudes corresponding to the planar parent
graphs in \fig{ParentScalarFigure} are,
$$
\eqalign{
A^{\P_1}_{S1234} & =
2 \rho \biggl\{
  s_{12} \, \I_4^\P[ \mud_p^2 \, \mud_{p+q}^2](s_{12},s_{23})  \cr
& \hskip 1 cm
+  {(D_s-2) \over s_{12}} \, \I_4^\bowtie[
      \mud_p^2 \, \mud_q^2 \, ( (p+q)^2 + s_{12} ) ] (s_{12},s_{23}) \cr
& \hskip 1 cm
+ 4 \, \I_4^\bowtie[ \mud_p^2 \, (\mud_p \cdot \mud_q) ] (s_{12}) \biggr\}
 \,, \cr
A^{\P_2}_{S1234} & =
 2 \rho \biggl\{
  s_{12} \, \I_4^\P[\mud_q^2 \, \mud_{p+q}^2](s_{12},s_{23})  \cr
& \hskip 1 cm
+  {(D_s-2) \over s_{12}} \, \I_4^\bowtie[
      \mud_p^2 \, \mud_q^2 \, ( (p+q)^2 + s_{12} ) ] (s_{12},s_{23}) \cr
& \hskip 1 cm
+ 4 \, \I_4^\bowtie[ \mud_q^2 \, (\mud_p \cdot \mud_q) ] (s_{12})
  \biggr\}
 \,, \cr
A^{\P_3}_{S1234} & =
2 \rho \biggl\{
s_{12} \, \I_4^\P[\mud_p^2 \mud_q^2](s_{12}, s_{23})
      + {1\over 2} \I_4^\bowtie[\mud_p^2 \mud_q^2](s_{12}) \biggr\}
\,,\cr
A^{\P_4}_{S1234} & =
   {4 \rho \over s_{12}} \,
   \I_4^\bowtie\Bigl[ \mud_p^2 \, \mud_{q}^2 \,
    ( (p+q)^2 + \hf s_{12} ) \Bigr](s_{12},s_{23})
\,, \cr
A^{\P_5}_{S1234} & =
 \rho \,
   \I_4^\bowtie[\mud_p^2 \, \mud_{q}^2 ](s_{12})
\,. }
\equn\label{PlanarPrimitiveScalar}
$$
Similarly, the  scalar loop primitive amplitudes corresponding
to the non-planar parent graphs in \fig{ParentScalarFigure} are,
$$
\eqalign{
A^{\NP_1}_{S1234} &=
2 \rho \, s_{12} \, \I_4^\NP[ \mud_p^2 \, \mud_{p+q}^2] (s_{12},s_{23})
 \,, \cr
A^{\NP_2}_{S1234} & =
2 \rho \, s_{12} \, \I_4^\NP[\mud_q^2 \, \mud_{p+q}^2](s_{12},s_{23})
\,, \cr
A^{\NP_3}_{S1234} & =
2 \rho \, s_{12} \, \I_4^\NP[\mud_p^2 \mud_q^2](s_{12}, s_{23})
\,.\cr}
\equn\label{NonPlanarPrimitiveScalar}
$$
The integrals appearing in
\eqns{PlanarPrimitiveScalar}{NonPlanarPrimitiveScalar} appear in the
pure gluon case as well; see appendix A of ref.~\cite{AllPlusTwo} for their
explicit values through $\Ord(\e^0)$.  We remind the reader that the
number of scalar states propagating in the above primitive amplitudes is
$2 \times (N_c^2-1)$ in the adjoint representation case (when they are
dressed with color using \eqn{FColor}), and $4 \times N_c$ states for the
$N_c + {\bar N}_c$ fundamental representation case (when they are dressed
with color using \eqn{FundamentalColor}).  In either case, this number
matches the number of states of the fermion superpartners in the fermion
loop primitive amplitudes presented in \sect{FermionLoopPrimitiveSubsection}.

A convenient way of organizing the numerators of the planar and non-planar
double box integrals is in terms of the labels of particles circulating in
the loops.  Because the external momenta $k_1, k_2, k_3, k_4$ are defined
to have vanishing extra dimensional components, as one follows a scalar
line around only two different extra dimensional momenta appear, which we
can label by $\lambda_{s_1}$ and $\lambda_{s_2}$. For example, in
\fig{ParentScalarFigure}($\P_3$) the two distinct extra dimensional
scalar momenta are $\lambda_p$ and $\lambda_q$.  Using these labels, the
numerator arguments of the planar and non-planar double box integrals in
~\eqns{PlanarPrimitiveScalar}{NonPlanarPrimitiveScalar} all have a uniform 
structure,
$$
\lambda_{s_1}^2 \lambda_{s_2}^2 \,.
\equn
$$
This observation simplifies the bookkeeping when evaluating the 
three-particle cuts.

\subsection{Fermion loop primitive amplitudes}
\label{FermionLoopPrimitiveSubsection}

The structure of the fermion loop primitive amplitudes is quite similar to
that of the scalar loop primitive amplitudes.  The parent diagrams
are described by \fig{ParentFermionFigure}, and the results are,
$$
\eqalign{
A^{\P_1}_{F1234} & =
\rho \biggl\{
s_{12} \, \I_4^\P \Bigl[
-{\mbox{$\frac{1}{2}$}}(D_s-2) ( \mud_{p+q} \cdot \mud_p \mud_q^2 )
-2\mud_p^2 \mud_{p+q}^2 -4((\mud_p \cdot \mud_q)^2 -
   \mud_p^2 \mud_q^2) \Bigr](s_{12}, s_{23}) \cr
& \hskip 1 cm
  - 2\frac{D_s-2}{s_{12}}
  \, \I_4^\bowtie[\mud_p^2 \mud_q^2((p+q)^2+s_{12})](s_{12},s_{23}) \cr
& \hskip 1 cm
 - (D_s+6) \, \I_4^\bowtie[\mud_q^2\mud_p \cdot \mud_q](s_{12}) \biggr\}
 \,,\cr
A^{\P_2}_{F1234} & =
 \rho \biggl\{
s_{12} \, 
\I_4^\P\Bigl[ 
- {\mbox{$\frac{1}{2}$}}(D_s-2) ( \mud_{p+q} \cdot \mud_q \mud_p^2 )
-2\mud_q^2 \mud_{p+q}^2 -
4((\mud_p \cdot \mud_q)^2 - \mud_p^2 \mud_q^2) \Bigr](s_{12}, s_{23}) \cr
& \hskip 1 cm
  - 2\frac{D_s-2}{s_{12}}
  \, \I_4^\bowtie[\mud_p^2 \mud_q^2((p+q)^2+s_{12})](s_{12},s_{23}) \cr
& \hskip 1 cm
 - (D_s+6) \, \I_4^\bowtie[\mud_p^2\mud_p \cdot \mud_q](s_{12}) \biggr\}
\,,\cr
A^{\P_3}_{F1234} &=
\rho \biggl\{
s_{12} \,
\I_4^\P\Bigl[ 
{\mbox{$\frac{1}{2}$}}(D_s-2) ( \mud_p \cdot \mud_q \mud_{p+q}^2 )
-2\mud_p^2 \mud_q^2 -4 ((\mud_p \cdot \mud_q)^2 - \mud_p^2 \mud_q^2) \Bigr]
(s_{12}, s_{23})  \cr
& \hskip 1 cm
  +  \Bigl(\mbox{$\frac{1}{2}$}(D_s-2)-2 \Bigr)
    s_{12} \, \I_4^\bowtie[\mud_p \cdot \mud_q ](s_{12})
+ (D_s-2) \, \I_4^\bowtie[\mud_p^2 \mud_q^2](s_{12}) 
\biggr\}
      \,, \cr
A^{\P_4}_{F1234} & =
 \rho \biggl\{
 \I_4^\bowtie\Bigl[ (\mud_p \cdot \mud_q)
    [ 2 (\mud_p^2 +\mud_q^2) - \hf s_{12} ] \Bigr](s_{12}) \cr
& \hskip 1 cm
  + \frac{4}{s_{12}} \I_4^\bowtie\Bigl[ \mud_p^2 \mud_q^2
  ((p+q)^2 + \mbox{$\frac{1}{2}$}s_{12})\Bigr] (s_{12},s_{23})  \biggr\}
\,,\cr}
\equn\label{PlanarPrimitiveFermion}
$$
$$
\eqalign{
A^{\NP_1}_{F1234} & =
\rho \, s_{12} \,
\I_4^\NP[ -{\mbox{$\frac{1}{2}$}}(D_s-2) ( \mud_{p+q} \cdot \mud_p \mud_q^2 )
-2\mud_p^2 \mud_{p+q}^2 -4((\mud_p \cdot \mud_q)^2 -
  \mud_p^2 \mud_q^2) ](s_{12}, s_{23})
\,, \cr
A^{\NP_2}_{F1234} & =
\rho \, 
s_{12} \,
\I_4^\NP[ - {\mbox{$\frac{1}{2}$}}(D_s-2) ( \mud_{p+q} \cdot \mud_q \mud_p^2 )
-2\mud_q^2 \mud_{p+q}^2 -
4((\mud_p \cdot \mud_q)^2 - \mud_p^2 \mud_q^2) ](s_{12}, s_{23})
   \,, \cr
A^{\NP_3}_{F1234} & =
\rho \,
s_{12} \,
\I_4^\NP[ {\mbox{$\frac{1}{2}$}}(D_s-2) ( \mud_p \cdot \mud_q \mud_{p+q}^2 )
-2\mud_p^2 \mud_q^2 -4 ((\mud_p \cdot \mud_q)^2 - \mud_p^2 \mud_q^2)]
(s_{12}, s_{23})
 \,.\cr}
\equn\label{NonPlanarPrimitiveFermion}
$$
As was the case for the scalar loop amplitudes, the normalization of the 
fermion loop amplitudes (when
dressed with color using \eqns{FColor}{FundamentalColor})
corresponds to $2 \times (N_c^2-1)$ states
for the adjoint representation case and $4 \times N_c$ states for the
$N_c + {\overline N}_c$ fundamental representation case.
Most of the integrals in
\eqns{PlanarPrimitiveFermion}{NonPlanarPrimitiveFermion} are again the
same ones as for the pure glue case~\cite{AllPlusTwo}.  There are, however, a
few new integrals.  In ref.~\cite{ggggpaper} these amplitudes, together
with the other helicity configurations, are given directly in terms 
of polylogarithms, after subtracting off certain universal pole terms in 
$\e$~\cite{CataniDiv}.

Just as we did for the scalar loop amplitudes, we can write the 
numerator arguments of the loop integrals 
in terms of the two distinct extra-dimensional momenta
flowing in the fermion propagators, which we call $\lambda_{f_1}$ and
$\lambda_{f_2}$, and the one in the internal gluon propagator, labeled as
$\lambda_g$.  In this case, the numerators appearing in the planar and
non-planar double-box integrals all have the form,
$$
-{\mbox{$\frac{1}{2}$}}(D_s-2)(\mud_{f_1} \cdot \mud_{f_2} \mud_g^2 )
-2\mud_{f_1}^2 \mud_{f_2}^2
-4 ((\mud_{f_1} \cdot \mud_{f_2})^2 - \mud_{f_1}^2 \mud_{f_2}^2) \,,
\equn\label{FLoopLabel}
$$
using the fact that
$$
(\mud_p \cdot \mud_q)^2 - \mud_p^2 \mud_q^2
\ = \ (\mud_{p+q} \cdot \mud_p)^2 - \mud_{p+q}^2 \mud_p^2
\ = \ (\mud_{p+q} \cdot \mud_q)^2 - \mud_{p+q}^2 \mud_q^2 \,.
\equn
$$


\subsection{Mixed scalar and fermion amplitudes}
\label{MixedPrimitiveSubsection}

In order to fully study the supersymmetry Ward identities, we need
also the ``mixed'' contributions where both fermions and scalars
appear in the loops.  The parent diagrams associated with these
contributions are shown in figure \ref{ParentMixedFigure}. In these
diagrams, the solid line represents fermions, while the dashed line
represents scalars.  These diagrams, and primitive amplitudes,
describe the pure gauge coupling contributions
$\A^{\rm fund}_M$, for which one of the fermions is a gluino, 
via~\eqn{PrimitiveContrib}.
They simultaneously describe the superpotential ($\xi$-dependent)
contributions, for which both of the fermions are matter fermions,
via~\eqn{YukawaPrimitiveContrib}.

Following the notation of the previous sections we find that the
explicit forms of the planar mixed primitive amplitudes are
$$
\eqalign{
A^{\P_1}_{M1234} & =
 \rho \Bigl\{ s_{12} \, \I_4^\P[ \mud_p \cdot \mud_q \mud_{p+q}^2
                                - \mud_q \cdot \mud_{p+q} \mud^2_p]
                              (s_{12}, s_{23})  
    +  s_{12} \, \I_4^\bowtie[\mud_p \cdot  \mud_q]
   (s_{12})          \Bigr\}
\,,\cr
A^{\P_2}_{M1234} & =
 \rho \Bigl\{ s_{12} \, 
\I_4^\P[  \mud_p \cdot \mud_q \mud_{p+q}^2
        - \mud_p \cdot \mud_{p+q} \mud^2_q] (s_{12}, s_{23})  
    +  s_{12} \, \I_4^\bowtie[\mud_p \cdot  \mud_q]
   (s_{12}) 
                 \Bigr\}
\,,\cr
A^{\P_3}_{M1234} & =
- \rho \Bigl\{
  s_{12} \, \I_4^\P[\mud_p \cdot \mud_{p+q} \mud^2_q
                   + \mud_q \cdot \mud_{p+q} \mud^2_p](s_{12}, s_{23})
   + 4 \, \I_4^\bowtie[\mud_p^2 \mud_q^2](s_{12})              \Bigr\}
    \,,\cr}
\equn
$$
corresponding to the first three diagrams in \fig{ParentMixedFigure}.
For the case where the fermion and scalar loops are separated from
each other, which does not contribute in the superpotential case, 
we have
$$
\eqalign{
A^{\P_4}_{M1234} & =
- 2 \rho \biggl\{
\I_4^\bowtie[ \mud_p \cdot \mud_q (\mud_p^2 + \mud_q^2) ]
   (s_{12}) 
+ \frac{4}{s_{12}} \I_4^\bowtie\Bigl[\mud_p^2 \mud_q^2
  ((p+q)^2 + \hf s_{12})\Bigr] (s_{12},s_{23}) \biggr\}
  \,. \cr}
\equn
$$
The mixed contributions corresponding to the three non-planar
parent diagrams in \fig{ParentMixedFigure} are
$$
\eqalign{
A^{\NP_1}_{M1234}  & =
 \rho \, s_{12} \, \I_4^\NP[ \mud_p \cdot \mud_q \mud_{p+q}^2
                           - \mud_q \cdot \mud_{p+q} \mud^2_p ]
   (s_{12}, s_{23})
\,, \cr
A^{\NP_2}_{M1234} & =
 \rho \, s_{12} \, \I_4^\NP[ \mud_p \cdot \mud_q \mud_{p+q}^2
                           - \mud_p \cdot \mud_{p+q} \mud^2_q  ]
   (s_{12}, s_{23})
   \,, \cr
A^{\NP_3}_{M1234} & =
- \rho \, s_{12} \, \I_4^\NP[ \mud_p \cdot \mud_{p+q} \mud^2_q
                            + \mud_q \cdot \mud_{p+q} \mud^2_p ] 
   (s_{12}, s_{23})
\,. }
\equn
$$
As in the previous cases, in the double box integrals 
we can label the numerator momentum factors 
in terms of which types of particles carry the loop momenta, {\it i.e.,}
$$
- \mud_{f_1} \cdot \mud_{f_2} \mud_s^2 \,.
\equn\label{MLoopLabel}
$$
In this case there are two contributions to each primitive
amplitude above, since interchanging matter fermions with scalars
in \fig{ParentMixedFigure} does not alter the color factor.


\subsection{Amplitudes with both a gluino and matter loop}
\label{GluinoMatterPrimitiveSubsection}

Up to color factors, the parent diagrams 
with both a gluino and matter loop depicted in
\figs{ClosedFermionGluinoFigure}{ClosedScalarGluinoFigure} are 
identical to the parent diagrams 
in \fig{ParentFermionFigure}($\P_4$) and \fig{ParentMixedFigure}($\P_4$).  
Thus, the associated primitive amplitudes are also simply related,
$$
\eqalign{
& A^{\P_1}_{\tildeF 1234} = A^{\P_2}_{\tildeF 1234} = A^{\P_4}_{F 1234} 
\,, \cr
& A^{\P_1}_{\tildeS 1234} =  
- 2 \rho \biggl\{
\I_4^\bowtie[ \mud_p \cdot \mud_q \mud_q^2 ] (s_{12}) 
+ \frac{2}{s_{12}} \I_4^\bowtie\Bigl[\mud_p^2 \mud_q^2
  ((p+q)^2 + \hf s_{12})\Bigr] (s_{12},s_{23}) \biggr\} \,, \cr
& A^{\P_2}_{\tildeS 1234} =  
- 2 \rho \biggl\{
\I_4^\bowtie[ \mud_p \cdot \mud_q \mud_p^2 ] (s_{12}) 
+ \frac{2}{s_{12}} \I_4^\bowtie\Bigl[\mud_p^2 \mud_q^2
  ((p+q)^2 + \hf s_{12})\Bigr] (s_{12},s_{23}) \biggr\}  \,. \cr}
\equn
$$
Hence
$$
A^{\P_1}_{\tildeS 1234} + A^{\P_2}_{\tildeS 1234} = A^{\P_4}_{M 1234} \,,
\equn
$$
accounting for the fact that 
$A^{\P_4}_{M 1234}$ incorporates the cases were the scalar loop is on
the left or right in diagram ($\P_4$) of \fig{ParentMixedFigure}, while
the diagrams in \fig{ClosedScalarGluinoFigure} separate the two cases.

\section{Two-loop supersymmetry identities}
\label{SuperVerificationSection}

We now verify that the supersymmetry Ward identity
(\ref{AllPlusSusyIdentity}) does in fact hold for the amplitudes
presented in \sect{AmplitudesSection} when using the FDH scheme.

\subsection{$N=1$ Identities}

The two-loop $N=1$ SWI are obtained by applying
\eqn{AllPlusSusyIdentity} to \eqns{N=1FullAmplitude}{YukawaAmplitude}.
For convenience, we subdivide the SWI according to the independent
parameters appearing in the amplitudes, $\Nf$ and $\xi$,
$$
\A_G^{\rm adjoint} +  \A_F^{\rm adjoint}  = 0 \,, \hskip 2. cm
\equn\label{FullIdentitiesAdjoint}
$$
$$
 \A_F^{\rm fund (1)} + \A_S^{\rm fund (1)} +
                  \,\A_M^{\rm fund(1)} + 
   \A_\tildeF^{\rm fund (1)} + \A_\tildeS^{\rm fund (1)} = 0 \,, \hskip -.1 cm
\equn\label{FullIdentitiesFund1}
$$
$$
 \A_F^{\rm fund (2)} + \A_S^{\rm fund (2)} +
                 \A_M^{\rm fund(2)}  = 0 \,,\hskip .1 cm
\equn\label{FullIdentitiesFund2}
$$
$$
 \A_S^\yukawa + \A_M^\yukawa = 0 \,, \hskip 1.5 cm
\equn\label{FullIdentitiesYukawa}
$$
corresponding to the four sets of diagrams represented in
\fig{TwoLoopSusyFigure}.  These identities must hold if supersymmetry is
preserved by a regularization scheme.

In general, we can further subdivide the identities by considering
separately the coefficient of each linearly independent color factor.  It
turns out, however, that for the identical-helicity amplitude a slightly
stronger subdivision is possible directly in terms of the primitive
amplitudes presented in \sect{AmplitudesSection}.  In other words, we
shall show that the identities hold for each coefficient of the different
color factors appearing in \eqns{NewFColor}{FundamentalColor}, even though
they are not all linearly independent.

The following two integrals, which will appear on the right-hand-side 
of several of the identities, have a simple representation in
terms of double-box integrals evaluated in $6-2\e$ dimensions,
$$
\eqalign{
\R^\P &\equiv
\rho \, s_{12} \, 
\I_4^\P\Bigl[(\mud_p\cdot\mud_q)^2-\mud_p^2\mud_q^2\Bigr](s_{12},s_{23}) 
 = - {\eps\over 2} (1+2 \eps) \, (4\pi)^2 \,
\rho \, s_{12} \, 
  \I_4^{\P, D=6-2\eps}(s_{12},s_{23})  \,,\cr
\R^\NP &\equiv
\rho \, s_{12} \, 
\I_4^\NP\Bigl[(\mud_p\cdot\mud_q)^2-\mud_p^2\mud_q^2\Bigr](s_{12},s_{23}) 
 = - {\eps\over 2} (1+2 \eps) \, (4\pi)^2 \,
\rho \, s_{12} \, 
  \I_4^{\NP, D=6-2\eps}(s_{12},s_{23}) \,, \cr}
\equn\label{D6Integrals}
$$
where
$$
\eqalign{
& \I_4^{\P, D=6-2\eps} (s_{12},s_{23}) \cr
& \hskip 1 cm \equiv \int
{d^{6 - 2\eps}p\over (2\pi)^{6-2\eps}} \,
{d^{6-2\eps} q\over (2\pi)^{6-2\eps}}\,
{ 1 \over  p^2\, q^2\, (p+q)^2 (p - k_1)^2 \,(p - k_1 - k_2)^2 \,
                  (q - k_4)^2 \, (q - k_3 - k_4)^2 }\,,  \cr
& \I_4^{\NP, D=6-2\eps} (s_{12},s_{23}) \cr
& \hskip 1 cm \equiv \int \! {d^{6-2\eps} p \over (2\pi)^{6-2\eps}} \,
        {d^{6-2\eps} q \over (2\pi)^{6-2\eps}}\,
{ 1 \over p^2\, q^2\, (p+q)^2 \,
         (p-k_1)^2 \,(q-k_2)^2\,
   (p+q+k_3)^2 \, (p+q+k_3+k_4)^2}\,. \cr}
\equn
$$
The planar and non-planar double box integrals have neither infrared nor
ultraviolet divergences in six dimensions.  Therefore,
due to the explicit $\eps$ in front of the integrals
in~eqs.~(\ref{D6Integrals}), $\R^\P$ and $\R^\NP$ are both 
of $\Ord(\eps)$ as $\e \to 0$.

We begin our inspection of the supersymmetry identities 
with~\eqn{FullIdentitiesAdjoint}.  This identity holds in the FDH
scheme ($D_s=4$), because the combinations
$$
\eqalignno{
& A^\P_{G1234} + \sum_{i=1}^4 A^{\P_i}_{F1234}
  = 6 \, \R^\P  = \Ord(\e) \,, 
&\equnno\label{AdjointPrimitiveIdentityP} \cr
& A^\NP_{G1234} + \sum_{i=1}^3 A^{\NP_i}_{F1234}
  = 6 \, \R^\NP = \Ord(\e) \,, 
&\equnno\label{AdjointPrimitiveIdentityNP} \cr}
$$
vanish as $\eps \rightarrow 0$.  The right-hand sides in these
equations are obtained by inserting the explicit values of the
primitive amplitudes given in \sect{AmplitudesSection}.
We dropped the integrals that vanish identically, {\it i.e.} the
bow-tie integrals that are odd in $\lambda_p$ or $\lambda_q$.  
The same equations~(\ref{AdjointPrimitiveIdentityP}) and
(\ref{AdjointPrimitiveIdentityNP}) obviously also hold 
after performing any permutation of the external legs.

Similarly, the identity (\ref{FullIdentitiesFund1}) holds because
$$
\eqalignno{
& \sum_{i=1}^2 \Bigl( A^{\P_i}_{F1234}
+  A^{\P_i}_{S1234}
+  A^{\P_i}_{M1234} \Bigr)
 =  - 4 \, \R^\P = \Ord(\e) \,, 
&\equnno\label{FundPrimitiveP1} \cr
& A^{\P_3}_{F1234}
+ \sum_{i \in \{3,5 \}} A^{\P_i}_{S1234}
+  A^{\P_3}_{M1234}   =
 - 2 \, \R^\P = \Ord(\e) \,, 
&\equnno\label{FundPrimitiveP2} \cr
& \sum_{i=1}^2 \Bigl( A^{\NP_i}_{F1234}
+  A^{\NP_i}_{S1234}
+  A^{\NP_i}_{M1234} \Bigr) =
 - 4 \, \R^\NP = \Ord(\e) \,, 
&\equnno\label{FundPrimitiveNP1} \cr
& A^{\NP_3}_{F1234} +  A^{\NP_3}_{S1234}
 +   A^{\NP_3}_{M1234} =
 - 2 \, \R^\NP = \Ord(\e)\,, 
&\equnno\label{FundPrimitiveNP2} \cr
& A^{\P_1}_{\tildeS 1234} + A^{\P_1}_{\tildeF 1234} = 
A^{\P_2}_{\tildeS 1234} + A^{\P_2}_{\tildeF 1234} = 0\,, 
&\equnno\label{FundPrimitiveTilde} \cr}
$$
and the identity (\ref{FullIdentitiesFund2}) holds because
$$
\eqalignno{
& A_{F1234}^{P_4} - A_{S1234}^{P_4} = 0 \,, 
&\equnno\label{DoubleOneLoopIdentityA} \cr
& 2 \, A_{F1234}^{P_4} + A_{M1234}^{P_4} = 0 \,. 
&\equnno\label{DoubleOneLoopIdentityB} \cr}
$$
The latter three equations hold to all orders in $\eps$.

Finally, \eqn{FullIdentitiesYukawa}, the identity for the amplitudes 
depending on the Yukawa coupling $\xi$, holds because
$$
\eqalignno{
& {1\over 2}\sum_{i=1}^3 A^{\P_i}_{M1234}
+ 2 A_{S1234}^{\P_5} = 2 \, \R^\P = \Ord(\eps) \,, 
&\equnno\label{YukawaPrimitiveP} \cr
& {1\over 2} \sum_{i=1}^3 A^{\NP_i}_{M1234}
 = 2 \, \R^\NP =  \Ord(\eps) \,,  
&\equnno\label{YukawaPrimitiveNP} \cr}
$$
where the factors of $1/2$ and $2$ on the left-hand side of the
equations are relative combinatoric factors compared to the gauge
case.

The same relations hold for any permutation of the external legs.  
We conclude that for $D_s=4$ (FDH scheme) the identical-helicity 
(++++) primitive amplitudes satisfy the SWI~(\ref{AllPlusSusyIdentity}) 
as $\e\to0$, at least through two loops.
In the course of computing the two-loop amplitude for the helicity 
configuration $-$+++ in QCD~\cite{ggggpaper}, we have also verified 
that the pure super-Yang-Mills identity~(\ref{FullIdentitiesAdjoint}) 
is satisfied for $-$+++ up to $\Ord(\eps)$ corrections, again provided that
$D_s=4$.   For either of these two helicity amplitudes, in the HV scheme 
where $D_s = 4-2\eps$ the identities~(\ref{AdjointPrimitiveIdentityP})
and (\ref{AdjointPrimitiveIdentityNP}) with gluon loops are {\it not} 
satisfied, even at $\Ord(\eps^{-1})$.

Even in the FDH scheme, the above two-loop $N=1$ identities are
generally satisfied only for $\eps \rightarrow 0$, because the
quantities $\R^\P$ and $\R^\NP$ are known at order $\e$, and they are
nonvanishing at this order. If one thinks about constructing
three-loop amplitudes via their unitarity cuts, which include the
product of a two-loop amplitude with a tree amplitude, one might be
concerned that the $\Ord(\eps)$ breaking of the identities at two
loops could lead to a non-vanishing breaking of the SWI at three loops
even as $\eps \rightarrow 0$.  Of course there will also be
contributions to the product of two-loop and tree amplitudes from
intermediate momenta in $(-2\e)$ dimensions, which may well cancel the
above contributions.  Clearly this issue warrants further
investigation.

\subsection{Exactness of $N=2$ Identities}
\label{Neq2IdentitiesSubsection}

An $\Ord(\eps)$ breaking in $N=1$ supersymmetry Ward identities in the FDH
scheme is perhaps to be expected since the scheme is defined by
``dimensional expansion'', {\it i.e.}, $D > 4$.  Supermultiplets become
larger as $D$ increases; hence for $D>4$ an $N=1$ supersymmetric theory in
$D=4$ does not provide enough states to form a supermultiplet for the
intermediate unobserved states with momenta in the extra $(-2\e)$
dimensions.  Thus, one might expect some kind of violation of
supersymmetry for finite $\e$.  On the other hand, the $N=2$ vector
multiplet, consisting of a gluon, two gluinos and a complex scalar, may be
viewed as coming from the compactification of a higher-dimensional
supersymmetric theory, for example $N=1$ supersymmetric gauge theory in
$D=5$.  So, under a dimensional expansion of the $N=2$ theory away from
$D=4$ there are sufficiently many states to form a supermultiplet, and we
might expect the SWI to hold to all orders in $\eps$ in the $N=2$ case.
The supersymmetry identity that should be satisfied in pure $N=2$ gauge
theory is given in \eqn{N=2FullAmplitude}.  This identity does hold to all
orders in $\e$, because of the following relations between primitive
amplitudes:
$$
\eqalign{
A^\P_{G1234}
+ \sum_{i=1}^5 A^{\P_i}_{S1234}
+  \sum_{i=1}^3 ( 2 A^{\P_i}_{F1234} + A^{\P_i}_{M1234} )
+ 4 A^{\P_4}_{F1234} + 2 A^{\P_4}_{M1234} 
& = 0 \,, \cr}
\equn\label{PlanarN=2MultipletSusy}
$$
which is the sum of eqs.~(\ref{AdjointPrimitiveIdentityP}),
(\ref{FundPrimitiveP1}), and (\ref{FundPrimitiveP2}); and
$$
\eqalign{
A^\NP_{G1234} 
+ \sum_{i=1}^3 
(  A^{\NP_i}_{S1234} + 2 A^{\NP_i}_{F1234} + A^{\NP_i}_{M1234} )
& = 0 \,, \cr}
\equn\label{NonplanarN=2MultipletSusy}
$$
which is the sum of eqs.~(\ref{AdjointPrimitiveIdentityNP}),
(\ref{FundPrimitiveNP1}), and (\ref{FundPrimitiveNP2}).

Now consider adding $\Nf$ hypermultiplets in the $N_c + \bar{N}_c$
representation to the $N=2$ gauge theory.  The $N=2$ hypermultiplets
have the same field content as the pairs of $N=1$ chiral matter fields, 
$Q_i + \tilde{Q}_i$.
The $\Nf^2$ terms in the SWI for $D_s=4$ clearly hold to all orders in 
$\e$, by virtue of \eqns{DoubleOneLoopIdentityA}{DoubleOneLoopIdentityB}.
The same statement is true for the $\Nf^1$ terms, however here the
cancellation is more subtle.  One can split the $N=2$ gauge multiplet
into an $N=1$ gauge multiplet plus an $N=1$ adjoint matter multiplet $\Phi$.  
There is a set of $\Nf^1$ contributions from coupling the hypermultiplet fields
to the $N=1$ gauge fields, which is given by precisely the combinations
(\ref{FundPrimitiveP1})--(\ref{FundPrimitiveNP2}) encountered above.
A second set of $\Nf^1$ contributions comes from coupling the 
hypermultiplet fields to $\Phi$.  This coupling is through an $N=1$
superpotential term, $\tilde{W} \propto \sum_i \tilde{Q}_i \Phi Q_i$.
The strength of $\tilde{W}$ is dictated by the gauge coupling, because
the Yukawa interactions with the $N=1$ gaugino $\lambda$ in~\eqn{LgFund}
have to be duplicated by couplings to the $N=1$ matter adjoint fermion,
which is the second gaugino of $N=2$ supersymmetry.
These Yukawa interactions just double the mixed fermion-scalar terms,
$A^{\P_i}_{M1234}$ and $A^{\NP_i}_{M1234}$.  However, there are additional
mixed terms due to the Yukawa vertices from $\tilde{W}$ which contain the
adjoint scalar particle.  These terms are dressed a bit differently with 
color than the mixed contributions considered 
in~\sect{FundTwoLoopColorSubsection}.  However, they still
can be written in terms of $A^{\P_i}_{M1234}$ and $A^{\NP_i}_{M1234}$.
Finally, there are new $\P_5$ terms from scalar four-point interactions 
generated by $\tilde{W}$.  Adding all the terms together, we find that 
the SWI hold to all orders in $\e$ by virtue of
$$
\eqalignno{
& \sum_{i=1}^2 \Bigl( A^{\P_i}_{F1234}
+  A^{\P_i}_{S1234}
+  A^{\P_i}_{M1234} \Bigr)
+ \biggl( \sum_{i=1}^3 A^{\P_i}_{M1234}
+ 4 A_{S1234}^{\P_5} \biggr) 
= 0 \,, 
&\equnno\label{HyperPrimitiveP1} \cr
& \biggl( A^{\P_3}_{F1234}
+ \sum_{i \in \{3,5 \}} A^{\P_i}_{S1234}
+  A^{\P_3}_{M1234} \biggr)  
+ {1\over2} \biggl( \sum_{i=1}^3 A^{\P_i}_{M1234}
                   + 4 A_{S1234}^{\P_5} \biggr)
= 0 \,, 
&\equnno\label{HyperPrimitiveP2} \cr
& \sum_{i=1}^2 \Bigl( A^{\NP_i}_{F1234}
+  A^{\NP_i}_{S1234}
+  A^{\NP_i}_{M1234} \Bigr)
+ \biggl( \sum_{i=1}^3 A^{\NP_i}_{M1234} \biggr) 
= 0 \,,
&\equnno\label{HyperPrimitiveNP1} \cr
& \biggl( A^{\NP_3}_{F1234} +  A^{\NP_3}_{S1234}
 +   A^{\NP_3}_{M1234} \biggr)
+ {1\over2} \biggl( \sum_{i=1}^3 A^{\NP_i}_{M1234} \biggr) 
= 0 \,,
&\equnno\label{HyperPrimitiveNP2} \cr}
$$
which follow from adding appropriate pairs of 
eqs.~(\ref{FundPrimitiveP1})--(\ref{YukawaPrimitiveNP}).

\section{Renormalization}
\label{RenormalizationSection}

The identical-helicity amplitudes presented in \sect{AmplitudesSection}
have not been renormalized.  Since the corresponding tree-level amplitudes
vanish for this helicity configuration, the renormalization procedure is
similar to that for a typical one-loop amplitude.  In particular, the
modified minimal subtraction counterterm to be subtracted from
$\A_4^{\twoloop}(1^+, 2^+, 3^+, 4^+)$ is
$$
4 g^2 b_0 \, c_\Gamma \, {1 \over \eps}\,
\A_4^{\oneloop}(1^+, 2^+, 3^+, 4^+) \,,
\equn\label{CounterTerm}
$$
where 
$$
c_\Gamma \equiv {\Gamma(1+\eps) \Gamma^2(1-\eps)
\over \Gamma(1-2 \eps) (4 \pi)^{2-\eps}} \,,
\equn\label{cGammaDef}
$$
and $b_0$ is the one-loop $\beta$-function coefficient.
For example, in QCD with $N_c$ colors and $\Nf$ quark flavors,
$$
b_0^{\rm QCD} =
 {{11 N_c - 2 \Nf} \over 6} \,,
\equn
$$
while for pure $N=1$ super-Yang-Mills theory,
$$
b_0^{N=1} = {3 \over 2} N_c \,.
\equn
$$
More generally, the counterterms 
depend on the scheme choice through the scheme dependence
of the one-loop amplitude.

\subsection{Coupling constant shift}
\label{CouplingShiftSubsection}

When comparing physical results obtained in the FDH scheme to ones
obtained in the 't Hooft-Veltman (HV) scheme, one must account for the
shift in the coupling.  In general, the coupling constant of the
theory is scheme dependent.  The one-loop relation between the QCD coupling
constant in dimensional regularization and that in dimensional reduction 
has been computed 
previously~\cite{NonSusyDimRed,AKT,MartinVaughn,KSTfourparton}.
Here we shall give the relation to two loops.  We also allow for
an arbitrary value of the variable controlling the number of gluon states
in loops, $D_s \equiv 4 - 2 \e \delta_R$, 
so that one can consider schemes other than the FDH/DR scheme 
($\delta_R = 0$) or the HV scheme ($\delta_R = 1$) if one desires.

A convenient method for calculating the scheme dependence of the coupling
constant is to evaluate the shift in the background field gauge
method~\cite{Background,AbbottPureGlue,AbbottFermion,PGJ} by computing
the vacuum polarization, following the one-loop discussion of
ref.~\cite{KSTfourparton}.  The background field gauge possesses a
Ward identity $Z_g = Z_A^{-1/2}$ which allows one to obtain the
coupling constant renormalization $Z_g$ from the wave-function
renormalization $Z_A$.  Our two-loop computation follows closely
the ones in refs.~\cite{AbbottPureGlue,AbbottFermion}, except that
we have to keep track of the finite terms, not just the poles in $\e$, 
when the terms are scheme dependent.

The coupling constants of the 't Hooft-Veltman and CDR schemes are
identical because the vacuum polarizations are computed using
exactly the same rules for internal states.  The HV and CDR regularization
schemes, together with the widely used modified minimal subtraction 
($\MSbar$) renormalization scheme, define the coupling constant
$\alpha_\MSbar$.  The modified minimally subtracted versions of the 
FDH and DR schemes, which we call $\FDHbar$ and $\DRbar$, define the
coupling constant $\alpha_\FDHbar = \alpha_\DRbar$.
For any coupling $\alpha$, let the reduced coupling be $a = \alpha/(2\pi)$.  
Then the one-loop coupling constant relation 
is~\cite{NonSusyDimRed,AKT,MartinVaughn,KSTfourparton}
$$
a_\DRbar = a_\FDHbar =
a_\MSbar \, \Bigl(1 + {C_A\over 6 } a_\MSbar + \cdots \Bigr) \,,
$$
where $C_A$ is the quadratic Casimir in the adjoint representation,
$C_A=N_c$ for gauge group $SU(N_c)$.
Our goal here is to compute the two-loop corrections to this relation.

{\it A priori}, there is a difference between FDH and
DR, in that DR divides a gluon Lorentz index into a
$D=4-2\e$ index and a residual $2\e$ index.  The latter components are
referred to as $\e$-scalars.  Imposing $D$-dimensional gauge
invariance still permits the $\e$-scalars, and their associated
couplings, to renormalize differently from the $D$-dimensional gauge
fields.  For example, when fermions are present, their coupling $g$ to
the gauge fields gives rise to a Yukawa coupling $\hat{g}$ of the fermions
to the $\e$-scalars.  In a non-supersymmetric theory, when one also
requires the $\e$-scalar Green functions to be finite, one finds that
$g$ and $\hat{g}$ have different $\beta$ functions even at one
loop~\cite{JJR}.

A nice feature of the background field method is that
none of the quantum fields have to be
renormalized~\cite{AbbottPureGlue}.  Quantum fields do not appear on
the external (background) legs, so $Z$ factors cancel between
propagators and vertices.  Therefore, one does not have to worry in
the present calculation about whether the $\e$-scalars that are
present in DR should be renormalized differently from the
$d$-dimensional gauge fields.  Thus, the counterterm graphs
are identical between DR and FDH.  The non-counterterm graphs
also turn out to be identical; the $D$ and $2\e$ ranges of the indices
in DR combine to give exactly the four-dimensional range
used in FDH.  Thus the $\FDHbar$ and $\DRbar$ gauge couplings are
equivalent, at least in any theory (supersymmetric or not) with only a
gauge coupling present.

\begin{figure}[ht]
\begin{center}
\begin{picture}(240,300)(0,0)
%
\Gluon(105,20)(120,20){1.5}{2}
\GlueArc(130,20)(10,0,180){1.5}{3}
\GlueArc(130,20)(10,180,360){1.5}{3}
\GBox(118,18)(122,22){0}
\Gluon(140,20)(155,20){1.5}{2}
\Gluon(160,20)(175,20){1.5}{2}
\GlueArc(185,20)(10,0,180){1.5}{3}
\GlueArc(185,20)(10,180,360){1.5}{3}
\GBox(193,18)(197,22){0}
\Gluon(195,20)(210,20){1.5}{2}
\Text(155,0)[c]{(m)}
\Gluon(30,20)(45,20){1.5}{2}
\GlueArc(55,20)(10,0,180){1.5}{2}
\GlueArc(55,20)(10,180,360){1.5}{3}
\GBox(53,28)(57,32){0}
\Gluon(65,20)(80,20){1.5}{2}
\Text(55,0)[c]{(l)}
\Gluon(00,70)(15,70){1.5}{2}
\DashCArc(25,70)(10,270,360){3}
\DashCArc(25,70)(10,180,270){3}
\DashCArc(25,70)(10,90,180){3}
\DashArrowArcn(25,70)(10,90,0){2}
\Gluon(35,70)(50,70){1.5}{2}
\Gluon(25,60)(25,80){1.5}{3}
\Text(25,50)[c]{(i)}
\Gluon(68,70)(83,70){1.5}{2}
\DashArrowArcn(93,70)(10,90,0){2}
\DashCArc(93,70)(10,270,360){3}
\GlueArc(93,70)(10,90,180){1.5}{1}
\GlueArc(93,70)(10,180,270){1.5}{1}
\Gluon(103,70)(118,70){1.5}{2}
\DashLine(93,60)(93,80){3}
\Gluon(122,70)(137,70){1.5}{2}
\DashCArc(147,70)(10,180,270){3}
\DashArrowArcn(147,70)(10,180,90){2}
\GlueArc(147,70)(10,270,360){1.5}{1}
\GlueArc(147,70)(10,0,90){1.5}{1}
\Gluon(157,70)(172,70){1.5}{2}
\DashLine(147,80)(147,60){3}
\Text(120,50)[c]{(j)}
\Gluon(190,70)(205,70){1.5}{2}
\GlueArc(215,70)(10,90,180){1.5}{1}
\GlueArc(215,70)(10,180,270){1.5}{1}
\GlueArc(215,70)(10,270,360){1.5}{1}
\GlueArc(215,70)(10,0,90){1.5}{1}
\Gluon(225,70)(240,70){1.5}{2}
\Gluon(215,60)(215,80){1.5}{3}
\Text(215,50)[c]{(k)}
\Gluon(00,120)(15,120){1.5}{2}
\GlueArc(25,120)(10,0,90){1.5}{1}
\GlueArc(25,120)(10,-180,0){1.5}{3}
\DashArrowArcn(25,120)(10,180,90){2}
\DashCArc(15,130)(10,270,360){2}
\Gluon(35,120)(50,120){1.5}{2}
\Gluon(55,120)(70,120){1.5}{2}
\GlueArc(80,120)(10,90,180){1.5}{1}
\GlueArc(80,120)(10,180,360){1.5}{3}
\DashArrowArcn(80,120)(10,90,0){2}
\DashCArc(90,130)(10,180,270){2}
\Gluon(90,120)(105,120){1.5}{2}
\Text(55,100)[c]{(g)}
\Gluon(135,120)(150,120){1.5}{2}
\GlueArc(160,120)(10,0,90){1.5}{1}
\GlueArc(160,120)(10,-180,0){1.5}{3}
\GlueArc(160,120)(10,90,180){1.5}{1}
\GlueArc(150,130)(10,270,360){1.5}{2}
\Gluon(170,120)(185,120){1.5}{2}
\Gluon(190,120)(205,120){1.5}{2}
\GlueArc(215,120)(10,90,180){1.5}{1}
\GlueArc(215,120)(10,180,360){1.5}{3}
\GlueArc(215,120)(10,0,90){1.5}{1}
\GlueArc(225,130)(10,180,270){1.5}{2}
\Gluon(225,120)(240,120){1.5}{2}
\Text(185,100)[c]{(h)}
\Gluon(05,170)(20,170){1.5}{2}
\DashCArc(30,170)(10,0,90){3}
\DashCArc(30,170)(10,180,360){3}
\DashArrowArcn(30,170)(10,180,90){2}
\GlueArc(20,180)(10,270,360){1.5}{2}
\Gluon(40,170)(55,170){1.5}{2}
\Gluon(65,170)(80,170){1.5}{2}
\DashArrowArcn(90,170)(10,180,0){2}
\DashCArc(90,170)(10,270,360){3}
\DashCArc(90,170)(10,180,270){3}
\GlueArc(80,160)(10,0,90){1.5}{2}
\Gluon(100,170)(115,170){1.5}{2}
\Gluon(125,170)(140,170){1.5}{2}
\DashCArc(150,170)(10,90,180){3}
\DashArrowArcn(150,170)(10,90,0){2}
\DashCArc(150,170)(10,180,360){3}
\GlueArc(160,180)(10,180,270){1.5}{2}
\Gluon(160,170)(175,170){1.5}{2}
\Gluon(185,170)(200,170){1.5}{2}
\DashArrowArcn(210,170)(10,180,0){2}
\DashCArc(210,170)(10,270,360){3}
\DashCArc(210,170)(10,180,270){3}
\GlueArc(220,160)(10,90,180){1.5}{2}
\Gluon(220,170)(235,170){1.5}{2}
\Text(120,150)[c]{(f)}
\Gluon(30,220)(45,220){1.5}{2}
\DashArrowArcn(55,220)(10,180,0){2}
\GlueArc(55,220)(10,180,360){1.5}{3}
\Gluon(65,220)(80,220){1.5}{2}
\DashLine(65,220)(45,220){3}
\Text(55,200)[c]{(c)}
\Gluon(95,220)(110,220){1.5}{2}
\GlueArc(120,220)(10,0,180){1.5}{3}
\GlueArc(120,220)(10,180,360){1.5}{3}
\Gluon(110,220)(130,220){1.5}{3}
\Gluon(130,220)(145,220){1.5}{2}
\Text(120,200)[c]{(d)}
\Gluon(160,220)(175,220){1.5}{2}
\GlueArc(180,220)(5,0,180){1.5}{2}
\GlueArc(180,220)(5,180,360){1.5}{2}
\GlueArc(190,220)(5,0,180){1.5}{2}
\GlueArc(190,220)(5,180,360){1.5}{2}
\Gluon(195,220)(210,220){1.5}{2}
\Text(185,200)[c]{(e)}
\Gluon(00,270)(15,270){1.5}{2}
\GlueArc(25,280)(7.5,205,335){1.5}{3}
\DashArrowArcn(25,270)(10,180,0){2}
\DashCArc(25,270)(10,180,360){3}
\Gluon(35,270)(50,270){1.5}{2}
\Gluon(55,270)(70,270){1.5}{2}
\GlueArc(80,260)(7.5,25,155){1.5}{3}
\DashArrowArcn(80,270)(10,180,0){2}
\DashCArc(80,270)(10,180,360){3}
\Gluon(90,270)(105,270){1.5}{2}
\Text(55,250)[c]{(a)}
\Gluon(135,270)(150,270){1.5}{2}
\GlueArc(160,270)(10,0,50){1.5}{1}
\GlueArc(160,270)(10,-180,0){1.5}{3}
\GlueArc(160,270)(10,130,180){1.5}{1}
\DashArrowArcn(160,277)(6,180,0){2}
\DashCArc(160,277)(6,180,360){2}
\Gluon(165,270)(180,270){1.5}{2}
\Gluon(190,270)(205,270){1.5}{2}
\GlueArc(215,270)(10,0,50){1.5}{1}
\GlueArc(215,270)(10,-180,0){1.5}{3}
\GlueArc(215,270)(10,130,180){1.5}{1}
\GlueArc(215,277)(5,0,180){1.5}{2}
\GlueArc(215,277)(5,180,360){1.5}{2}
\Gluon(225,270)(240,270){1.5}{2}
\Text(185,250)[c]{(b)}
\end{picture}
\caption[]{
\label{BackgroundGluonFigure}
\small Feynman diagrams for the pure glue contributions to the
vacuum polarization, following the labeling of
ref.~\cite{AbbottPureGlue}. Dashed lines represent ghosts, and square dots
represent counterterms.}
\end{center}
\end{figure}
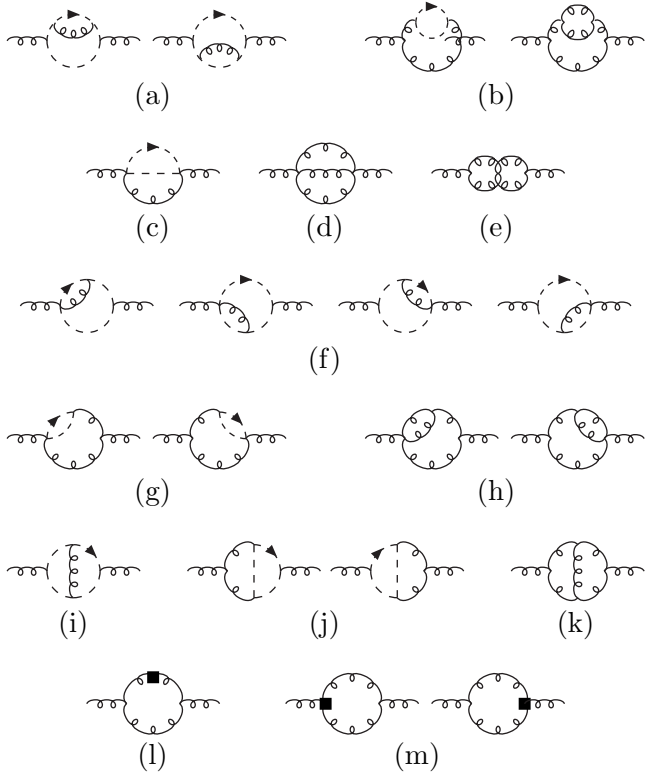

Following the computational rules for the FDH scheme, we
write the Lorentz contraction that tracks the number of physical states as
$$
\eta^{\mu}_{\mu} = D_s = 4-2\e \, \delta_R \, ,
\equn\label{deltaR}
$$
where $\eta$ is the Minkowski metric,
$\delta_R = 1$ for $\MSbar$, $\delta_R = 0$ for $\DRbar$ or $\FDHbar$.
In ref.~\cite{AbbottPureGlue,AbbottFermion} the divergent terms in the
two-loop vacuum polarization were computed in $\MSbar$.  Here we also
need the portion of the finite terms that depends on $\delta_R$.
For pure gauge theory, Table~1 of ref.~\cite{AbbottPureGlue} gives the
divergent parts of the Feynman graphs shown in \fig{BackgroundGluonFigure}.
Of the graphs a--m, only b, h, k, and the counterterm graphs l and m, 
are $\delta_R$ dependent, although a little inspection is necessary to
show that graphs d and e are independent of $\delta_R$.

The counterterm contributions l and m come from one-loop renormalization
of the gauge-fixing terms.  For our purposes, this renormalization
has to be performed including the finite level, whereas in obtaining the
$\beta$ function in ref.~\cite{AbbottPureGlue} it was sufficient to
carry out the renormalization only through $\Ord(1/\e)$.  We find that the
gauge-fixing renormalization factor $Z_\alpha$ is given by
$$
 Z_\alpha
= 1 + \biggl[ \Bigl( {5\over3\e} + {\delta_R\over3} + {28\over9} \Bigr) C_A
              - \Bigl( {4\over3\e} + {20\over 9} \Bigr) T_F \biggr]
                    { g^2 \over (4\pi)^2 } \,,
\equn\label{Zalpha}
$$
where the fermion generators have trace $\Tr(T^aT^b) = T_F \delta^{ab}$
(for QCD with $\Nf$ flavors, $T_F = {1\over 2} \Nf$).

In the notation of refs.~\cite{AbbottPureGlue,AbbottFermion}, 
the vacuum polarization is
$$
\Pi_{\mu\nu}(k) = {i g^4 C_A^2 \delta^{ab} \over (4\pi)^4 }
\Bigl[ A g_{\mu\nu} -B k_\mu k_\nu \Bigr] \,.
\equn\label{Bdef}
$$
In the background field method the final sum over diagrams gives $A=B$, 
ensuring the transversality of the vacuum polarization.  Here we give just
the $B$ terms, dropping for convenience the combination
$\gamma_E-\ln4\pi+\ln(k^2/\mu^2)$, as well as all $\delta_R$-independent
finite contributions.  For the $B$ terms of \eqn{Bdef} we obtain,
$$
\eqalign{
 \hbox{graph b:} & \hskip1cm
{25\over6\e^2} \Bigl( 1 + {22\over5}  \e \Bigr)
         + {3\over2\e} (\delta_R-1)
         + {49\over6} \delta_R \,, \cr
 \hbox{graph h:} & \hskip1cm
-{9\over8\e^2} \Bigl( 1 + {31\over6} \e \Bigr)
         - {3\over4\e} (\delta_R-1)
         - {27\over8} \delta_R \,, \cr
 \hbox{graph k:} & \hskip1cm
{27\over8\e^2} \Bigl( 1 + {245\over54} \e \Bigr)
         - {3\over4\e} (\delta_R-1)
         - {25\over8} \delta_R \,, \cr
 \hbox{graphs l$+$m:} & \hskip1cm
{10\over3\e} + {2\over3} \delta_R \,, \cr
 \hbox{a--m total:} & \hskip1cm
{17\over3\e} + {7\over3} \delta_R \,. \cr}
\equn\label{puregluetable}
$$
The values of the remaining graphs with no $\delta_R$ dependence may
be obtained from ref.~\cite{AbbottPureGlue}.  

\begin{figure}[ht]
\begin{center}
\begin{picture}(240,150)(0,0)
%
\Gluon(105,20)(120,20){1.5}{2}
\GlueArc(130,20)(10,0,180){1.5}{3}
\GlueArc(130,20)(10,180,360){1.5}{3}
\GBox(118,18)(122,22){0}
\Gluon(140,20)(155,20){1.5}{2}
\Gluon(160,20)(175,20){1.5}{2}
\GlueArc(185,20)(10,0,180){1.5}{3}
\GlueArc(185,20)(10,180,360){1.5}{3}
\GBox(193,18)(197,22){0}
\Gluon(195,20)(210,20){1.5}{2}
\Text(155,0)[c]{(f)}
\Gluon(35,20)(50,20){1.5}{2}
\GlueArc(60,20)(10,0,180){1.5}{2}
\GlueArc(60,20)(10,180,360){1.5}{3}
\GBox(58,28)(62,32){0}
\Gluon(70,20)(85,20){1.5}{2}
\Text(60,0)[c]{(e)}
\Gluon(30,70)(45,70){1.5}{2}
\GlueArc(55,70)(10,90,180){1.5}{1}
\GlueArc(55,70)(10,180,270){1.5}{1}
\CArc(55,70)(10,270,360)
\ArrowArc(55,70)(10,0,90)
\Gluon(65,70)(80,70){1.5}{2}
\Line(55,80)(55,60)
\Gluon(85,70)(100,70){1.5}{2}
\GlueArc(110,70)(10,270,360){1.5}{1}
\GlueArc(110,70)(10,0,90){1.5}{1}
\ArrowArc(110,70)(10,90,180)
\CArc(110,70)(10,180,270)
\Gluon(120,70)(135,70){1.5}{2}
\Line(110,60)(110,80)
\Text(85,50)[c]{(c)}
\Gluon(160,70)(175,70){1.5}{2}
\ArrowArc(185,70)(10,0,90)
\CArc(185,70)(10,90,180)
\CArc(185,70)(10,180,270)
\CArc(185,70)(10,270,360)
\Gluon(195,70)(210,70){1.5}{2}
\Gluon(185,60)(185,80){1.5}{3}
\Text(185,50)[c]{(d)}

\Gluon(30,120)(45,120){1.5}{2}
\ArrowArc(55,120)(10,0,180)
\CArc(55,120)(10,180,360)
\GlueArc(55,130)(7.5,205,335){1.5}{3}
\Gluon(65,120)(80,120){1.5}{2}
\Gluon(85,120)(100,120){1.5}{2}
\ArrowArc(110,120)(10,0,180)
\CArc(110,120)(10,180,360)
\GlueArc(110,110)(7.5,25,155){1.5}{3}
\Gluon(120,120)(135,120){1.5}{2}
\Text(85,100)[c]{(a)}
\Gluon(160,120)(175,120){1.5}{2}
\ArrowArc(185,127)(6,0,180)
\CArc(185,127)(6,180,360)
\GlueArc(185,120)(10,180,360){1.5}{3}
\GlueArc(185,120)(10,0,50){1.5}{1}
\GlueArc(185,120)(10,130,180){1.5}{1}
\Gluon(195,120)(210,120){1.5}{2}
\Text(185,100)[c]{(b)}
\end{picture}
\caption[]{
\label{BackgroundFermionFigure}
\small Fermion loop contributions to the vacuum polarization,
following ref.~\cite{AbbottFermion}. Solid lines represent fermions, 
and square dots represent counterterms.}
\end{center}
\end{figure}
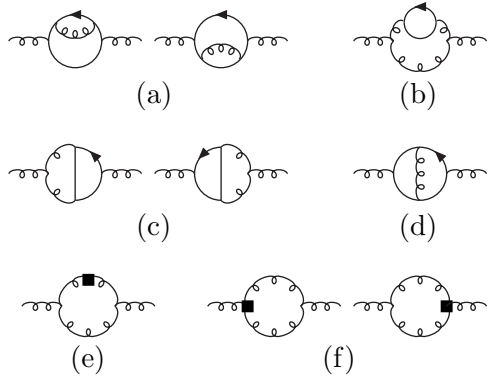

For the fermion-loop contributions shown in
\fig{BackgroundFermionFigure}, we find that the only finite
$\delta_R$-dependent terms that survive are in the $T_F C_F$ color
structure, and the total in Table~1 from ref.~\cite{AbbottFermion}
becomes
$$
\eqalign{
 \hbox{total:} & \hskip1cm
 -{1\over\e} \Bigl( {10\over3} T_F C_A + 2 T_F C_F \Bigr)
- 2 T_F C_F \delta_R \,. \cr}
\equn\label{fermionlooptable}
$$
Here $C_F$ is the quadratic Casimir in the fermion representation,
$C_F = (N_c^2-1)/(2N_c)$ for the fundamental representation of $SU(N_c$).
It is rather simple to see that adding scalars to the theory does not
induce any additional $\delta_R$ dependence; the corresponding diagrams 
do not contain any $\eta^\mu_\mu$ contractions.

The background field Ward identity~\cite{AbbottPureGlue} relates
the inverse coupling directly to the vacuum polarization.  Using this
relation, we find that the bare inverse coupling is given in terms
of the renormalized coupling 
$a_{\delta_R} \equiv \alpha_{\delta_R}/(2\pi)$ by
$$
\eqalign{
{1\over a_{\rm bare}}
&= {1\over a_{\delta_R}} \Biggl\{ 1
- a_{\delta_R} \biggl[
   {1\over2\e} \Bigl( {11\over3} C_A - {4\over3} T_F \Bigr)
    + {1\over 6} C_A \delta_R + c_1 \biggr] 
  \Bigl( {e^{\gamma_E} k^2 \over 4\pi} \Bigr)^{-\e}  \cr
& \hskip1cm
- a_{\delta_R}^2 \biggl[
    {1\over8\e} \Bigl( {34\over3} C_A^2 - {20\over3} T_F C_A
                                        - 4 T_F C_F \Bigr)
    + \Bigl( {7\over12} C_A^2 - {1\over2} T_F C_F \Bigr) \delta_R
    + c_2 \biggr] 
  \Bigl( {e^{\gamma_E} k^2 \over 4\pi} \Bigr)^{-2\e} \Biggr\} \,, \cr}
\equn\label{invcoupling}
$$
where $c_1$ and $c_2$ are independent of $\delta_R$.  To relate
$a_{\delta_R}$ to the standard $\MSbar$ coupling, 
$a \equiv \alpha_{\MSbar}/(2\pi)$ with $\delta_R = 1$, we equate the 
two versions of the bare inverse coupling, which leads to 
$$
\eqalign{
  a_{\delta_R} &= a \biggl[ 1 + {1\over6} C_A (1 - \delta_R) \, a
   + \Bigl({7\over 12} C_A^2 (1 - \delta_R)
           + {1\over 36} C_A^2 (1 - \delta_R)^2
            - {1\over 2} T_F C_F (1 - \delta_R) \Bigr) a^2  \biggr]
  \, . \cr}
\equn\label{couplingrel}
$$
In particular, for a gauge theory with a general fermion content,
the $\DRbar$ coupling, $\ta \equiv \alpha_{\DRbar}/(2\pi)$, is expressed
in terms of the $\MSbar$ coupling by setting $\delta_R = 0$
in~\eqn{couplingrel}, yielding
$$
\eqalign{
  \ta &= a \biggl[ 1 + {1\over6} C_A \, a
       + \Bigl( {11\over18} C_A^2 - {1\over2} T_F C_F \Bigr) a^2 \biggr]
+ \Ord(a^3)  \, . \cr}
\equn\label{couplingrelDR}
$$

\subsection{Three-loop $\beta$ function}
\label{ThreeLoopBetaSubsection}

An interesting side-benefit of the above computation is that it
allows us to obtain the {\it three}-loop $\beta$ function in the
$\FDHbar$ and $\DRbar$ schemes from the known
$\beta$ function in the $\MSbar$ scheme~\cite{MSbarb2}.

In general, knowledge of the scheme dependence of the coupling
allows one to convert the $\beta$ function between schemes
at one higher loop order. For example, the $\beta$-function coefficients 
in four different momentum-subtraction (MOM) schemes were computed at 
{\it four} loops in similar fashion~\cite{ChetRetey}; see also
refs.~\cite{ChetSeid} for three-loop MOM computations.  There are
similar computations on the lattice~\cite{Latticed2}.

Consider two different renormalization schemes, and hence two different
coupling constants, for a single theory,
$$
a \equiv {\alpha\over2\pi} \,, \qquad \ta \equiv {\tilde{\alpha}\over2\pi} \,,
\equn\label{coupdef}
$$
and suppose they are related through two loops by
$$
\eqalign{
\ta &= a \Bigl[ 1 + d_1 a + d_2 a^2 + \ldots \Bigr] \,, \cr
a &= \ta \Bigl[ 1 - d_1 \ta + (2 d_1^2 - d_2 ) \ta^2 + \ldots \Bigr] \,.
\cr}
\equn\label{couprel}
$$
Let the $\beta$ functions in the two schemes be
$$
\eqalign{
\mu {d a \over d\mu} \equiv \beta(a) &\equiv 
- 2 \Bigl[ b_0 a^2 + b_1 a^3 + b_2 a^4 + \ldots \Bigr], \cr
\mu {d \tilde{a} \over d\mu} \equiv \tilde\beta(\ta) &\equiv 
- 2 \Bigl[ \tb_0 \ta^2 + \tb_1 \ta^3 + \tb_2 \ta^4 + \ldots\Bigr]. \cr}
\equn\label{betadef}
$$

One can also calculate the $\beta$ function in the second scheme as
$$
\tilde\beta(\ta) 
= \mu {d \tilde{a} \over d\mu}
= \mu {d a \over d\mu} \times {d \ta \over da}
= \beta(a) \times [ 1 + 2d_1 a + 3 d_2 a^2 + \ldots ].
\equn\label{tbetaeqn}
$$
Comparing this expression with that in \eqn{betadef}, after substituting
for $\ta$ using \eqn{couprel}, we have
$$
\eqalign{
& [ b_0 a^2 + b_1 a^3 + b_2 a^4 + \ldots ] \times
[ 1 + 2d_1 a + 3 d_2 a^2 + \ldots ]  \cr
& = \tb_0 a^2 [1 + d_1 a + d_2 a^2 + \ldots ]^2
  + \tb_1 a^3 [1 + d_1 a + \ldots ]^3
  + \tb_2 a^4 [1 + \ldots ]^4 \cr } \,.
\equn\label{comparebeta}
$$
Equating coefficients of $a$ gives the standard results that $\tb_0=b_0$
and $\tb_1=b_1$, and also
$$
\tb_2 = b_2 - d_1^2 b_0 - d_1 b_1 + d_2 b_0.
\equn\label{shiftbeta2}
$$
This gives the shift in the three-loop $\beta$-function coefficient in
terms of the one- and two-loop coupling constant shifts $d_1$ and $d_2$.

To compute the three-loop $\beta$-function coefficient in $\DRbar$ scheme,
we now use \eqn{shiftbeta2}, in conjunction with the 
well-known three-loop $\beta$ function in 
$\MSbar$~\cite{MSbarb0,MSbarb1,MSbarb2,PGJ},
$$
\eqalign{
  b_0 &=   {11\over6} C_A - {2\over3} T_F \,, \cr
  b_1 &=   {17\over6} C_A^2 - {5\over3} C_A T_F -  C_F T_F \,, \cr
  b_2 &=   {2857\over432} C_A^3 - {1415\over216} C_A^2 T_F
         - {205\over72} C_A C_F T_F + {1\over4} C_F^2 T_F
         + {79\over108} C_A T_F^2 + {11\over18} C_F T_F^2 \,.  \cr}
\equn\label{betadata}
$$
Hence the three-loop $\beta$-function coefficient in $\DRbar$ is
$$
\eqalign{
\tb_2 &=   {3115\over432} C_A^3 - {1439\over216} C_A^2 T_F
         - {259\over72} C_A C_F T_F + {1\over4} C_F^2 T_F
         + {79\over108} C_A T_F^2 + {17\over18} C_F T_F^2  \,. \cr}
\equn\label{shiftbeta2qcd}
$$
For QCD, $SU(3)$ with $\Nf$ flavors of quarks, the result is
$$
\eqalign{
\tb_2 &=   {3115\over16} - {5321\over144} \Nf + {373\over432} \Nf^2 \,. \cr}
\equn\label{shiftbeta2realqcd}
$$
For a generic number of flavors, this value is significantly closer to
the $\MSbar$ value than are the MOM versions of
$b_2$ reported in refs.~\cite{ChetSeid,ChetRetey}, although for five
flavors the value in the scheme labeled MOMggg in 
ref.~\cite{ChetSeid} happens to be the closest to
$\MSbar$.  The result~(\ref{shiftbeta2qcd}) should be obtainable as an
extension of the work of ref.~\cite{PGJ}.  In that paper, three-loop
$\beta$ functions were calculated for a variety of theories, in a way
that could be applied to both $\MSbar$ and $\DRbar$ schemes, but only
the $\MSbar$ results were presented.

For the special case of pure super-Yang-Mills theory, with a single
Majorana gluino, we set $T_F = C_A/2$, $C_F = C_A$, to obtain
$$
\eqalign{
 b_0 &= {3\over 2} C_A \, ,  \hskip 1.3 cm b_1 = {3\over 2} C_A^2 \, , \cr
   b_2 &= {19\over 8} C_A^3 \,, \hskip 1 cm \tb_2 = {21\over 8} C_A^3 \,,\cr}
\equn\label{puresym}
$$
in agreement with ref.~\cite{AvdeevTarasov}.
(The $\MSbar$ value, $b_2$, was first calculated in ref.~\cite{MSbarb2}.)

\subsection{Scheme dependence of the primitive amplitudes}
\label{SchemeDepSubsection}

Since our expressions for the amplitudes in \sect{AmplitudesSection}
smoothly interpolate between the FDH ($D_s = 4 $) and HV ($D_s = 4 -2
\eps$) schemes we obtain the scheme dependence of
the amplitudes by subtracting the two forms.
First consider the pure glue amplitude. From
\eqns{PlanarPrimitiveGlue}{NonPlanarPrimitiveGlue},
that the difference between the unrenormalized pure glue
primitive amplitudes in the two schemes is:
$$
\eqalign{
\delta A^\P_{G1234} \equiv
A^{\P, \rm FDH}_{G1234}
- A^{\P, \rm HV}_{G1234}
& = - 4  \, {c_\Gamma \over \e} 
\biggl({\mu^2 \over - s} \biggr)^{\e} \, {\rho \over i} \,
 \I_4^\oneloop[\lambda_p^4] + \Ord(\e)  \cr
& =  -2  \, {c_\Gamma \over \e} 
\biggl({\mu^2 \over - s} \biggr)^{\e}
A_{4}^{\rm gluon \ loop, FDH}(1_g^+,2_g^+,3_g^+,4_g^+) + \Ord(\e)
\,, \cr} \equn
$$
where we used eq.~(A.11) of ref.~\cite{AllPlusTwo} to obtain the
first line and \eqn{OneLoopAllPlusAmplitudes} to obtain the second.
Similarly, for the non-planar pure glue primitive amplitudes
$$
\delta A^\NP_{G1234} \equiv
A^{\NP, \rm FDH}_{G1234}
- A^{\NP, \rm HV}_{G1234} =
-
{ c_\Gamma \over \e}  \biggl({\mu^2 \over - s} \biggr)^{\e}
 A_{4}^{\rm gluon\ loop,FDH}(1_g^+,2_g^+,3_g^+,4_g^+) + \Ord(\e) \, , \equn
$$
where we used eq.~(A.14) of ref.~\cite{AllPlusTwo}.
In these expressions, the coupling
associated with the HV amplitude is the standard $\MSbar$ one, and the
coupling associated with the FDH amplitude is the $\DRbar$ one.  This
shift does not include the coupling shift (\ref{couplingrel}) or the
shift in the counterterm, which are straightforward to incorporate.  
For the remaining cases with fermion,
scalar, or mixed loops we find that the differences are all of
$\Ord(\e)$.  For example,
$$
\eqalign{
\delta A^{\P_1}_{F1234} & = - \e \rho
\biggl\{ s_{12} \I_4^\P[\mud_p^2 \mud_q^2 + \mud_p \cdot \mud_q \mud_q^2]
 + 2 \I_4^\bowtie\Bigl[\mud_q^2
 \Bigl( \mud_p \cdot \mud_q  
  + {2\over s_{12}}  \mud_p^2 ((p + q)^2 + s_{12}) \Bigr) \Bigr]
\biggr\}
= \Ord(\e) \,, \cr
\delta A^{\NP_1}_{F1234} & = - \e \rho
  s_{12} \I_4^\NP[\mud_p^2 \mud_q^2 + \mud_p \cdot \mud_q \mud_q^2]
 = \Ord(\e) \,.}
\equn
$$

The simple structure of the primitive amplitude scheme shift is,
of course, not accidental.  It may be understood in terms of the
infrared divergences~\cite{CataniDiv} appearing in the amplitudes. As one
shifts between the schemes the integrands undergo shifts of
$\Ord(\eps)$.  These shifts can contribute only when they
are multiplied by $\eps^{-1}$ divergences. The complete scheme
dependence through $\Ord(\eps^0)$ therefore depends only on the scheme
dependence of the universal divergences.  The scheme shift for
two-loop helicity amplitudes for which the tree amplitudes vanish are 
especially simple, because the divergences are rather like those of
typical one-loop amplitudes. One may also shift the amplitudes of
\sect{AmplitudesSection} to the CDR scheme following the same
strategy used at one loop~\cite{KSTfourparton,CST}. For more general
helicity amplitudes, the scheme shift is more complicated because one
must account for non-trivial scheme dependence in the 
${\bom I}^{(2)}(\e,\mu^2;\{p\})$ function of Catani~\cite{CataniDiv}.
These issues are presented in more detail in ref.~\cite{ggggpaper}.

\section{Conclusions}
\label{ConclusionSection}

In this paper we presented detailed rules for the four-dimensional
helicity (FDH) regularization scheme to higher loops.  The scheme is
designed to preserve the number of bosonic and fermionic states at
their four-dimensional values.  Computationally, it amounts to a 
relatively minor
modification of the 't~Hooft-Veltman~\cite{HV} dimensional
regularization scheme.  As an initial check of the supersymmetric
properties at higher loops, we explicitly evaluated a number of
two-loop examples, illustrating that the scheme preserves the required
supersymmetry Ward identities, while the 't~Hooft-Veltman scheme fails
(as expected) to do so.  The identities checked in this paper were
relatively simple to evaluate directly, because they involved 
identical-helicity four-gluon amplitudes which have an especially simple
analytic structure.  Although we did not present it here, we
verified that for $N=1$ super-Yang-Mills theory, the $-$+++ helicity
amplitude~\cite{ggggpaper} also vanishes in the
FDH scheme, in accordance with supersymmetry.  It would be interesting
to systematically investigate the supersymmetric properties of
the remaining helicity amplitudes.  In these cases the Ward identities relate 
amplitudes with four external gluons to amplitudes with external fermions
as well; the former have been computed~\cite{ggggpaper} but the latter
have not.

By keeping track of a parameter $\delta_R$ which smoothly interpolates 
between the 't~Hooft-Veltman and four-dimensional helicity schemes, one can
conveniently verify that supersymmetry Ward Identities hold even in
non-supersymmetric theories such as QCD --- after modifying
color factors and multiplicities so that states fall into supermultiplets.  
This approach is of some interest in non-trivial QCD calculations, 
where it can serve as a cross check.  Indeed,
the fact that the two-loop identical-helicity amplitude presented here 
satisfies the required supersymmetry Ward identities provides one additional
check on the four-gluon QCD amplitudes of refs.~\cite{GOTYgggg,ggggpaper}.
We mention that the more complicated helicity amplitudes in 
ref.~\cite{ggggpaper} were decomposed into supersymmetric and 
non-supersymmetric pieces.  The FDH scheme was applied to the 
supersymmetric pieces in order to preserve their supersymmetry;
the same functions will thus appear in a supersymmetric decomposition
of the two-quark-two-gluon and four-quark amplitudes.

In this paper we also evaluated the two-loop relation between the standard
$\MSbar$ coupling and the couplings in the four-dimensional helicity
and dimensional reduction schemes.  It turns out that the couplings in
the latter two schemes are identical, at least in the case of a single
gauge coupling.  The two-loop coupling shift was then used to obtain
the three-loop gauge theory $\beta$ function in the dimensional
reduction and four-dimensional helicity schemes, using the previously
calculated~\cite{MSbarb0,MSbarb1,MSbarb2} $\MSbar$ $\beta$-function.

It would be interesting to investigate the supersymmetry-preserving
properties of the FDH scheme more generally, including amplitudes with
external fermions.  Some open issues are whether the FDH scheme
continues to preserve supersymmetry beyond two loops, and whether
there are any subtleties analogous to the evanescent couplings that
appear in the dimensional reduction scheme.

\vskip .4 cm

\noindent{\bf Acknowledgments} We thank David Kosower for
collaboration at early stages of this work. We also thank Adrian
Ghinculov, Tim Jones and Steve Martin for helpful discussions.
Z.B. thanks SLAC, and L.D. thanks UCLA, for hospitality during the paper's
completion.



\newpage 
\begin{thebibliography}{99}

\bibitem{CDR}
J.C.~Collins, {\it Renormalization: an introduction to the
renormalization group, and the operator-product expansion},
Cambridge Univ. Press, 1984 (Cambridge Monographs on Mathematical
Physics).

\bibitem{HV}
G.~'t Hooft and M.~Veltman,
Nucl.\ Phys.\ {\bf B44}, 189 (1972).

\bibitem{TreeHelicity}
F.A.~Berends, R.~Kleiss, P.~De Causmaecker, R.~Gastmans and
T.T.~Wu,
Phys.\ Lett.\ B {\bf 103}, 124 (1981);\\
%
P.~De Causmaecker, R.~Gastmans, W.~Troost and T.T.~Wu,
Phys.\ Lett.\ B {\bf 105}, 215 (1981);\\
%
R.~Kleiss and W.J.~Stirling,
Nucl.\ Phys.\ B {\bf 262}, 235 (1985);\\
%
J.F.~Gunion and Z.~Kunszt,
Phys.\ Lett.\ B {\bf 161}, 333 (1985);\\
%
Z.~Xu, D.~Zhang and L.~Chang,
Nucl.\ Phys.\ B {\bf 291}, 392 (1987).

\bibitem{ManganoReview}
M.L.~Mangano and S.J.~Parke,
Phys.\ Rept.\ {\bf 200}, 301 (1991);\\
L.J.~Dixon,
in {\it Proceedings of Theoretical Advanced Study Institute in
Elementary Particle Physics (TASI 95)}, ed.\ D.E.\ Soper, 
World Scientific, 1996
[arXiv:hep-ph/9601359].

\bibitem{BKgggg}
Z.~Bern and D.A.~Kosower,
Nucl.\ Phys.\ {\bf B379}, 451 (1992).

\bibitem{Review}
Z.~Bern, L.J.~Dixon and D.A.~Kosower,
Ann.\ Rev.\ Nucl.\ Part.\ Sci.\ {\bf 46}, 109 (1996)
[arXiv:hep-ph/9602280].

\bibitem{DR}
W.~Siegel,
Phys.\ Lett.\  {\bf B84}, 193 (1979);\\
%
D.~M.~Capper, D.R.T.~Jones and P.~van Nieuwenhuizen,
Nucl.\ Phys.\  {\bf B167}, 479 (1980);\\
%
I.~Jack, D.R.T.~Jones and K.L.~Roberts,
Z.\ Phys.\  {\bf C63}, 151 (1994)
[arXiv:hep-ph/9401349].

\bibitem{NonSusyDimRed}
G.~Altarelli, G.~Curci, G.~Martinelli and S.~Petrarca,
Nucl.\ Phys.\ B {\bf 187}, 461 (1981).

\bibitem{JJR}
I.~Jack, D.R.T.~Jones and K.L.~Roberts,
Z.\ Phys.\ C {\bf 62}, 161 (1994)
[arXiv:hep-ph/9310301].

\bibitem{NSVZ}
V.A.~Novikov, M.A.~Shifman, A.I.~Vainshtein and V.I.~Zakharov,
Nucl.\ Phys.\ B {\bf 229}, 407 (1983);\\
%
V.A.~Novikov, M.A.~Shifman, A.I.~Vainshtein and V.I.~Zakharov,
Phys.\ Lett.\ B {\bf 166}, 329 (1986)
[Sov.\ J.\ Nucl.\ Phys.\  {\bf 43}, 294 (1986); Yad.\ Fiz.\ {\bf 43},
459 (1986)];\\
%
M.A.~Shifman and A.I.~Vainshtein,
Nucl.\ Phys.\ B {\bf 277}, 456 (1986)
[Sov.\ Phys.\ JETP {\bf 64}, 428 (1986)];\\
%
I.~Jack, D.R.T.~Jones and C.G.~North,
Nucl.\ Phys.\ B {\bf 486}, 479 (1997)
[arXiv:hep-ph/9609325];\\
%
I.~Jack, D.R.T.~Jones and A.~Pickering,
Phys.\ Lett.\ B {\bf 435}, 61 (1998)
[arXiv:hep-ph/9805482].

\bibitem{KSTfourparton}
Z.~Kunszt, A.~Signer and Z.~Tr\'ocs\'anyi,
Nucl.\ Phys.\ {\bf B411}, 397 (1994)
[arXiv:hep-ph/9305239].

\bibitem{SWIOld}
M.T.~Grisaru, H.N.~Pendleton and P.~van Nieuwenhuizen,
Phys.\ Rev.\ {\bf D15}, 996 (1977);\\
%
M.T.~Grisaru and H.N.~Pendleton,
Nucl.\ Phys.\ {\bf B124}, 81 (1977).

\bibitem{SWINew}
S.J.~Parke and T.R.~Taylor,
Phys.\ Lett.\ {\bf B157}, 81 (1985),
err. {\it ibid.} {\bf B174}, 465 (1986).

\bibitem{BRY}
Z.~Bern, J.S.~Rozowsky and B.~Yan,
Phys.\ Lett.\ {\bf B401}, 273 (1997)
[arXiv:hep-ph/9702424];

\bibitem{AllPlusTwo}
Z.~Bern, L.J.~Dixon and D.A.~Kosower,
JHEP {\bf 0001}, 027 (2000)
[arXiv:hep-ph/0001001].

\bibitem{Bhabha}
Z.~Bern, L.J.~Dixon and A.~Ghinculov,
Phys.\ Rev.\ D {\bf 63}, 053007 (2001) [arXiv:hep-ph/0010075].

\bibitem{GOTY2to2}
C.~Anastasiou, E.W.N.~Glover, C.~Oleari and M.E.~Tejeda-Yeomans,
Nucl.\ Phys.\ B {\bf 601}, 318 (2001) [arXiv:hep-ph/0010212];
%
Nucl.\ Phys.\ B {\bf 601}, 341 (2001) [arXiv:hep-ph/0011094];
%
Nucl.\ Phys.\ B {\bf 605}, 486 (2001) [arXiv:hep-ph/0101304].

\bibitem{GOTYgggg}
E.W.N.~Glover, C.~Oleari and M.E.~Tejeda-Yeomans,
Nucl.\ Phys.\ B {\bf 605}, 467 (2001)
[arXiv:hep-ph/0102201].

\bibitem{ggggpaper}
Z.~Bern, A.~De Freitas and L.J.~Dixon,
preprint arXiv:hep-ph/0201161.

\bibitem{GGGamGam}
Z.~Bern, A.~De~Freitas and L.J.~Dixon,
JHEP {\bf 0109}, 037 (2001) [arXiv:hep-ph/0109078].

\bibitem{Lbyl}
Z.~Bern, A.~De~Freitas, L.J.~Dixon, A.~Ghinculov and H.L.~Wong,
JHEP {\bf 0111}, 031 (2001) [arXiv:hep-ph/0109079].

\bibitem{PBScalar}
V.A.~Smirnov,
Phys.\ Lett.\  {\bf B460}, 397 (1999)
[arXiv:hep-ph/9905323].

\bibitem{NPBScalar}
J.B.~Tausk,
Phys.\ Lett.\  {\bf B469}, 225 (1999)
[arXiv:hep-ph/9909506].

\bibitem{PBReduction}
V.A.~Smirnov and O.L.~Veretin,
Nucl.\ Phys.\  {\bf B566}, 469 (2000)
[arXiv:hep-ph/9907385].

\bibitem{NPBReduction}
C.~Anastasiou, T.~Gehrmann, C.~Oleari, E.~Remiddi and J.B.~Tausk,
Nucl.\ Phys.\  {\bf B580}, 577 (2000)
[arXiv:hep-ph/0003261].

\bibitem{GRReduction}
T.~Gehrmann and E.~Remiddi,
Nucl.\ Phys.\  {\bf B580}, 485 (2000)
[arXiv:hep-ph/9912329].

\bibitem{AGO}
C.~Anastasiou, E.W.N.~Glover and C.~Oleari,
Nucl.\ Phys.\  {\bf B565}, 445 (2000) [arXiv:hep-ph/9907523];
%
Nucl.\ Phys.\  {\bf B575}, 416 (2000)
[arXiv:hep-ph/9912251].

\bibitem{TwoloopOneMassIntegrals}
V.A.~Smirnov,
Phys.\ Lett.\ B {\bf 491}, 130 (2000) [arXiv:hep-ph/0007032];\\
%
T.~Gehrmann and E.~Remiddi,
Nucl.\ Phys.\ B {\bf 601}, 248 (2001) [arXiv:hep-ph/0008287];
%
Nucl.\ Phys.\ B {\bf 601}, 287 (2001) [arXiv:hep-ph/0101124];
%
Comput.\ Phys.\ Commun.\  {\bf 141}, 296 (2001)
[arXiv:hep-ph/0107173];
%
arXiv:hep-ph/0111255.

\bibitem{TwoLoopee3Jets}
L.W.~Garland, T.~Gehrmann, E.W.N.~Glover, A.~Koukoutsakis and
E.~Remiddi,
arXiv:hep-ph/0112081.

\bibitem{CST}
S.~Catani, M.H.~Seymour and Z.~Tr\'ocs\'anyi,
Phys.\ Rev.\ {\bf D55}, 6819 (1997)
[arXiv:hep-ph/9610553].

\bibitem{AKT}
I.~Antoniadis, C.~Kounnas and K.~Tamvakis,
Phys.\ Lett.\ B {\bf 119}, 377 (1982).

\bibitem{MartinVaughn}
S.P.~Martin and M.T.~Vaughn,
Phys.\ Lett.\ B {\bf 318}, 331 (1993) [arXiv:hep-ph/9308222].

\bibitem{MSbarb2}
O.V.~Tarasov, A.A.~Vladimirov and A.Y.~Zharkov,
Phys.\ Lett.\ B {\bf 93}, 429 (1980);\\
%
S.A.~Larin and J.A.M.~Vermaseren, 
Phys.\ Lett.\ B {\bf 303}, 334 (1993) [arXiv:hep-ph/9302208].

\bibitem{TasiZvi}
Z.~Bern,
in {\it Proceedings of Theoretical Advanced Study Institute in
High Energy Physics (TASI 92)}, eds.\ J. Harvey and J. Polchinski,
World Scientific, 1993 [arXiv:hep-ph/9304249].

\bibitem{AllNSusy}
Z.~Bern, L.J.~Dixon, D.C.~Dunbar and D.A.~Kosower,
Nucl.\ Phys.\ {\bf B425}, 217 (1994) [arXiv:hep-ph/9403226];\\
%
Nucl.\ Phys.\ {\bf B435}, 59 (1995)
[arXiv:hep-ph/9409265].

\bibitem{SuperLagrangian}
S.~Ferrara and B.~Zumino,
Nucl.\ Phys.\ B {\bf 79}, 413 (1974);\\
%
M.F.~Sohnius,
Phys.\ Rept.\  {\bf 128}, 39 (1985);\\
%
A.~Bilal,
arXiv:hep-th/0101055.

\bibitem{WessBagger}
J.~Wess and J.~Bagger, {\it Supersymmetry and Supergravity},
second edition, Princeton Univ. Press, 1992 (Princeton Series in Physics).

\bibitem{Superspace}
S.J.~Gates, Jr., M.T.~Grisaru, M.~Rocek and W.~Siegel, {\it
Superspace: or one thousand and one lessons in supersymmetry},
Benjamin/Cummings, 1983 (Frontiers in Physics, 58).

\bibitem{FiveParton}
Z.~Bern, L.J.~Dixon and D.A.~Kosower,
Phys.\ Rev.\ Lett.\  {\bf 70}, 2677 (1993)
[arXiv:hep-ph/9302280];\\
Z.~Kunszt, A.~Signer and Z.~Tr\'ocs\'anyi,
Phys.\ Lett.\ B {\bf 336}, 529 (1994)
[arXiv:hep-ph/9405386];\\
Z.~Bern, L.J.~Dixon and D.A.~Kosower,
Nucl.\ Phys.\ B {\bf 437}, 259 (1995)
[arXiv:hep-ph/9409393].

\bibitem{Z4Partons}
Z.~Bern, L.J.~Dixon and D.A.~Kosower,
Nucl.\ Phys.\  {\bf B513}, 3 (1998)
[arXiv:hep-ph/9708239].

\bibitem{BernMorgan}
Z.~Bern and A.G.~Morgan,
Nucl.\ Phys.\ {\bf B467}, 479 (1996)
[arXiv:hep-ph/9511336].

\bibitem{OtherSewing}
Z.~Bern, L.J.~Dixon, M.~Perelstein and J.S.~Rozowsky,
Phys.\ Lett.\ {\bf B444}, 273 (1998)
[arXiv:hep-th/9809160];
%
Nucl.\ Phys.\ {\bf B546}, 423 (1999)
[arXiv:hep-th/9811140].

\bibitem{N8Susy}
Z.~Bern, L.J.~Dixon, D.C.~Dunbar, M.~Perelstein and J.S.~Rozowsky,
Nucl.\ Phys.\ {\bf B530}, 401 (1998)
[arXiv:hep-th/9802162].

\bibitem{OldCutting}
L.D.~Landau, Nucl.\ Phys.\ {\bf 13}, 181 (1959);\\
S.~Mandelstam, Phys.\ Rev.\ {\bf 112}, 1344 (1958),
{\bf 115}, 1741 (1959);\\
R.E.~Cutkosky, J.\ Math.\ Phys.\ {\bf 1}, 429 (1960).

\bibitem{TreeColor}
J.E.~Paton and H.~Chan,
Nucl.\ Phys.\ B {\bf 10}, 516 (1969);\\
%
P.~Cvitanovi\'c, P.G.~Lauwers and P.N.~Scharbach,
Nucl.\ Phys.\ B {\bf 186}, 165 (1981);\\
%
M.~Mangano, S.~Parke and Z.~Xu,
Nucl.\ Phys.\ B {\bf 298}, 653 (1988);\\
%
M.~Mangano,
Nucl.\ Phys.\ B {\bf 309}, 461 (1988).

\bibitem{OneloopColor}
Z.~Bern and D.A.~Kosower,
Nucl.\ Phys.\ B {\bf 362}, 389 (1991).

\bibitem{DDFColor}
V.~Del Duca, L.J.~Dixon and F.~Maltoni,
Nucl.\ Phys.\ B {\bf 571}, 51 (2000)
[arXiv:hep-ph/9910563].

\bibitem{OneLoopAllPlus}
G.~Mahlon,
Phys.\ Rev.\ {\bf D49}, 4438 (1994)
[arXiv:hep-ph/9312276];
%
Z.~Bern, G.~Chalmers, L.J.~Dixon and D.A.~Kosower,
Phys.\ Rev.\ Lett.\ {\bf 72}, 2134 (1994)
[arXiv:hep-ph/9312333];\\
%
Z.~Bern, L.J.~Dixon, D.C.~Dunbar and D.A.~Kosower,
Phys.\ Lett.\ B {\bf 394}, 105 (1997)
[arXiv:hep-th/9611127].

\bibitem{Lewin}
L.~Lewin, {\it Dilogarithms and Associated Functions},
Macdonald, London, 1958.

\bibitem{CataniDiv}
S.~Catani,
Phys.\ Lett.\ {\bf B427}, 161 (1998)
[arXiv:hep-ph/9802439].

\bibitem{Background}
B.S.~DeWitt, Phys. Rev. {\bf 162}, 1195 (1967);\\
G. 't Hooft, in {\it Acta Universitatis Wratislavensis no.\
38}, 12th Winter School of Theoretical Physics in Karpacz, {\it
Functional and Probabilistic Methods in Quantum Field Theory},
Vol. 1 Wroclaw University Press (1975); 
B.S.\ DeWitt, in {\it Quantum gravity II}, eds. C. Isham, R.\ Penrose and
D.\ Sciama, Oxford, 1981.

\bibitem{AbbottPureGlue}
L.F.~Abbott,
Nucl.\ Phys.\ B {\bf 185}, 189 (1981).

\bibitem{AbbottFermion}
L.F.~Abbott, M.T.~Grisaru and R.K.~Schaefer,
Nucl.\ Phys.\ B {\bf 229}, 372 (1983).

\bibitem{PGJ}
A.G.~Pickering, J.A.~Gracey and D.R.T.~Jones,
Phys.\ Lett.\ B {\bf 510}, 347 (2001) [Phys.\ Lett.\ B {\bf 512},
230 (2001)] [arXiv:hep-ph/0104247].

\bibitem{ChetRetey}
K.G.~Chetyrkin and A.~Retey,
arXiv:hep-ph/0007088.

\bibitem{ChetSeid}
P.~Boucaud, J.P.~Leroy, J.~Micheli, O.~Pene and C.~Roiesnel,
JHEP {\bf 9812}, 004 (1998)
[arXiv:hep-ph/9810437]; \\
K.G.~Chetyrkin and T.~Seidensticker,
Phys.\ Lett.\ B {\bf 495}, 74 (2000)
[arXiv:hep-ph/0008094].

\bibitem{Latticed2}
M.~L\"uscher and P.~Weisz,
Nucl.\ Phys.\ B {\bf 452}, 234 (1995)
[arXiv:hep-lat/9505011]; \\
C.~Christou, A.~Feo, H.~Panagopoulos and E.~Vicari,
Nucl.\ Phys.\ B {\bf 525}, 387 (1998)
[arXiv:hep-lat/9801007],
err. {\it ibid.} {\bf B608}, 479 (2001);\\
A.~Bode and H.~Panagopoulos,
Nucl.\ Phys.\ B {\bf 625}, 198 (2002)
[arXiv:hep-lat/0110211].

\bibitem{MSbarb0}
D.J.~Gross and F.~Wilczek,
Phys.\ Rev.\ Lett.\  {\bf 30}, 1343 (1973);\\
H.D.~Politzer,
Phys.\ Rev.\ Lett.\  {\bf 30}, 1346 (1973).

\bibitem{MSbarb1}
W.E.~Caswell,
Phys.\ Rev.\ Lett.\  {\bf 33}, 244 (1974);\\
D.R.T.~Jones,
Nucl.\ Phys.\ B {\bf 75}, 531 (1974);\\
%
E.S.~Egorian and O.V.~Tarasov,
Theor.\ Math.\ Phys.\  {\bf 41}, 863 (1979)
[Teor.\ Mat.\ Fiz.\  {\bf 41}, 26 (1979)].

\bibitem{AvdeevTarasov}
L.V.~Avdeev and O.V.~Tarasov,
Phys.\ Lett.\ B {\bf 112}, 356 (1982).

\end{thebibliography}
\end{document}